\documentclass[a4paper,11pt, openany]{book} %Remove draft option to hide ToDos

\pdfoutput=1

\def\tikzitcache{1}

\include{header/header}
\newcommand{\setauthor}[1]{\author{#1}\def\theauthor{#1}}
\newcommand{\settitle}[1]{\title{#1}\def\thetitle{#1}}

\settitle{Quantum Machine Learning using the ZXW-Calculus}
\setauthor{Mark Koch}

\DeclareTextFontCommand{\emph}{\bfseries}

\fancypagestyle{plain}{
	\fancyhf{}

}

\makeglossaries

\begin{document}

\renewcommand\cite{\citep}

\thispagestyle{empty}

%\begin{titlepage}
%	\begin{center}
%		{\fontfamily{qag}\selectfont\fontsize{40}{40}\selectfont Quantum Machine \\[10pt] Learning }
%		\\[25pt]
%		{\fontfamily{qag}\selectfont \fontsize{20}{40}\selectfont using the} 
%		\\[5pt]
%		{\usefont{T1}{cambridge}{m}{n}\fontsize{110}{40}\selectfont z}%
%		{\usefont{T1}{titlefont}{m}{n}\fontsize{50}{40}\selectfont XW-Calculus}
%		\\[1.5cm]
%		\includegraphics[width=11cm,draft=false]{logo}			
%		\\[1.5cm]
%		
%		\fontfamily{ppl}\selectfont
%		\large
%	
%		{\LARGE Candidate no. 1064973}
%		
%		\vspace{1cm}
%		
%		A thesis submitted for the degree of 
%		\\[10pt]
%		{\textit{M.Sc. in Advanced Computer Science}}
%		\\[1cm]
%		Trinity 2022
%		
%	\end{center}
%\end{titlepage}

\begin{titlepage}
	\begin{center}
		\vspace*{-3ex}
		{ \LARGE \bfseries {Quantum Machine Learning \\ using the ZXW-Calculus} \par}
		{\large \vspace*{25mm} {\includegraphics[draft=false]{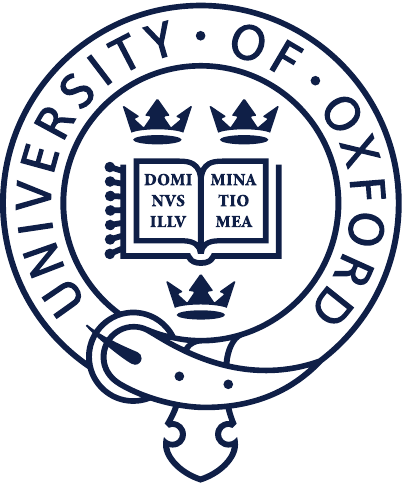} \par} \vspace*{20mm}}
		{{\Large Mark Koch} \par}
		\vspace*{0.5ex}
		{Lady Margaret Hall \par}
		{University of Oxford \par}
		\vspace*{30mm}
		{{A thesis submitted for the degree of} \par}
		%\vspace*{0.5ex}
		{\it {Master of Science in Advanced Computer Science} \par}
		\vspace*{3ex}
		{Trinity 2022}
	\end{center}
\end{titlepage}

% perl ../texcount/texcount.pl -1 chapter1/chapter1.tex chapter2/chapter2.tex chapter3/chapter3.tex chapter4/chapter4.tex chapter5/chapter5.tex chapter6/chapter6.tex appendix-w/appendix-w.tex appendix-contraction/appendix-contraction.tex appendix-proofs/appendix-proofs.tex

\thispagestyle{empty}
\vspace*{\fill}
\begin{alignat*}{3}
	&\text{Word count:} &\quad& 15,793
	\\[5pt]
	&\text{Diagram count:} &\quad& 806 \\[5pt]
\end{alignat*}
\begin{adjustwidth*}{1.25cm}{1.25cm}
	The word count was calculated using \texttt{texcount} via \texttt{perl texcount.pl -1 thesis.tex}.
	Note that in diagram equations, each step is counted as a separate diagram.
\end{adjustwidth*}
\vspace*{\fill}

\doublespacing
\raggedbottom  % Don't strech unfilled pages to the bottom

%Word Count: 10000
%Diagram Count: 500

\chapter*{Abstract}
\addcontentsline{toc}{chapter}{\numberline{}Abstract}

The field of \textit{quantum machine learning} (QML) explores how quantum computers can be used to more efficiently solve machine learning problems.
As an application of hybrid quantum-classical algorithms, it promises a potential quantum advantages in the near term.
In this thesis, we use the ZXW-calculus to \textit{diagrammatically} analyse two key problems that QML applications face.

First, we discuss algorithms to compute gradients on quantum hardware that are needed to perform gradient-based optimisation for QML.
Concretely, we give new diagrammatic proofs of the common 2- and 4-term parameter shift rules used in the literature.
Additionally, we derive a novel, generalised parameter shift rule with $2n$ terms that is applicable to gates that can be represented with $n$ parametrised spiders in the ZXW-calculus.
Furthermore, to the best of our knowledge, we give the first proof of a conjecture by Anselmetti et al. by proving a no-go theorem ruling out more efficient alternatives to the 4-term shift rule.

Secondly, we analyse the gradient landscape of quantum ansätze for barren plateaus using both empirical and analytical techniques.
Concretely, we develop a tool that automatically calculates the variance of gradients and use it to detect likely barren plateaus in commonly used quantum ansätze.
Furthermore, we formally prove the existence or absence of barren plateaus for a selection of ansätze using diagrammatic techniques from the ZXW-calculus.

\newpage

\chapter*{Acknowledgements}
\addcontentsline{toc}{chapter}{\numberline{}Acknowledgements}

First and foremost, I would like to thank my advisors Quanlong Wang and Richie Yeung for their invaluable support and guidance throughout the writing of this thesis. 
I am very grateful for their advice and many helpful discussions and ideas.
I would also like to thank Aleks Kissinger, as well as John van de Wetering and Stephano Gogioso for sparking my interest in quantum computing and the ZX-calculus through their lectures.
In particular, I am thankful for the opportunity to write this thesis under Aleks' supervision.

Furthermore, I am very grateful to my family and friends both in Germany and Oxford, who supported me throughout my studies.
Together with the academic community at my wonderful college Lady Margaret Hall, they provided a great intellectual atmosphere that made the past year a truly unique experience.
In particular, I would like to thank Nikhil Khatri for many inspiring discussions and for proofreading this thesis.

Finally, I would like to thank the German Academic Exchange Service (DAAD) for financially supporting me during this year at Oxford.

\addtocontents{toc}{\vspace{5mm}}

\tableofcontents

%\listoftodos

\clearpage
\pagenumbering{arabic}

\chapter{Introduction}

It is widely believed that quantum computers are capable of solving certain computational problems that are intractable for classical computers.
While this potential quantum advantage was already recognised in the 1980s, the quantum devices available today still lack the scale and reliability to tackle many practical problems, with anticipated algorithms like Grover's search~\cite{grover1996fast} or Shor’s factorisation algorithm~\cite{shor1999polynomial} remaining out of reach.
Because of those limitations, there is increasing interest in \textit{hybrid quantum-classical algorithms}.
The rationale behind hybrid approaches is that the required quantum resources can be significantly reduced by implementing some subroutines on classical hardware.
As a result, those algorithms are runnable on the \textit{noisy intermediate-scale quantum} (NISQ) devices available today.
%, and promise a quantum advantage in the near term.

One area where hybrid algorithms promise a quantum advantage is the field of \textit{machine learning} (ML).
Roughly, ML is concerned with recognising and generalising patterns in statistical data.
It has been shown that even relatively small quantum circuits can represent functions that are highly complex and difficult to express via classical means~\cite{bremner2017achieving}.
Hence, the hope is that quantum computers can capture certain data patterns more efficiently than classical computers, yielding a quantum advantage in ML.
\enlargethispage{\baselineskip}
This field of study is commonly referred to as \textit{quantum machine learning} (QML)~\cite{biamonte2017quantum}.

Typically, hybrid QML algorithms rely on \textit{parametrised quantum circuits}, i.e. circuits that depend on some tunable parameters.
An optimisation algorithm running on a classical computer is used to find a parameter assignment such that the output of the quantum circuit minimises some cost function.
For example, circuits can be trained to solve ML tasks like classification, regression, or generative modelling.
There are many classical optimisation techniques that can be used to train quantum circuits.
In the field of QML, one commonly uses gradient-based techniques like gradient descent, which have already been very successfully used in classical ML, especially for the training of neural networks.
Notably, gradient-based methods have also been proven to improve convergence in the quantum domain~\cite{harrow2021low}.
However, compared to classical neural networks, training quantum circuits using gradient descent comes with a set of unique challenges.

First, one has to determine the gradient of parametrised circuits, i.e. compute how the output of a circuit changes when the parameters are altered.
As it turns out, it is not feasible to perform this computation classically.
Instead, gradients need to be evaluated on the quantum device itself.
The quantum algorithms used for those gradient computations are called \textit{gradient recipes} and are subject to a lot of research interest~\cite{mitarai2018quantum,schuld2019evaluating,anselmetti2021local,wierichs2022general}.
Secondly, it has been shown that the gradient landscape of many quantum circuits is not amenable to learning.
Concretely, the landscape is often exponentially flat~\cite{mcclean2018barren}, making gradient descent difficult or even impossible.
Naturally, there is a lot of interest in determining which circuits exhibit those so-called \textit{barren plateaus}~\cite{holmes2022connecting,cerezo2021cost,zhao2021analyzing}.

This thesis is concerned with analysing both of these problems using \textit{diagrammatic} means.
The ZX-calculus~\cite{coecke2008interacting} is a graphical language for reasoning about quantum computation that has been successfully applied to a wide range of tasks in the quantum domain, including circuit optimisation~\cite{duncan2020graph}, compilation~\cite{de2020architecture}, and simulation~\cite{kissinger2022simulating}.
The ZXW-calculus~\cite{shaikh2022sum} is a variant of ZX that has recently been used to diagrammatically represent gradients and integrals~\cite{wang2022differentiating}.
Thus, it is particularly well-suited for our diagrammatic analysis of gradient based optimisation for QML.
\enlargethispage{\baselineskip}

\section{Main Contributions}

Below are the main contributions of this thesis with regard to gradient recipes:
\begin{itemize}
	\item 
	We derive a simplified version of Wang and Yeung's diagrammatic differentiation~\cite{wang2022differentiating} for the special case of parametrised circuits (\Cref{thm:circ-diff}).
	
	\item
	We give a diagrammatic proof of the most general version of Schuld et al.'s~\cite{schuld2019evaluating} two-term parameter shift rule (\Cref{thm:shift-2}) and Anselmetti et al.'s~\cite{anselmetti2021local} four-term shift rule (\Cref{thm:shift-4}).
	
	\item
	We derive a novel generalised $2n$-term shift rule for gates that can be represented with $n$ parametrised spiders (\Cref{thm:shift-general})
	
	\item
	To the best of our knowledge, we give the first proof of a conjecture by Anselmetti et al.~\cite{anselmetti2021local} showing that their shift rule is optimal.
	Concretely, we prove a no-go theorem ruling out shift rules with less than four terms for all gates whose Hermitian generators have eigenvalues of shape $-\lambda,0,\lambda$ (\Cref{thm:nogo}).
\end{itemize}

On the topic of barren plateaus we make the following contributions:

\begin{itemize}
	\item
	We develop a tool that automatically computes $\text{Var}\left({\frac{\partial \langle H\rangle}{\partial \theta_i}}\right)$ and use it to empirically show that barren plateaus likely appear in 7 ansätze studied by Sim et al.~\cite{sim2019expressibility} when measuring in the computational basis (Figures \ref{fig:sim-single-layer} and \ref{fig:sim-single-layer-parameters}).
	
	\item
	We formally prove the existence of barren plateaus in three of the Sim ansätze and give necessary conditions on the measurement Hamiltonian for when they occur (Theorems \ref{fact:sim1-barren}, \ref{fact:sim2-barren}, and \ref{thm:sim9-barren}).
	
	\item
	We give a general framework for the barren plateau analysis of IQP circuits (\Cref{thm:IQP-variance}) and use it to prove that the main circuit used by the quantum natural language processing library \texttt{lambeq}~\cite{kartsaklis2021lambeq} has barren plateaus when measuring in the computational basis (\Cref{thm:iqp4-barren}).
\end{itemize}

\section{Structure of this Thesis}

We begin by discussing some of the background necessary to follow this thesis in \Cref{chap:background} and introduce diagrammatic differentiation in \Cref{chap:diagrammatic-differentiation}.
\Cref{chap:gradient-recipes} is concerned with deriving gradient recipes using this diagrammatic technique.
Subsequently, we study the gradient landscape of parametrised circuit with regard to barren plateaus in \Cref{chap:barren-plateaus}.
Finally, we discuss our results and comment on future work in \Cref{chap:discussion}.

For presentation purposes, we move some of the proofs throughout the thesis to the appendix.
This is remarked on underneath each such lemma.
In the PDF version of this thesis one can easily jump to the corresponding proof by clicking on the arrow symbol ($\downarrow$) on the right-hand side of the page.

The code to reproduce all numerical results and graphs in this thesis is available at
\begin{center}
	\url{https://github.com/mark-koch/msc-code}
\end{center}

\chapter{Background} \label{chap:background}

\def\tikzitpath{chapter2/figs/}

In this chapter we give the necessary background to follow the thesis.
Concretely, we give a brief introduction to quantum theory in \Cref{sec:quantum-intro} and discuss quantum machine learning in \Cref{sec:QML}.
Finally, we introduce the ZXW-calculus in \Cref{sec:zxw-caclulus}.

\section{An Introduction to Quantum Theory} \label{sec:quantum-intro}

\subsection{States}

\begin{figure}
	\begin{center}
		\input{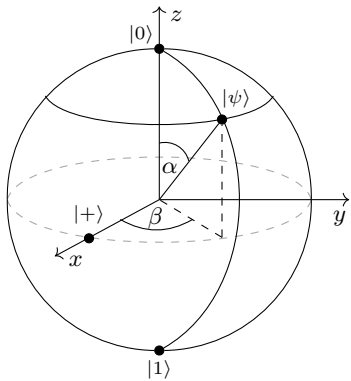}
	\end{center}
	\caption{Visualisation of a qubit state $\ket{\psi} = x\ket{0} + y\ket{1}$ as a point on the Bloch sphere. We have $x = \cos(\frac{\alpha}{2})$ and $y = e^{i\beta}\sin(\frac{\alpha}{2})$.}
	\label{fig:bloch-sphere}
\end{figure}

The states of quantum systems are given by normalised vectors in a complex Hilbert space $\mathcal H$.
We exclusively work within $\mathcal H = \mathbb C^{2^n}$ for this thesis, where states are given by column vectors of complex numbers.
The \textit{adjoint} $\psi^\dagger$ of a state $\psi$ in this case is given by the conjugate-transpose of $\psi$.
States and their adjoints are usually written in the \textit{Dirac bra-ket notation}:
\[ 
	\psi \quad\rightsquigarrow\quad \ket{\psi} 
	\qquad\qquad 
	\psi^\dagger \quad\rightsquigarrow\quad \bra{\psi} 
\]
The symbol $\ket{\psi}$ is called \textit{ket} and $\bra{\psi}$ is called \textit{bra}.
Plugging a bra into a ket yields the inner product of the two vectors which we denote by $\braket{\psi}{\phi} := \bra{\psi}\ket{\phi}$ and call \textit{bra-ket}.
The most elementary state is given by a single quantum bit, or \textit{qubit}, which belongs to the two-dimensional Hilbert space $\mathbb C^2$ spanned by the standard basis
\[ 
	\ket{0} := \begin{pmatrix} 1 \\ 0 \end{pmatrix} 
	\qquad\qquad
	\ket{1} := \begin{pmatrix} 0 \\ 1 \end{pmatrix}.
\]
The states $\ket{0}$ and $\ket{1}$ are the quantum analogues of classical bits.
Therefore, the basis $\{\ket{0}, \ket{1}\}$ is usually called \textit{computational basis}.
However, unlike classical bits, qubits can represent any linear combination of $\ket{0}$ and $\ket{1}$:
\[ \ket{\psi} = x\ket{0} + y\ket{1} \]
for some $x,y\in\mathbb C$ with $|x|^2+|y|^2=1$.
We can picture the state $\ket{\psi}$ as a point on the so-called \textit{Bloch sphere} as illustrated in \Cref{fig:bloch-sphere}.
We refer to those states \enquote{in-between} 0 and 1 as \textit{superpositions}.

In order to unleash the full power of quantum computation, we describe interactions between multiple systems using the \textit{tensor product} operation $\otimes$ corresponding to the Kronecker product.
For example, the two-qubit system $\mathbb C^2 \otimes \mathbb C^2 = \mathbb C^4$ is spanned by the basis
\begin{align*}
	\ket{00} &:= \ket{0} \otimes \ket{0} = (1,0,0,0)^T &
	\ket{10} &:= \ket{0} \otimes \ket{1} = (0,1,0,0)^T \\
	\ket{01} &:= \ket{1} \otimes \ket{0} = (0,0,1,0)^T &
	\ket{11} &:= \ket{1} \otimes \ket{1} = (0,0,0,1)^T
\end{align*}
where $\ket{\psi} \otimes \ket{\phi}$ is the \textit{product state} of $\ket{\psi}$ and $\ket{\phi}$.
We sometimes also write the computational basis vectors for $\mathbb C^{2^n}$ as $\ket{j}$ for $j = 0,1,...,2^{n}-1$.

\subsection{Unitary Evolution}

\begin{definition}
	A square matrix $U$ is unitary if $UU^\dagger = U^\dagger U = I$.
\end{definition}

Computation on a quantum state $\ket{\psi} \in \mathbb C^{2^n}$ is done using \textit{unitary evolutions}, i.e. acting on $\ket{\psi}$ according to a unitary matrix $U \in \mathbb C^{2^n \times 2^n}$.
The resulting state is given by $\ket{\psi'} = U\ket{\psi}$.
An example of a single-qubit action is the Hadamard operation
\begin{equation} \label{eqn:had-def}
	H = \frac{1}{\sqrt 2} \begin{pmatrix} 1 & 1 \\ 1 & -1 \end{pmatrix}
\end{equation}
that maps the computational basis to the so-called \textit{X-basis} $\{\ket{+}, \ket{-}\}$:
\[ 
	H\ket{0} = \ket{+} := \frac{\ket{0} + \ket{1}}{\sqrt 2}
	\qquad\qquad
	H\ket{1} = \ket{-} := \frac{\ket{0} - \ket{1}}{\sqrt 2}
\]
Another example is the single-qubit $R_Z(\alpha)$ operation that corresponds to a Z-rotation on the Bloch sphere by an angle of $\alpha$:
\begin{equation} \label{eqn:RZ-def}
	R_Z(\alpha) := \begin{pmatrix}
		e^{-i\frac{\alpha}{2}} & 0 \\
		0 & e^{i\frac{\alpha}{2}}
	\end{pmatrix}
\end{equation}
$R_Z(\alpha)$ is an example of a parametrised unitary:
\begin{definition}
	A (strongly continuous) one-parameter unitary group is a family $\{U(\alpha)\}_{\alpha\in\mathbb R}$ of single-parameter unitary matrices that are strongly continuous $(\lim\limits_{\alpha\to\alpha_0} U(\alpha) = U(\alpha_0)$ for all $\alpha_0\in\mathbb R)$ and homomorphisms $(U(\alpha+\beta) = U(\alpha)\,U(\beta))$.
\end{definition}
When speaking of (single-)parametrised unitaries, we generally refer to one-parameter unitary groups.

\begin{definition}
	A matrix $H$ is self-adjoint, or Hermitian, if $H^\dagger = H$.
\end{definition}

Remarkably, there is a one-to-one correspondence between single-parameter unitaries and Hermitian operators:

\begin{theorem}[Stone~\cite{stone1932one}] \label{thm:stone}
	Every strongly continuous one-parameter unitary group $\{U(\alpha)\}_{\alpha\in\mathbb R}$ is generated by a Hermitian operator $H$ via $U(\alpha) = e^{i\alpha H}$.
\end{theorem}

The matrix exponentials $e^A$ for square matrices $A$ used in this theorem are defined by $e^{iA} := \sum_{k=0}^\infty \frac{A^k}{k!}$ and satisfy
\begin{equation}
e^{\text{diag}(a_1,...,a_n)} = \text{diag}(e^{a_1},...,e^{a_n})
\qquad\qquad
e^{U^\dagger AU} = U^\dagger e^A U^\dagger
\end{equation}
for all unitaries $U$.

\subsection{Measurements} \label{sec:measurements}

In order to extract information from quantum systems, we need to perform \textit{measurements}.
Importantly, measuring a system usually also alters its state, making measurement a somewhat destructive process.
Note that there are many different kinds of measurements one can perform.
Mathematically, a measurement is specified by a set $\mathcal M = \{P_1,...,P_k\}$ of projectors that sum up to the identity $\sum_i P_i = I$.

\begin{definition}
	A square matrix $P$ is a projector if $P = P^\dagger = P^2$.
\end{definition}

Each projector represents a measurement outcome.
Since measurement is a non-deterministic process, we get a probability distribution over the outcomes.
When measuring $\ket{\psi}$, the probability of outcome $P_i$ can be computed using the \textit{Born rule}:
\begin{equation} \label{eqn:born-rule}
	\text{Prob}(i|\psi) = \bra{\psi} P_i \ket{\psi}
\end{equation}

\begin{example}[ONB Measurements]
	The orthonormal basis measurement corresponding to a basis $\mathcal B = \{ \ket{\phi_i} \}_{i}$ is given by $\mathcal M_{\mathcal B} = \{ \ket{\phi_i}\bra{\phi_i} \}_i$.
	For example, the two-dimensional computational basis yields $\mathcal M = \{ \ket{0}\bra{0}, \ket{1}\bra{1} \}$.
	In that case, we have $\text{Prob}(i|\psi) = \braket{\psi}{i}\braket{i}{\psi}$.
	We can think of this as a measure of how \enquote{close} $\ket{\psi}$ is to $\ket{0}$ or $\ket{1}$:
	If $\ket{\psi} = x\ket{0} + y\ket{1}$ then $\text{Prob}(0|\psi) = (\overline x\braket{0}{0} + \overline y\braket{1}{0}) (x\braket{0}{0} + y\braket{0}{1}) = x\overline x = |x|^2$.
\end{example}

An important observation is that states that are equal up to a \textit{global phase} of $e^{i\alpha}$ behave exactly the same with regard to measurement:
Let $\ket{\phi} := e^{i\alpha}\ket{\psi}$, then
\[ \text{Prob}(j|\phi) = \bra{e^{i\alpha}\psi} P_j \ket{e^{i\alpha}\psi} = e^{i\alpha}\bra{\psi} P_j e^{-i\alpha}\ket{\psi} = \bra{\psi} P_j \ket{\psi} = \text{Prob}(j|\psi). \]
Thus, there is no measurable difference between $\ket{\phi}$ and $\ket{\psi}$.
Hence, states are not just vectors, but equivalence classes of vectors that are equal up to a global phase.
One way to remove this redundancy is the \textit{doubling} construction where we represent the states as $\ket{\phi}\bra{\phi}$ and $\ket{\psi}\bra{\psi}$ instead, which are actually equal.
%\[ \ket{\phi}\bra{\phi} = \ket{e^{i\alpha}\psi}\bra{e^{i\alpha}\psi} = e^{i\alpha}\ket{\psi} e^{-i\alpha}\bra{\psi} = \ket{\psi}\bra{\psi} \]
We will make heavy use of this when describing gradients of parametrised quantum circuits later.

Performing a single measurement corresponds to sampling from the distribution (\ref{eqn:born-rule}).
However, often we are not necessarily interested in a single sample, but want to understand the broader distribution of outcomes.
A useful tool for this is the \textit{expectation value}.
To motivate its definition, suppose we associate a real number $x_j$ with each projector $P_j$.
Then, we define random variable $X$ that takes the value $x_j$ whenever we get the measurement outcome $j$.
The expectation value of our state $\ket{\psi}$ w.r.t. this operator then corresponds to the mean value of $X$:
\[
	\mathbf E(X)
	= \sum_{j=1}^k x_j \cdot \text{Prob}(j|\psi)
	\eqq{\ref{eqn:born-rule}} \sum_{j=1}^k x_j \cdot \bra{\psi} P_j \ket{\psi}
	= \bra{\psi} \left(\sum_{j=1}^k x_j P_j\right) \ket{\psi}
\]
In order to estimate the expectation value on a quantum computer, one can compute the statistical mean of $X$ by preparing and measuring the state $\ket{\psi}$ for a large number of executions.
One commonly refers to the different executions as \textit{shots}.

Interestingly, $\sum_{j=1}^k x_j P_j$ is self-adjoint.
Conversely, every self-adjoint matrix $H$ with eigenvectors $\lambda_1,...,\lambda_k$ gives rise to a unique set of projectors $\mathcal M_H = \left\{ \sum_{\phi \in \Phi_i} \ket{\phi}\bra{\phi} \right\}_{i=1}^k$ where $\Phi_i$ is the set of eigenvectors of $H$ corresponding to the eigenvalue $\lambda_i$.
Because of this duality, it is often more convenient to describe measurements via Hermitian operators instead of projectors.
In this context, $H$ is commonly referred to as an \textit{observable}, or \textit{Hamiltonian} and the expectation value is denoted by
\[ \langle H\rangle := \bra{\psi}H\ket{\psi}. \]
Interestingly, every Hermitian matrix $H \in \mathbb C^{2^n}$ can be written as a real combination of Pauli operators $P \in \{X,Y,Z,I\}^{\otimes n}$.
We will use this in \Cref{chap:barren-plateaus} to simplify our barren plateau analysis.

\subsection{The Quantum Circuit Model}

The \textit{quantum circuit model} is a model to describe quantum computation that is inspired by classical circuits.
After preparing $n$ qubits in a fixed state (usually $\ket{0}^{\otimes n}$) we apply \textit{gates} that correspond to unitary operations on the qubits.
Finally, we measure one or more qubits.
Circuits are read from left to right and qubits are drawn as wires with gates on them:

\bigskip
\begin{figure}[h]
	\begin{center}
		\input{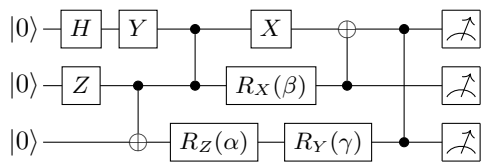}
	\end{center}
	\caption{Example of a 3-qubit quantum circuit.}	\label{fig:circuit-example}
\end{figure}

We have already seen the Hadamard gate $H$ and the $Z$-rotation $R_Z$ in (\ref{eqn:had-def}) and (\ref{eqn:RZ-def}), respectively.
Similarly, the single-qubit gates $R_X$ and $R_Y$ correspond to $X$- and $Y$-rotations on the Bloch sphere:
\[
	R_X(\alpha) := \begin{pmatrix} 
		\cos(\frac{\alpha}{2}) & -i\sin(\frac{\alpha}{2}) \\
		-i\sin(\frac{\alpha}{2}) & \cos(\frac{\alpha}{2})
	\end{pmatrix}
	\qquad\qquad
	R_Y(\alpha) := \begin{pmatrix}
		\cos(\frac{\alpha}{2}) & -\sin(\frac{\alpha}{2}) \\
		\sin(\frac{\alpha}{2}) & \cos(\frac{\alpha}{2})
	\end{pmatrix}
\]
The special cases for $\alpha=180^\circ$ rotations around the Bloch sphere give rise to the so-called \textit{Pauli matrices} (up to a global phase):
\[
	X := \begin{pmatrix}
		0 & 1 \\
		1 & 0
	\end{pmatrix}
	\qquad
	Y := \begin{pmatrix}
		0 & -i \\
		i & 0
	\end{pmatrix}
	\qquad
	Z := \begin{pmatrix}
		1 & 0 \\
		0 & -1
	\end{pmatrix}
\]
Finally, controlled gates are gates where the first qubit controls whether a unitary $U$ is applied to the remaining gates.
They can be constructed via
\[ 
	C_U 
	= ~\input{\tikzitpathgates/CU.tikz}~
	:= (\ket{0}\bra{0} \otimes I) + (\ket{1}\bra{1} \otimes U)
	=  \begin{pmatrix}
		I & 0 \\
		0 & U
	\end{pmatrix}.
\]
In \Cref{fig:circuit-example}, we have controlled $X$ and $Z$ gates that are usually called $\mathit{CNOT}$\footnote{This is because the Pauli $X$ acts like negation on the computational basis.} and $\mathit{CZ}$, respectively.
They have a special notation:
\begin{align*}
	\mathit{CNOT} &= \input{\tikzitpathgates/CNOT-2.tikz} = \input{\tikzitpathgates/CNOT-1.tikz} := \begin{pmatrix}
		1 & 0 & 0 & 0 \\
		0 & 1 & 0 & 0 \\
		0 & 0 & 0 & 1 \\
		0 & 0 & 1 & 0
	\end{pmatrix}
	\\[\linesep]
	\mathit{CZ} &= \input{\tikzitpathgates/CZ-2.tikz} = \input{\tikzitpathgates/CZ-1.tikz} := \begin{pmatrix}
		1 & 0 & 0 & 0 \\
		0 & 1 & 0 & 0 \\
		0 & 0 & 1 & 0 \\
		0 & 0 & 0 & -1
	\end{pmatrix}
\end{align*}
The $\mathit{CNOT}$ gate is drawn with a $\oplus$ symbol since it acts like $\ket{x,y} \mapsto \ket{x,x\oplus y}$ on the computational basis where $\oplus$ denotes XOR.
The $\mathit{CZ}$ gate is drawn with two black dots since it is symmetric in which qubit is the control.
Both $\mathit{CNOT}$ and $\mathit{CZ}$ are used to \textit{entangle} the two qubits to which they are applied.

%Finally, there are also controlled versions of the rotation gates.
%For this thesis, the controlled $R_Z$ gates is particularly relevant:
%\[
%	CR_Z(\alpha) := \begin{pmatrix}
%		1 & 0 & 0 & 0 \\
%		0 & 1 & 0 & 0 \\
%		0 & 0 & e^{-i\frac{\alpha}{2}} & 0 \\
%		0 & 0 & 0 & e^{i\frac{\alpha}{2}}
%	\end{pmatrix}
%\]

\section{Quantum Machine Learning} \label{sec:QML}

The goal of \textit{quantum machine learning} (QML) is to achieve a quantum advantage using the current \textit{noisy intermediate-scale quantum} (NISQ) hardware.
%This is because in today's NISQ era, anticipated algorithms such as Shor’s or Grover’s are still out of reach because current quantum hardware lacks the necessary scale and reliability.
Typically, QML algorithms employ a hybrid approach where a quantum processor works in tandem with a classical computer.
In this thesis, we focus on \textit{variational algorithms} for QML.
This approach relies on \textit{parametrised quantum circuits} (PQCs), i.e. circuits that depend on tunable parameters.
For example, the circuit in \Cref{fig:circuit-example} is a PQC if the parameters $\alpha,\beta,\gamma$ are not fixed.
Given a PQC that depends on some parameters $\vec\theta$, machine learning techniques are used to find an optimal parameter assignment $\vec\theta^\star$ for which the circuit exhibits some desired behaviour.
This could for example be fitting a dataset in a supervised classification or regression task~\cite{schuld2019quantum,mitarai2018quantum}, or modelling a probability distribution for a generative task~\cite{benedetti2019generative,liu2018differentiable,lloyd2018quantum}.
Other applications of variational algorithms include simulating quantum chemistry~\cite{kandala2017hardware,cao2019quantum}, solving combinatorial optimisation problems~\cite{farhi2014quantum}, and performing natural language processing tasks~\cite{meichanetzidis2020quantum,kartsaklis2021lambeq}.

\begin{figure}
	\begin{center}
		\input{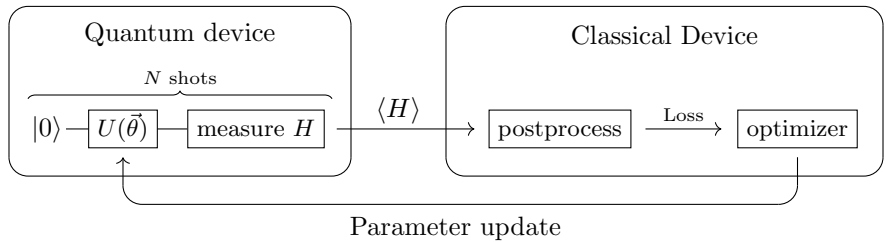}
	\end{center}
	\caption{Pipeline for variational algorithms (adapted from Figure 1 in~\cite{zhao2021analyzing}).}
	\label{fig:variational-pipeline}
\end{figure}

\Cref{fig:variational-pipeline} shows the schematic pipeline used by variational algorithms.
Essentially, the PQC is trained using a classical optimiser in order to minimise some loss calculated based on the expectation value $\langle H\rangle$ produced by the quantum device.
Because of the current NISQ hardware, this process is generally noisy.
However, many optimisers developed for machine learning are resilient to a certain amount of noise which makes variational algorithms applicable in the NISQ era.

\subsection{Types of Ansätze}

\begin{figure}[t]
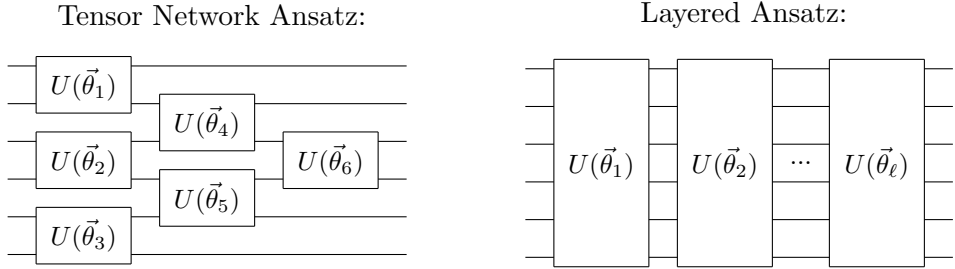

	\begin{minipage}{.5\textwidth}
		\centering
		Tensor Network Ansatz: \\[10pt]
		\input{\tikzitpathtensor-network-ansatz.tikz}
	\end{minipage}%
	\begin{minipage}{.5\textwidth}
		\centering
		Layered Ansatz: \\[10pt]
		\input{\tikzitpathlayered-ansatz.tikz}
	\end{minipage}%
	\caption{Different ansatz layouts.}
	\label{fig:ansatz-layouts}
\end{figure}

The PQCs used for variational algorithms are typically referred to as \textit{ansätze}. The term ansatz comes from mathematics and physics where it describes an initial strategy or approach to express a solution.
Broadly, one can distinguish two different kinds of ansatz designs commonly used for QML which are depicted in \Cref{fig:ansatz-layouts}.

\textit{Tensor network ansätze} arrange gates in a fixed layout inspired by tensor networks~\cite{huggins2019towards,qian2022tree}.
For example, the blocks in \Cref{fig:ansatz-layouts} are laid out in a tree architecture.
\textit{Layered ansätze} on the other hand consist of layers that are repeated one after the other for a fixed number of times.
Commonly, each layer is made up of single qubit unitaries, preceded or followed by a block of entangling gates.
Another commonly used type of layered ansatz is the \textit{alternating operator ansatz} used in the \textit{quantum approximate optimization algorithm} (QAOA)~\cite{farhi2014quantum}. 
There, the layers are defined in terms of two Hamiltonians that encode a combinatorial optimisation problem which can be solved by training the circuit.

In this thesis, we focus on layered ansatz designs that have been shown to be more expressive than tensor network ansätze~\cite{du2018expressive}.
In particular, see \Cref{fig:sim-circuits} in \Cref{chap:barren-plateaus} for layered ansätze that are used in practice.

\subsection{Gradient-Based Optimisation}

There is a wide range of optimisation algorithms that can be used to train PQCs~\cite{bonet2021performance,spall1992multivariate,powell1994direct,kubler2020adaptive,nakanishi2020sequential,wilson2021optimizing}.
In this thesis, we focus on gradient-based optimisation approaches which are commonly used in QML and provably improve convergence in variational algorithms~\cite{harrow2021low}.

Gradient-based optimisation techniques such as gradient descent have been proven to be widely successful in the domain of classical machine learning, in particular neural networks.
Given the output $\vec y$ of a neural network, the gradient $\frac{\partial \mathcal L}{\partial w_i}(\vec y)$ of some loss function $\mathcal L$ with respect to the weight $w_i$ is computed via backpropagation and the weight is updated in the opposite direction of this gradient.
We can transfer this approach to the quantum realm:
Instead of the weights of a neural network, we train the parameters of an ansatz.
The \enquote{output} of the quantum circuit is an expectation value $\langle H\rangle$.
Hence, we want to compute $\frac{\partial \mathcal L}{\partial \theta_i}(\langle H\rangle)$ which by the chain rule depends on $\frac{\partial \langle H\rangle}{\partial \theta_i}$.
Unlike individual measurements, expectation values are continuous variables such that this gradient is well-defined.
Finally, we update the circuit parameters according to the loss gradient.

However, gradient descent on quantum computers comes with a set of unique challenges.
First, it is not feasible to compute $\frac{\partial \langle H\rangle}{\partial \theta_i}$ classically.
In particular, the backpropagation algorithm is not available since quantum circuits have a fundamentally different structure than neural networks.
Instead, the gradient must be computed on the quantum device itself.
Quantum algorithms that solve this task are commonly referred to as \textit{gradient recipes} and are subject of a lot research interest at the moment~\cite{mitarai2018quantum,schuld2019evaluating,anselmetti2021local,wierichs2022general}.
We contribute to this in \Cref{chap:barren-plateaus} by giving diagrammatic interpretations and proofs of existing recipes, and by proving a conjecture by Anselmetti et al.~\cite{anselmetti2021local} establishing the optimality of a certain recipe.

The second issue lies with the geometry of the gradient landscape.
It is hypothesised that gradient descent performs well on classical neural networks because their loss surface has few bad local minima~\cite{choromanska2015loss}.
The same can unfortunately not be said for PQCs~\cite{you2021exponentially}.
Even worse, it has been shown that the gradient landscape of many ansätze is exponentially flat with respect to circuit size, making gradient descent difficult or even impossible~\cite{mcclean2018barren}.
Thus, there is a lot of interest in analysing which ansätze exhibit those \textit{barren plateaus}.
In \Cref{chap:barren-plateaus} we apply a diagrammatic method to analyse ansätze for this problem.

\section{The ZXW-Calculus} \label{sec:zxw-caclulus}

The \textit{ZX-calculus} is graphical language for reasoning about quantum computation originally developed by Coecke and Duncan~\cite{coecke2008interacting}.
It is universal and complete~\cite{ng2017universal} meaning that all quantum reasoning can be carried out in the realm of ZX diagrams.
The ZX-calculus has been applied in a variety of areas, including circuit optimisation~\cite{duncan2020graph,kissinger2019reducing}, compilation~\cite{de2020architecture,cowtan2020generic}, simulation~\cite{kissinger2022simulating}, measurement-based quantum computing~\cite{duncan2012graphical,kissinger2019universal} and surface codes~\cite{de2020zx}.

The \textit{ZXW-calculus}~\cite{shaikh2022sum} is a variant of ZX that has its roots in the \textit{algebraic ZX-calculus}~\cite{wang2020algebraic}.
It has recently been used to express derivates and integrals~\cite{wang2022differentiating} which makes it well-suited for our diagrammatic treatment of gradient-based QML.

\subsection{Generators and String Diagrams}

ZXW diagrams consist of generators that are wired together and connected to inputs and outputs.
Following the circuit notation, we put the inputs on the left side and the outputs on the right.
While diagrams can be studied as mathematical objects in their own right, for this thesis we are mainly interested in their interpretation as linear maps.
Concretely, a diagram with $n$ inputs and $m$ outputs represents a $2^m \times 2^n$ complex matrix.
There are also diagrams with zero inputs and outputs which thus represent single complex numbers.

We now give the three main generators of the ZXW-calculus:
\[
	\input{\tikzitpathzx/generators/box.tikz} := \ket{0^m}\bra{0^n} + a\ket{1^m}\bra{1^n}
	\qquad
	\input{\tikzitpathzx/generators/had.tikz} := \frac{1}{\sqrt 2} \begin{pmatrix}
		1 & 1 \\
		1 & -1
	\end{pmatrix}
	\qquad
	\input{\tikzitpathzx/generators/W.tikz} := \begin{pmatrix}
		1 & 0 \\
		0 & 1 \\
		0 & 1 \\
		0 & 0
	\end{pmatrix}
\]
where $a\in\mathbb C$.
We call the generators the \textit{green box}, \textit{Hadamard}, and \textit{black triangle}, respectively.
ZXW diagrams are formed by wiring these generators together.
For this, we also introduce generators that allow us to bend and cross wires:
\[
	\input{\tikzitpathzx/generators/wire.tikz} := \begin{pmatrix}
		1 & 0 \\
		0 & 1
	\end{pmatrix}
	\qquad
	\input{\tikzitpathzx/generators/cup.tikz} := \begin{pmatrix}
		1 \\ 0 \\ 0 \\ 1
	\end{pmatrix}
	\qquad
	\input{\tikzitpathzx/generators/cap.tikz} := \begin{pmatrix}
	1 & 0 & 0 & 1
	\end{pmatrix}
	\qquad
	\input{\tikzitpathzx/generators/swap.tikz} := \begin{pmatrix}
		1 & 0 & 0 & 0 \\
		0 & 0 & 1 & 0 \\
		0 & 1 & 0 & 0 \\
		0 & 0 & 0 & 1
	\end{pmatrix}
\]
We can wire the generators together using the sequential and parallel composition operators $\circ$ and $\otimes$, corresponding to matrix multiplication and tensor product on the underlying matrices.
For example, we write
$\input{\tikzitpathzx/ex-composition.tikz} := \input{\tikzitpathzx/generators/cap.tikz} \circ (\input{\tikzitpathzx/ex-composition-2.tikz} \otimes \input{\tikzitpathzx/generators/had.tikz}).$
Furthermore, the wires satisfy the \textit{yanking equations}
\[ 
	\input{\tikzitpathzx/yanking/1.tikz} ~=~ \input{\tikzitpathzx/generators/wire.tikz}
	\qquad\qquad
	\input{\tikzitpathzx/yanking/2.tikz} ~=~ \input{\tikzitpathzx/generators/cup.tikz}
\]
This means we can arbitrarily deform diagrams by moving the generators around the plane, bending and unbending wires as we go, without changing the underlying matrix.
We only have to make sure that the inputs and outputs stay in the same order.
This principle is summarised in the slogan \textit{only connectivity matters}.

\subsection{Additional Notation}

Based on the generators, we define some additional notation.
For example, the green spider from the original ZX-calculus can be defined via the green box:
\begin{equation} \label{eqn:green-spider-def} 
	\input{\tikzitpathzx/notation/green-spider-1.tikz} ~:=~ \input{\tikzitpathzx/notation/green-spider-2.tikz} 
	\qquad\qquad
	\input{\tikzitpathzx/notation/green-spider-3.tikz} ~:=~ \input{\tikzitpathzx/notation/green-spider-4.tikz}
\end{equation}
%In particular, if all green boxes in a ZXW diagram only have phases $e^{i\alpha}$ in them, then the diagram is equivalent to a ZX diagram since the other generators can all be encoded in ZX.\todo{The green box can be encoded as well. Find a better phrasing for this...}
%Thus, we will drop the \enquote{W} in these cases and refer to them as ZX-diagrams.
The red spiders from the original ZX-calculus can be defined by Hadamard conjugation:
\begin{equation} \label{eqn:red-spider-def}
	\input{\tikzitpathzx/notation/red-spider-1.tikz} ~:=~ \input{\tikzitpathzx/notation/red-spider-2.tikz} 
	\qquad\qquad
	\input{\tikzitpathzx/notation/red-spider-3.tikz} ~:=~ \input{\tikzitpathzx/notation/red-spider-4.tikz}
\end{equation}
If a diagram only contains spiders and no boxes or black triangles, we sometimes drop the \enquote{W} and speak of traditional \textit{ZX-diagrams}.
Often we only have spiders with phase $\alpha=0$ or $\alpha=\pi$.
For those cases, we define a special pink spider as a rescaled version of the red spider that only has integer components in its matrix:
\begin{equation} \label{eqn:pink-spider-def}
	\input{\tikzitpathzx/notation/pink-spider-1.tikz} ~:=~ 2^{\frac{n+m-2}{2}}~ \input{\tikzitpathzx/notation/pink-spider-2.tikz}
	\qquad\qquad
	\input{\tikzitpathzx/notation/pink-spider-3.tikz} ~:=~ 2^{\frac{n+m-2}{2}}~ \input{\tikzitpathzx/notation/red-spider-3.tikz}
\end{equation}
We give the scalars that are represented by commonly occurring diagrams below:
\[
	\input{\tikzitpathzx/scalars/dot-green.tikz} = 2
	\qquad
	\input{\tikzitpathzx/scalars/dot-pink.tikz} = 1
	\qquad
	\input{\tikzitpathzx/scalars/pi-green.tikz} = \input{\tikzitpathzx/scalars/pi-pink.tikz} = 0
	\qquad
	\input{\tikzitpathzx/scalars/conn-0.tikz} = 1
	\qquad
	\input{\tikzitpathzx/scalars/conn-1.tikz} = e^{i\alpha}
\]
Finally, we define the triangle and inverse triangle as well as their transposes:
\begin{equation} \label{eqn:tri-def}
\begin{aligned}
	\input{\tikzitpathzx/notation/tri-1.tikz} ~&:=~ \input{\tikzitpathzx/notation/tri-2.tikz} ~= \begin{pmatrix}
		1 & 1 \\
		0 & 1
	\end{pmatrix}
	\qquad&\qquad
	\input{\tikzitpathzx/notation/tri-inv-1.tikz} ~&:=~ \input{\tikzitpathzx/notation/tri-inv-2.tikz} ~= \begin{pmatrix}
		1 & 0 \\
		1 & 1
	\end{pmatrix}
	\\[\linesep]
	\input{\tikzitpathzx/notation/tri-3.tikz} ~&:=~ \input{\tikzitpathzx/notation/tri-4.tikz} ~= \begin{pmatrix}
		1 & -1 \\
		0 & 1
	\end{pmatrix}
	\qquad&\qquad
	\input{\tikzitpathzx/notation/tri-inv-3.tikz} ~&:=~ \input{\tikzitpathzx/notation/tri-inv-4.tikz} ~= \begin{pmatrix}
		1 & 0 \\
		-1 & 1
	\end{pmatrix}
\end{aligned}
\end{equation}

\subsection{Rules}

So far, we have only seen ZX(W) diagrams as graphical representations of matrices.
Their real power comes from the rewrite rules that allow us to do matrix calculations diagrammatically.
The rules of the ZXW-calculus are listed in \Cref{fig:ZXW-rules}.

\newcommand{\rulesep}{5pt}
\begin{figure}[t]
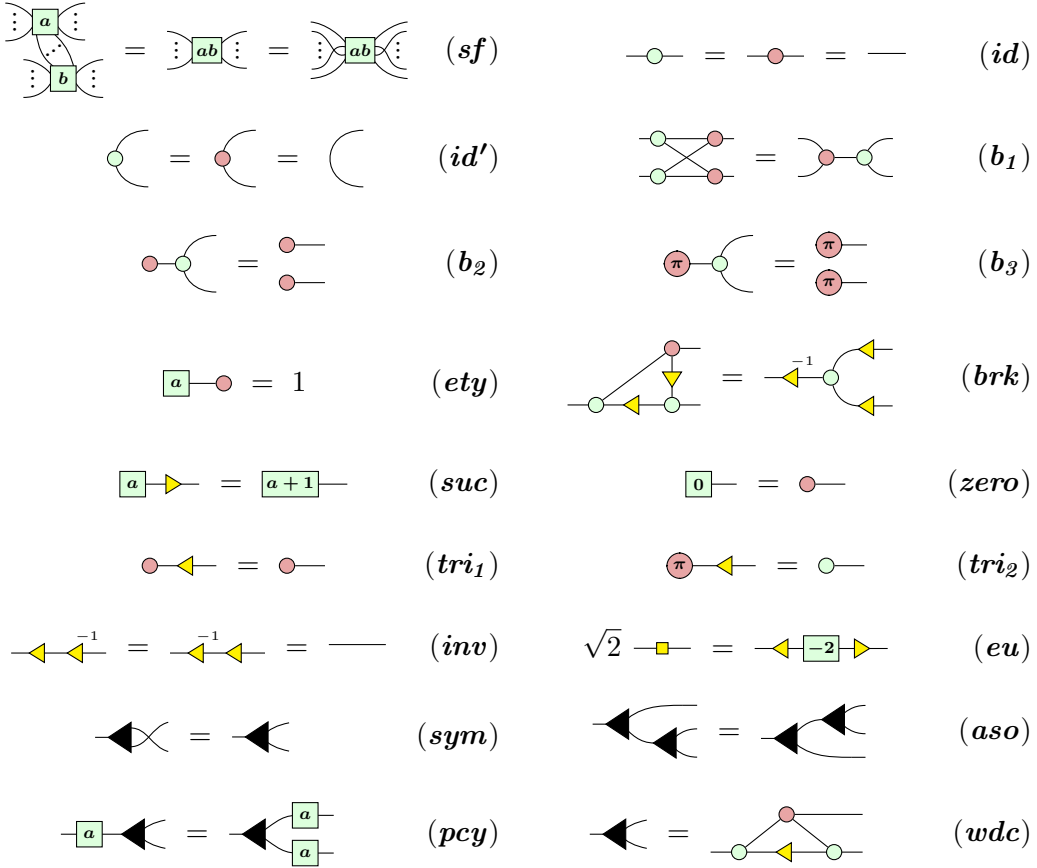

	\begin{minipage}{0.5\linewidth}
		\begin{equation} \tag{$\bm{\mathit{sf}}$}\label{eqn:sf}
		\input{\tikzitpathzx/rules/sf-1.tikz} ~=~ \input{\tikzitpathzx/rules/sf-2.tikz} ~=~ \input{\tikzitpathzx/rules/sf-3.tikz}
		\end{equation}
	\end{minipage}%
	\begin{minipage}{0.5\linewidth}
		\begin{equation} \tag{$\bm{\mathit{id}}$}\label{eqn:id}
		\input{\tikzitpathzx/rules/id-1.tikz} ~=~ \input{\tikzitpathzx/rules/id-2.tikz} ~=~ \input{\tikzitpathzx/rules/id-3.tikz} 
		\end{equation}
	\end{minipage}
	\\%[\rulesep]
	\begin{minipage}{0.5\linewidth}
		\begin{equation} \tag{$\bm{\mathit{id'}}$}\label{eqn:ic}
		\input{\tikzitpathzx/rules/idc-1.tikz} ~=~ \input{\tikzitpathzx/rules/idc-2.tikz} ~=~ \input{\tikzitpathzx/rules/idc-3.tikz}
		\end{equation}
	\end{minipage}%
	\begin{minipage}{0.5\linewidth}
		\begin{equation} \tag{$\bm{\mathit{b_1}}$}\label{eqn:b1}
		\input{\tikzitpathzx/rules/b2-1.tikz} ~=~ \input{\tikzitpathzx/rules/b2-2.tikz}
		\end{equation}
	\end{minipage}
	\\[\rulesep]
	\begin{minipage}{0.5\linewidth}
		\begin{equation} \tag{$\bm{\mathit{b_2}}$}\label{eqn:b2}
		\input{\tikzitpathzx/rules/b3-1.tikz} ~=~ \input{\tikzitpathzx/rules/b3-2.tikz}
		\end{equation}
	\end{minipage}%
	\begin{minipage}{0.5\linewidth}
		\begin{equation} \tag{$\bm{\mathit{b_3}}$}\label{eqn:b3}
		\input{\tikzitpathzx/rules/b1-1.tikz} ~=~ \input{\tikzitpathzx/rules/b1-2.tikz}
		\end{equation}
	\end{minipage}
	\\[\rulesep]
	\begin{minipage}{0.5\linewidth}
		\begin{equation} \tag{$\bm{\mathit{ety}}$}\label{eqn:ety}
		\input{\tikzitpathzx/rules/one.tikz} ~=~ 1
		\end{equation}
	\end{minipage}%
	\begin{minipage}{0.5\linewidth}
		\begin{equation} \tag{$\bm{\mathit{brk}}$}\label{eqn:brk}
		\input{\tikzitpathzx/rules/brk-1.tikz} ~=~ \input{\tikzitpathzx/rules/brk-2.tikz}
		\end{equation}
	\end{minipage}
	\\[\rulesep]
	\begin{minipage}{0.5\linewidth}
		\begin{equation} \tag{$\bm{\mathit{suc}}$}\label{eqn:suc}
		\input{\tikzitpathzx/rules/suc-1.tikz} ~=~ \input{\tikzitpathzx/rules/suc-2.tikz}
		\end{equation}
	\end{minipage}%
	\begin{minipage}{0.5\linewidth}
		\begin{equation} \tag{$\bm{\mathit{zero}}$}\label{eqn:zero}
		\input{\tikzitpathzx/rules/zero-1.tikz} ~=~ \input{\tikzitpathzx/rules/zero-2.tikz}
		\end{equation}
	\end{minipage}
	\\[\rulesep]
	\begin{minipage}{0.5\linewidth}
		\begin{equation} \tag{$\bm{\mathit{tri_1}}$}\label{eqn:tri1}
		\input{\tikzitpathzx/rules/tr0-1.tikz} ~=~ \input{\tikzitpathzx/rules/tr0-2.tikz}
		\end{equation}
	\end{minipage}%
	\begin{minipage}{0.5\linewidth}
		\begin{equation} \tag{$\bm{\mathit{tri_2}}$}\label{eqn:tri2}
		\input{\tikzitpathzx/rules/tr1-1.tikz} ~=~ \input{\tikzitpathzx/rules/tr1-2.tikz}
		\end{equation}
	\end{minipage}
	\\[\rulesep]
	\begin{minipage}{0.5\linewidth}
		\begin{equation} \tag{$\bm{\mathit{inv}}$}\label{eqn:inv}
		\input{\tikzitpathzx/rules/inv-1.tikz} ~=~ \input{\tikzitpathzx/rules/inv-2.tikz} ~=~ \input{\tikzitpathzx/rules/inv-3.tikz}
		\end{equation}
	\end{minipage}%
	\begin{minipage}{0.5\linewidth}
		\begin{equation} \tag{$\bm{\mathit{eu}}$}\label{eqn:eu}
		\sqrt{2}~ \input{\tikzitpathzx/rules/eu-1.tikz} ~=~ \input{\tikzitpathzx/rules/eu-2.tikz}
		\end{equation}
	\end{minipage}
	\\[\rulesep]
	\begin{minipage}{0.5\linewidth}
		\begin{equation} \tag{$\bm{\mathit{sym}}$}\label{eqn:sym}
		\input{\tikzitpathzx/rules/sym-1.tikz} ~=~ \input{\tikzitpathzx/rules/sym-2.tikz}
		\end{equation}
	\end{minipage}%
	\begin{minipage}{0.5\linewidth}
		\begin{equation} \tag{$\bm{\mathit{aso}}$}\label{eqn:aso}
		\input{\tikzitpathzx/rules/aso-1.tikz} ~=~ \input{\tikzitpathzx/rules/aso-2.tikz}
		\end{equation}
	\end{minipage}
	\\[\rulesep]
	\begin{minipage}{0.5\linewidth}
		\begin{equation} \tag{$\bm{\mathit{pcy}}$}\label{eqn:pcy}
		\input{\tikzitpathzx/rules/pcy-1.tikz} ~=~ \input{\tikzitpathzx/rules/pcy-2.tikz}
		\end{equation}
	\end{minipage}%
	\begin{minipage}{0.5\linewidth}
		\begin{equation} \tag{$\bm{\mathit{wdc}}$}\label{eqn:wdc}
		\input{\tikzitpathzx/rules/wdc-1.tikz} ~=~ \input{\tikzitpathzx/rules/wdc-2.tikz}
		\end{equation}
	\end{minipage}

	\caption{Rules of the ZXW-calculus for $a,b\in\mathbb C$.}
	\label{fig:ZXW-rules}
\end{figure}

Note that the equality signs in the rules mean that both sides represent exactly the same matrix.
In the original ZX-calculus, many rules like $(\bOne)$, $(\bTwo)$, or $(\bThree)$ only hold up to a (non-zero) scalar that is often ignored.
However, for the purposes of this thesis we need to be precise about scalars.
The fact that we can give many rules without them is thanks to the rescaled pink spider.
As a trade-off, the colour-change rule now introduces scalars for pink spiders\footnote{We prove this rule as well as other rules of the original ZX-calculus in \Cref{sec:zx-lemmas}.}:
\[ 
	\input{\tikzitpathzx/lem:cc/statement-1.tikz} \eqq{\cc} 2^{-\frac{n+m-2}{2}}~ \input{\tikzitpathzx/lem:cc/statement-2.tikz}
	\qquad\qquad
	\input{\tikzitpathzx/lem:cc/statement-3.tikz} \eqq{\cc} 2^{\frac{n+m-2}{2}}~ \input{\tikzitpathzx/lem:cc/statement-4.tikz}
\]
Furthermore, fusing pink spiders that are connected by multiple wires also introduces a scalar:
\[ \input{\tikzitpathzx/lem:pink-sf/statement-1.tikz} \eqq{\text{Lem. }\ref{lem:sf-pink}} 2^{n-1}~ \input{\tikzitpathzx/lem:pink-sf/statement-2.tikz} \]
We define a multi-legged version of the black triangle, which we call \textit{W spider}:
\begin{equation} \label{eqn:W-spider-def}
	\input{\tikzitpathzx/notation/W-spider-1.tikz} := \input{\tikzitpathzx/notation/W-spider-2.tikz} 
	\qquad\qquad
	\input{\tikzitpathzx/notation/W-spider-3.tikz} := \input{\tikzitpathzx/notation/W-spider-4.tikz}
\end{equation}
Because of the $(\aso)$ rule it actually does not matter in which order we plug the triangles together and it is easy to see that W spiders satisfy the following fusion rule:

\begin{equation} \tag{$\bm{\mathit{wf}}$}\label{eqn:wf}
	\input{\tikzitpathzx/lem:W-fuse/1.tikz} ~=~ \input{\tikzitpathzx/lem:W-fuse/2.tikz}
\end{equation}

On top of this, as we will prove in \Cref{lem:W-action}, they interact with pink spiders in the following way:
\[
	\input{\tikzitpathzx/lem:W-action/statement-1.tikz} \eqq{\w} \input{\tikzitpathzx/lem:W-action/statement-2.tikz}
	\qquad\qquad
	\input{\tikzitpathzx/lem:W-action/statement-3.tikz} \eqq{\w} \input{\tikzitpathzx/lem:W-action/statement-4.tikz} ~+~ \input{\tikzitpathzx/lem:W-action/statement-5.tikz} ~+~ ... ~+~ \input{\tikzitpathzx/lem:W-action/statement-6.tikz}
\]
This property will prove to be crucial when discussing diagrammatic differentiation in \Cref{chap:diagrammatic-differentiation}.

\subsection{Quantum Gates and Computation in ZXW} \label{sec:zx-quantum}

Next, we explain how quantum computation is expressed in ZX(W).
First, note that pink and green spiders can describe the computational and the X-basis:
\begin{equation} \label{eqn:zx-states}
	\input{\tikzitpathzx/quantum/basis-0.tikz} ~=~ \ket{0}
	\qquad
	\input{\tikzitpathzx/quantum/basis-1.tikz} ~=~ \ket{1}
	\qquad
	\input{\tikzitpathzx/quantum/basis-2.tikz} ~=~ \sqrt{2}~ \ket{+}
	\qquad
	\input{\tikzitpathzx/quantum/basis-3.tikz} ~=~ \sqrt{2}~ \ket{-}
\end{equation}
Many matrix operations commonly used in quantum computing have elegant representations in ZXW.
For example, transposing a matrix corresponds to mirroring the diagram horizontally and the conjugate matrix is obtained by conjugating the numbers in boxes and negating the phases in spiders.
Thus the adjoint of a ZXW diagram is constructed by combining those two operations.

The Hadamard gate is given as a generator. We introduce the following notation, denoting edges with a Hadamard on them as dashed blue lines:
\[ \input{\tikzitpathzx/quantum/H-1.tikz} \quad\rightsquigarrow\quad \input{\tikzitpathzx/quantum/H-2.tikz} \]
The Pauli matrices are represented by
\[
	X ~=~ \input{\tikzitpathzx/quantum/X.tikz}
	\quad\qquad
	Y ~=~ i~\input{\tikzitpathzx/quantum/Y.tikz}
	\qquad\qquad
	Z ~=~ \input{\tikzitpathzx/quantum/Z.tikz}
\]
the rotation gates can be written as
\begin{gather*}
	R_Z(\alpha) ~=~ e^{-i\frac{\alpha}{2}}~ \input{\tikzitpathzx/quantum/RZ.tikz}
	\qquad\qquad
	R_X(\alpha) ~=~ e^{-i\frac{\alpha}{2}}~ \input{\tikzitpathzx/quantum/RX.tikz}
	\\[\linesep]
	R_Y(\alpha) ~=~ e^{-i\frac{\alpha}{2}}~ \input{\tikzitpathzx/quantum/RY-1.tikz} ~=~ e^{-i\frac{\alpha}{2}}~ \input{\tikzitpathzx/quantum/RY-2.tikz}
\end{gather*}
and common two-qubit gates are given by
\begin{gather*}
	\mathit{CNOT} ~=~ \input{\tikzitpathzx/quantum/CNOT-1.tikz} ~=~ \input{\tikzitpathzx/quantum/CNOT-2.tikz}
	\qquad\qquad
	\mathit{CZ} ~=~ \input{\tikzitpathzx/quantum/CZ-1.tikz} ~=~ \sqrt 2~ \input{\tikzitpathzx/quantum/CZ-2.tikz}
	\\[\linesep]
	\mathit{CR}_Z(\alpha) ~=~ \input{\tikzitpathzx/quantum/CRZ-1.tikz} ~=~ \input{\tikzitpathzx/quantum/CRZ-2.tikz} \numberthis\label{eqn:CRZ-zx-def}
\end{gather*}
Using those building blocks, we can easily turn quantum circuits into ZXW diagrams.
However, recall from our discussion in \Cref{sec:measurements} that the matrix representation of quantum states has a certain redundancy in that states that only differ by a global phase behave exactly the same.
To deal with this problem, we use the \textit{doubling construction} to represent quantum circuits in ZXW.
Concretely, whenever we want to express quantum circuits in ZXW, we first construct a diagram capturing the circuit structure, and then we double it.
Doubling means tensoring the diagram with its complex conjugate, i.e.
\[ \doubled\left( \input{\tikzitpathzx/double-1.tikz} \right) := \input{\tikzitpathzx/double-2.tikz} \]
This way, all global phases cancel out.
See~\cite{coecke2017picturing} for a more detailed description of doubling.

\subsection{Pauli Boxes}

A useful ZX construction related to Paulis are so-called \textit{Pauli boxes}~\cite{cowtan2019phase,kissinger2022quantumsoftware}:
\begin{definition}~\cite{kissinger2022quantumsoftware}
	The Pauli boxes are defined as
	\begin{align*}
		\input{\tikzitpathzx/def:pauli-boxes/I-1.tikz} ~&:=~ \input{\tikzitpathzx/def:pauli-boxes/I-2.tikz}
		&
		\input{\tikzitpathzx/def:pauli-boxes/X-1.tikz} ~&:=~ \input{\tikzitpathzx/def:pauli-boxes/X-2.tikz}
		\\
		\input{\tikzitpathzx/def:pauli-boxes/Y-1.tikz} ~&:=~ \input{\tikzitpathzx/def:pauli-boxes/Y-2.tikz}
		&
		\input{\tikzitpathzx/def:pauli-boxes/Z-1.tikz} ~&:=~ \input{\tikzitpathzx/def:pauli-boxes/Z-2.tikz}
	\end{align*}
\end{definition}
Note that we can treat the wire sticking out on top as either input or output.
Plugging in a green $\pi$ yields the corresponding Pauli:
\begin{lemma}~\cite{kissinger2022quantumsoftware}
	For all $P \in \{I,X,Y,Z\}$ we have
	\[ \input{\tikzitpathzx/lem:pauli-box-pi/statement-1.tikz} ~=~ \input{\tikzitpathzx/lem:pauli-box-pi/statement-2.tikz} \]
\end{lemma}
Pauli boxes can be used to define a type of gate we have not mentioned so far.
Given a \textit{Pauli string} $\vec P \in \{ I,X,Y,Z \}^{\otimes n}$, i.e. a tensor product of Paulis, we define the \textit{Pauli exponential gate} $\vec P(\alpha)$ by
\[
	\vec P(\alpha) := e^{-i\frac{\alpha}{2}\vec P} ~=~ \input{\tikzitpathzx/pauli-exp.tikz}
\]
In the special case where $\vec P \in \{I,Z\}^{\otimes n}$, we call $\vec P(\alpha)$ a \textit{phase gadget}.
Paul exponentials based on the same Pauli string fuse together:
\begin{lemma}~\cite{cowtan2019phase} \label{lem:pauli-exp-fuse}
	For all Pauli strings $\vec P$ we have
	\begin{equation} \label{eqn:pauli-exp-fuse}
		\input{\tikzitpathzx/lem:pauli-exp-fuse/statement-1.tikz} ~=~ \input{\tikzitpathzx/lem:pauli-exp-fuse/statement-2.tikz}
	\end{equation}
	In particular, this implies $\vec P(\alpha) \vec P(\beta) = \vec P(\alpha + \beta)$.
\end{lemma}
Pauli gadgets also have interesting commutation properties:
\begin{lemma}~\cite{cowtan2019phase} \label{lem:pauli-exp-comm}
	Let $\vec P,\vec Q$ be $n$-qubit Pauli strings.
	If the number of positions $i$ for which $P_i \neq Q_i$ and $P_i,Q_i \neq I$ is even, then
	\[ \input{\tikzitpathzx/lem:pauli-exp-comm/statement-1.tikz} ~=~ \input{\tikzitpathzx/lem:pauli-exp-comm/statement-2.tikz} \]
	Otherwise,
	\[ \input{\tikzitpathzx/lem:pauli-exp-comm/statement-1.tikz} ~=~ \frac{1}{\sqrt 2}~ \input{\tikzitpathzx/lem:pauli-exp-comm/statement-3.tikz} \]
\end{lemma}

\subsection{Useful Lemmas} \label{sec:zx-lemmas}

We close the chapter by stating and proving some basic results that we will use throughout the thesis.

\begin{lemma}~\cite{wang2020algebraic}
	Hadamard is involutive:
	\begin{equation} \tag{$\bm{\mathit{hh}}$}\label{eqn:hh}
		\input{\tikzitpathzx/lem:hh/statement-1.tikz} ~=~ \input{\tikzitpathzx/lem:hh/statement-2.tikz}
	\end{equation}
\end{lemma}

\begin{lemma}
	Hadamards switch colours up to a scalar. For $\tau \in \{0,\pi\}$:
	\begin{equation} \tag{$\bm{\mathit{cc}}$}\label{eqn:cc}
		\input{\tikzitpathzx/lem:cc/statement-1.tikz} ~=~ 2^{-\frac{n+m-2}{2}}~ \input{\tikzitpathzx/lem:cc/statement-2.tikz}
		\qquad\qquad
		\input{\tikzitpathzx/lem:cc/statement-3.tikz} ~=~ 2^{\frac{n+m-2}{2}}~ \input{\tikzitpathzx/lem:cc/statement-4.tikz}
	\end{equation}
	The only scalar-free colour change happens for two legs:
	\[
		\input{\tikzitpathzx/lem:cc/statement-5.tikz} ~=~ \input{\tikzitpathzx/lem:cc/statement-6.tikz}
		\qquad\qquad
		\input{\tikzitpathzx/lem:cc/statement-7.tikz} ~=~ \input{\tikzitpathzx/lem:cc/statement-8.tikz}
	\]
\end{lemma}
\begin{proof}
	\begin{gather*}
		\input{\tikzitpathzx/lem:cc/statement-1.tikz}
		\eqq{\ref{eqn:red-spider-def}} \input{\tikzitpathzx/lem:cc/proof-1.tikz}
		\eqq{\ref{eqn:pink-spider-def}} 2^{-\frac{n+m-2}{2}}~ \input{\tikzitpathzx/lem:cc/statement-2.tikz}
		\\[\linesep]
		\input{\tikzitpathzx/lem:cc/statement-3.tikz}
		\eqq{\ref{eqn:pink-spider-def}} 2^{\frac{n+m-2}{2}}~ \input{\tikzitpathzx/lem:cc/proof-2.tikz}
		\eqq{\ref{eqn:red-spider-def}} 2^{\frac{n+m-2}{2}}~ \input{\tikzitpathzx/lem:cc/proof-3.tikz} \\[\linesep]
		\eqq{\hh} 2^{\frac{n+m-2}{2}}~ \input{\tikzitpathzx/lem:cc/statement-4.tikz} \qedhere
	\end{gather*}
\end{proof}

\begin{lemma} \label{lem:sf-pink}
	Pink spiders fuse together. We also call this rule $(\bm{\mathit{sf}})$.
	\begin{equation} \tag{$\bm{\mathit{sf}}$}\label{eqn:sf-pink}
		\input{\tikzitpathzx/lem:pink-sf/statement-1.tikz} = 2^{n-1}~ \input{\tikzitpathzx/lem:pink-sf/statement-2.tikz}
	\end{equation}
\end{lemma}
\begin{proof}
	\begin{gather*}
		\input{\tikzitpathzx/lem:pink-sf/proof-1.tikz}
		\eqq{\ref{eqn:pink-spider-def}} 2^{\frac{a+b+n-2}{2}}2^{\frac{c+d+n-2}{2}}~ \input{\tikzitpathzx/lem:pink-sf/proof-2.tikz}
		\eqq{\ref{eqn:red-spider-def}} 2^{\frac{a+b+c+d+2n-4}{2}}~ \input{\tikzitpathzx/lem:pink-sf/proof-3.tikz} \\[\linesep]
		\eqq{\hh} 2^{\frac{a+b+c+d+2n-4}{2}}~ \input{\tikzitpathzx/lem:pink-sf/proof-4.tikz}
		\eqq{\sf} 2^{\frac{a+b+c+d+2n-4}{2}}~ \input{\tikzitpathzx/lem:pink-sf/proof-5.tikz} \\[\linesep]
		\eqq{\ref{eqn:red-spider-def}} 2^{\frac{a+b+c+d+2n-4}{2}}~ \input{\tikzitpathzx/lem:pink-sf/proof-6.tikz}
		\eqq{\ref{eqn:pink-spider-def}} 2^{n-1}~ \input{\tikzitpathzx/lem:pink-sf/statement-2.tikz} \qedhere
	\end{gather*}
\end{proof}

\begin{lemma}
	The zero box disconnects:
	\begin{equation} \label{eqn:box-zero}
		\input{\tikzitpathzx/lem:box-zero/statement-1.tikz} ~=~ \input{\tikzitpathzx/lem:box-zero/statement-2.tikz}
	\end{equation}
\end{lemma}
\begin{proof}
	\[
	\input{\tikzitpathzx/lem:box-zero/statement-1.tikz}
	\eqq{\sf} \input{\tikzitpathzx/lem:box-zero/proof-1.tikz}
	\eqq{\zero} \input{\tikzitpathzx/lem:box-zero/proof-2.tikz}
	\eqq{\cp} \input{\tikzitpathzx/lem:box-zero/statement-2.tikz}
	\]
\end{proof}

\begin{lemma}
	Pink spiders can be decomposed as follows:
	\begin{equation} \label{eqn:pink-decompose}
		\input{\tikzitpathzx/lem:pink-decompose/statement-1.tikz} ~=~ \frac{1}{2} \left( \input{\tikzitpathzx/lem:pink-decompose/statement-3.tikz} ~+~ \input{\tikzitpathzx/lem:pink-decompose/statement-4.tikz} \right)
		\qquad
		\input{\tikzitpathzx/lem:pink-decompose/statement-2.tikz} ~=~ \frac{1}{2} \left( \input{\tikzitpathzx/lem:pink-decompose/statement-3.tikz} ~-~ \input{\tikzitpathzx/lem:pink-decompose/statement-4.tikz} \right)
	\end{equation}
\end{lemma}
\begin{proof}
	\begin{gather*}
	\input{\tikzitpathzx/lem:pink-decompose/proof-1.tikz}
	\eqq{\cc} 2^{\frac{n+m-2}{2}}~ \input{\tikzitpathzx/lem:pink-decompose/proof-2.tikz}
	\eqq{\ref{eqn:green-spider-def}} 2^{\frac{n+m-2}{2}}~ \left( \input{\tikzitpathzx/lem:pink-decompose/proof-3.tikz} ~+~ (-1)^k~ \input{\tikzitpathzx/lem:pink-decompose/proof-4.tikz} \right) \\[\linesep]
	\eqq{\cc} 2^{\frac{n+m-2}{2}}~ \left( 2^{-\frac{n+m}{2}}~ \input{\tikzitpathzx/lem:pink-decompose/statement-3.tikz} ~+~ 2^{-\frac{n+m}{2}}(-1)^k~ \input{\tikzitpathzx/lem:pink-decompose/statement-4.tikz} \right) \\[\linesep]
	~=~ \frac{1}{2} \left( \input{\tikzitpathzx/lem:pink-decompose/statement-3.tikz} ~+~ (-1)^k~ \input{\tikzitpathzx/lem:pink-decompose/statement-4.tikz} \right) \qedhere
	\end{gather*}
\end{proof}

\begin{lemma}~\cite{wang2020algebraic}
	Hopf rule:
	\begin{equation} \tag{$\bm{\mathit{ho}}$}\label{eqn:hopf}
		\input{\tikzitpathzx/lem:hopf/statement-1.tikz} ~=~ \input{\tikzitpathzx/lem:hopf/statement-2.tikz}
	\end{equation}
\end{lemma}

\begin{lemma}~\cite{wang2020algebraic}
	Strong complementarity:
	\begin{equation} \tag{$\bm{\mathit{sc}}$}\label{eqn:sc}
		\input{\tikzitpathzx/lem:sc/statement-1.tikz} ~=~ \input{\tikzitpathzx/lem:sc/statement-2.tikz}
	\end{equation}
\end{lemma}

%\begin{lemma}
%	For $k \in \{0,1\}$ we have
%	\begin{equation}
%		\tikzfig{zx/lem:box-scalar/statement} ~=~ a^k
%	\end{equation}
%\end{lemma}
%\begin{proof}
%	The case $k = 0$ holds by $(\ety)$. For $k = 1$ we get $\tikzfig{zx/lem:box-scalar/proof} ~= \bra{1}(\ket{0} + a\ket{1}) = \braket{1}{0} + a\braket{1}{1} = 1$.
%\end{proof}

\begin{lemma}~\cite{wang2020algebraic}
	Pink $\pi$ copies through and negates phases:
	\begin{equation} \tag{$\bm{\mathit{\pi}}$}\label{eqn:pi}
	\input{\tikzitpathzx/lem:pi/statement-1.tikz} ~=~ e^{i\alpha}~\input{\tikzitpathzx/lem:pi/statement-2.tikz}
	\end{equation}
\end{lemma}

\begin{lemma}~\cite{wang2020algebraic}
	For $x,y \in \{0,1\}$ we have
	\begin{equation} \tag{$\bm{\mathit{cp}}$}\label{eqn:cp}
		\input{\tikzitpathzx/lem:cp/statement-1.tikz} ~=~ a^x~\input{\tikzitpathzx/lem:cp/statement-2.tikz}
		\qquad\qquad
		\input{\tikzitpathzx/lem:cp/statement-3.tikz} ~=~ (-1)^{xy}~\input{\tikzitpathzx/lem:cp/statement-4.tikz}
	\end{equation}
\end{lemma}

\begin{lemma}~\cite{wang2020algebraic}
	Pink $\pi$ transposes the triangle:
	\begin{equation} \label{eqn:tri-pi-transpose}
		\input{\tikzitpathzx/lem:tri-pi-transpose/statement-1.tikz} ~=~ \input{\tikzitpathzx/lem:tri-pi-transpose/statement-2.tikz}
	\end{equation}
\end{lemma}

\begin{lemma}
	The triangle acts as a change of bases:
	\begin{equation} \tag{$\bm{\mathit{tri}}$}\label{eqn:tri}
		\begin{aligned}[c]
			\input{\tikzitpathzx/rules/tr0-1.tikz} ~&=~ \input{\tikzitpathzx/rules/tr0-2.tikz}
			\qquad&\qquad
			\input{\tikzitpathzx/rules/tr1-1.tikz} ~&=~ \input{\tikzitpathzx/rules/tr1-2.tikz}
			\\
			\input{\tikzitpathzx/lem:tri/statement-1.tikz} ~&=~ \input{\tikzitpathzx/lem:tri/statement-2.tikz}
			\qquad&\qquad
			\input{\tikzitpathzx/lem:tri/statement-3.tikz} ~&=~ \input{\tikzitpathzx/lem:tri/statement-4.tikz}
		\end{aligned}
	\end{equation}
\end{lemma}
\begin{proof}
	The first two equations are just $(\triTipZero)$ and $(\triTipOne)$.
	The third equation has been proven in~\cite{wang2020algebraic}:
	\[
	\input{\tikzitpathzx/lem:tri/statement-1.tikz}
	\eqq{\zero} \input{\tikzitpathzx/lem:tri/proof-1.tikz}
	\eqq{\suc} \input{\tikzitpathzx/lem:tri/proof-2.tikz}
	~=~ \input{\tikzitpathzx/lem:tri/statement-2.tikz}
	\]
	Then, the second equations follow from
	\[
	\input{\tikzitpathzx/lem:tri/statement-3.tikz}
	\eqq{\sf} \input{\tikzitpathzx/lem:tri/proof-3.tikz}
	\eqq{\ref{eqn:tri-pi-transpose}} \input{\tikzitpathzx/lem:tri/proof-4.tikz}
	\eqq{\triTipZero} \input{\tikzitpathzx/lem:tri/statement-4.tikz} \qedhere
	\]
\end{proof}

\begin{lemma}
	The two-legged W spider satisfies
	\begin{equation} \label{eqn:W2-act}
		\input{\tikzitpathzx/lem:W-action2/statement-1.tikz} ~=~ \input{\tikzitpathzx/lem:W-action2/statement-2.tikz}
		\qquad\qquad
		\input{\tikzitpathzx/lem:W-action2/statement-3.tikz} ~=~ \input{\tikzitpathzx/lem:W-action2/statement-4.tikz}
	\end{equation}
\end{lemma}
\begin{proof}
	\[
	\input{\tikzitpathzx/lem:W-action2/statement-1.tikz}
	\eqq{\wdc} \input{\tikzitpathzx/lem:W-action2/a/1.tikz}
	\eqq{\cp,\sf} \input{\tikzitpathzx/lem:W-action2/a/2.tikz}
	\eqq{\tri} \input{\tikzitpathzx/lem:W-action2/a/3.tikz}
	\eqq{\cp,\sf} \input{\tikzitpathzx/lem:W-action2/statement-2.tikz}
	\]
	\[
	\input{\tikzitpathzx/lem:W-action2/statement-3.tikz}
	\eqq{\wdc} \input{\tikzitpathzx/lem:W-action2/b/1.tikz}
	\eqq{\cp,\sf} \input{\tikzitpathzx/lem:W-action2/b/2.tikz}
	\eqq{\tri} \input{\tikzitpathzx/lem:W-action2/b/3.tikz}
	\eqq{\id} \input{\tikzitpathzx/lem:W-action2/statement-4.tikz} \qedhere
	\]
\end{proof}

\begin{lemma} \label{lem:W-action}
	In general, the W spider acts on the computational basis as follows:
	\begin{equation} \tag{$\bm{\mathit{w}}$}\label{eqn:w}
		\input{\tikzitpathzx/lem:W-action/statement-1.tikz} ~=~ \input{\tikzitpathzx/lem:W-action/statement-2.tikz}
		\qquad\qquad
		\input{\tikzitpathzx/lem:W-action/statement-3.tikz} ~=~ \input{\tikzitpathzx/lem:W-action/statement-4.tikz} ~+~ \input{\tikzitpathzx/lem:W-action/statement-5.tikz} ~+~ ... ~+~ \input{\tikzitpathzx/lem:W-action/statement-6.tikz}
	\end{equation}
\end{lemma}
\begin{proof}
	We prove both equation simultaneously by induction on the number of outputs.
	If the W spider has a single output, the equations hold trivially:
	\[
	\input{\tikzitpathzx/lem:W-action/base/a/0.tikz} \eqq{\ref{eqn:W-spider-def}} \input{\tikzitpathzx/lem:W-action/base/a/1.tikz} 
	\qquad\qquad
	\input{\tikzitpathzx/lem:W-action/base/b/0.tikz} \eqq{\ref{eqn:W-spider-def}} \input{\tikzitpathzx/lem:W-action/base/b/1.tikz} 
	\]
	For the inductive step, we have
	\begin{gather*}
		\input{\tikzitpathzx/lem:W-action/rec/a/0.tikz}
		\eqq{\wf} \input{\tikzitpathzx/lem:W-action/rec/a/1.tikz}
		%\eqq{\wdc} \tikzfig{zx/lem:W-action/rec/a/2}
		%\eqq{\cp,\sf} \tikzfig{zx/lem:W-action/rec/a/3} \\[\linesep]
		%\eqq{\triTipZero} \tikzfig{zx/lem:W-action/rec/a/4}
		\eqq{\ref{eqn:W2-act}} \input{\tikzitpathzx/lem:W-action/rec/a/5.tikz}
		\eqq{\text{IH}} \input{\tikzitpathzx/lem:W-action/rec/a/6.tikz}
	\end{gather*}
	\begin{gather*}
		\input{\tikzitpathzx/lem:W-action/rec/b/0.tikz}
		\eqq{\wf} \input{\tikzitpathzx/lem:W-action/rec/b/1.tikz}
		%\eqq{\wdc} \tikzfig{zx/lem:W-action/rec/b/2}
		%\eqq{\cp,\sf} \tikzfig{zx/lem:W-action/rec/b/3} \\[\linesep]
		\eqq{\ref{eqn:W2-act}} \input{\tikzitpathzx/lem:W-action/rec/b/4.tikz}
		\eqq{*} \input{\tikzitpathzx/lem:W-action/rec/b/5-1.tikz} ~+~ \input{\tikzitpathzx/lem:W-action/rec/b/5-2.tikz} \\[\linesep]
		\eqq{IH} \input{\tikzitpathzx/lem:W-action/statement-4.tikz} ~+~ \input{\tikzitpathzx/lem:W-action/statement-5.tikz} ~+~ ... ~+~ \input{\tikzitpathzx/lem:W-action/statement-6.tikz}
	\end{gather*}
	where the step $(*)$ follows from
	\begin{align*}
		\input{\tikzitpathzx/lem:W-action/pi-1.tikz}
		~&= \ket{+}\bra{+} - \ket{-}\bra{-} \\
		&= \frac{1}{2}(\ket{0}+\ket{1})(\bra{0}+\bra{1}) - \frac{1}{2}(\ket{0}-\ket{1})(\bra{0}-\bra{1}) \\
		&= \ket{0}\bra{1} + \ket{1}\bra{0} \\
		&=~ \input{\tikzitpathzx/lem:W-action/pi-2.tikz} + \input{\tikzitpathzx/lem:W-action/pi-3.tikz} \qedhere
	\end{align*}
\end{proof}

\begin{lemma}
	Plugging a pink dot into a two-legged spider produces identity:
	\begin{equation} \label{eqn:W2-plug-leg}
		\input{\tikzitpathzx/lem:W2-plug-leg/statement-1.tikz} ~=~ \input{\tikzitpathzx/lem:W2-plug-leg/statement-2.tikz}
	\end{equation}
\end{lemma}
\begin{proof}
	\[ 
	\input{\tikzitpathzx/lem:W2-plug-leg/statement-1.tikz}
	\eqq{\wdc} \input{\tikzitpathzx/lem:W2-plug-leg/proof-1.tikz}
	\eqq{\cp,\sf} \input{\tikzitpathzx/lem:W2-plug-leg/proof-2.tikz}
	\eqq{\tri,\sf} \input{\tikzitpathzx/lem:W2-plug-leg/proof-3.tikz}
	\eqq{\id} \input{\tikzitpathzx/lem:W2-plug-leg/statement-2.tikz} \qedhere
	\]
\end{proof}

\begin{lemma}
	Plugging a pink dot into a W spider makes the leg disappear:
	\begin{equation} \label{eqn:W-plug-leg}
		\input{\tikzitpathzx/lem:W-plug-leg/statement-1.tikz} ~=~ \input{\tikzitpathzx/lem:W-plug-leg/statement-2.tikz}
	\end{equation}
\end{lemma}
\begin{proof}
	By induction on the number of outputs.
	The base case holds by (\ref{eqn:W2-plug-leg}).
	For the inductive step we have
	\[
	\input{\tikzitpathzx/lem:W-plug-leg/proof-1.tikz}
	\eqq{\wf} \input{\tikzitpathzx/lem:W-plug-leg/proof-2.tikz}
	\eqq{\text{IH}} \input{\tikzitpathzx/lem:W-plug-leg/proof-3.tikz}
	\eqq{\wf} \input{\tikzitpathzx/lem:W-plug-leg/proof-4.tikz} \qedhere
	\]
\end{proof}

\begin{lemma}
	The two-legged W spider adds boxes:
	\begin{equation} \label{eqn:W2-add}
		\input{\tikzitpathzx/lem:W2-add/statement-1.tikz} ~=~ \input{\tikzitpathzx/lem:W2-add/statement-2.tikz}
	\end{equation}
\end{lemma}
\begin{proof}
	If $a = 0$, we have
	\[
	\input{\tikzitpathzx/lem:W2-add/proof-1.tikz}
	\eqq{\zero} \input{\tikzitpathzx/lem:W2-add/proof-2.tikz}
	\eqq{\ref{eqn:W2-plug-leg}} \input{\tikzitpathzx/lem:W2-add/proof-3.tikz}
	\]
	If $a \neq 0$, we have
	\begin{gather*}
		\input{\tikzitpathzx/lem:W2-add/statement-1.tikz}
		\eqq{\sf,\pcy} \input{\tikzitpathzx/lem:W2-add/proof-4.tikz}
		%\eqq{\wdc} \tikzfig{zx/lem:W2-add/proof-5}
		\eqq{\ref{eqn:tri-def}} \input{\tikzitpathzx/lem:W2-add/proof-6.tikz}
		\eqq{\suc} \input{\tikzitpathzx/lem:W2-add/proof-7.tikz}
		\eqq{\sf} \input{\tikzitpathzx/lem:W2-add/statement-2.tikz} \qedhere
	\end{gather*}
\end{proof}

\begin{corollary}
	The W spider adds boxes:
	\begin{equation} \label{eqn:W-add}
		\input{\tikzitpathzx/lem:W-add/statement-1.tikz} ~=~ \input{\tikzitpathzx/lem:W-add/statement-2.tikz}
	\end{equation}
\end{corollary}
\begin{proof}
	Follows by induction on $n$ using (\ref{eqn:W2-add}).
\end{proof}

\chapter{Diagrammatic Differentiation} \label{chap:diagrammatic-differentiation}

\def\tikzitpath{chapter3/figs/}

For our diagrammatic analysis of gradient-based QML, we crucially need a graphical representation of derivatives.
This so-called \textit{diagrammatic differentiation} for ZX-calculus was first discovered in~\cite{yeung2020diagrammatic} and~\cite{zhao2021analyzing} and subsequently generalised to tensor calculi based on monoidal categories~\cite{toumi2021diagrammatic}.
Recently, Wang and Yeung~\cite{wang2022differentiating} developed a more compact graphical representation of derivatives avoiding sums of diagrams using the ZXW-calculus.

We give an overview on diagrammatic differentiation in \Cref{sec:diff-background}, following the treatment by Wang and Yeung~\cite{wang2022differentiating}.
In \Cref{sec:diff-circ}, we present a novel, simplified gradient representation for the special case of parametrised quantum circuits (\Cref{thm:circ-diff}) that we will use for the remainder of the thesis.
Finally, we discuss some properties of this representation in \Cref{sec:diff-facts}.

\section{Background} \label{sec:diff-background}

Recall that we can interpret every ZX diagram $D$ with $n$ inputs and $m$ outputs as a matrix $\mathbb C^{2^m\times 2^n}$.
The derivative of a parametrised diagram $D(\theta)$, written $\frac{\partial}{\partial\theta}D(\theta)$, is defined as the gradient of the matrix associated with $D(\theta)$.
Consider for example a single-legged green spider:
\[
\input{\tikzitpathex:single-leg/1.tikz}
~=~ \def\arraystretch{.8} \begin{pmatrix}
	1 \\
	e^{i\theta}
\end{pmatrix}
\qquad\Rightarrow\qquad
\frac{\partial}{\partial\theta} \left[ \input{\tikzitpathex:single-leg/1.tikz} \right]
= \begin{pmatrix}
	0 \\
	ie^{i\theta}
\end{pmatrix}
\]
The goal of diagrammatic differentiation is to represent those gradients as ZX(W) diagrams.
For example, by inspecting the gradient matrix of our single-legged spider, we observe that
\[ \frac{\partial}{\partial\theta} \left[ \input{\tikzitpathex:single-leg/1.tikz} \right] ~=~ ie^{i\theta} ~ \input{\tikzitpathex:single-leg/2.tikz} \eqq{\cp} i ~ \input{\tikzitpathex:single-leg/3.tikz} \]
In fact, this is one of the key equations of diagrammatic differentiation.
However, to cover the most general case, we should also to consider the possibility that the phase of the spider is a different function in $\theta$.
The resulting equation is very similar to the rule above:

\begin{lemma} \label{lem:spider-diff}
	Let $f$ be a differentiable real function. Then
	\begin{equation} \label{eqn:spider-diff}
		\frac{\partial}{\partial\theta} \left[ \input{\tikzitpathlem:spider-diff/spider.tikz} \right] ~=~ if'(\theta)~\input{\tikzitpathlem:spider-diff/spider-pi.tikz}
	\end{equation}
\end{lemma}
\begin{proof}
	We have
	\[ 
	\frac{\partial}{\partial\theta} \left[ \input{\tikzitpathlem:spider-diff/spider.tikz} \right]
	\eqq{\ref{eqn:green-spider-def}} \frac{\partial}{\partial\theta} \left( \ket{0} + e^{if(\theta)}\ket{1} \right)
	= if'(\theta) \cdot e^{if(\theta)}\ket{1}
	\eqq{\cp} if'(\theta)~\input{\tikzitpathlem:spider-diff/spider-pi.tikz} \qedhere
	\]
\end{proof}

Furthermore, we note that when differentiating a larger diagram, we can ignore the parts that do not depend on $\theta$.
This property is called \textit{linearity}~\cite{yeung2020diagrammatic}:

\begin{lemma}[Linearity] \label{lem:linearity}
	Let $D(\theta)$ be a parametrised ZX diagram depending on $\theta$ and let $E$ be a ZX diagram in which $\theta$ does not occur. Then
	\begin{gather*}
		\ddx{\theta} \left[ \input{\tikzitpathlem:linearity/ED.tikz} \right] = \ddx{\theta} \left[ \input{\tikzitpathlem:linearity/D.tikz} \right] \circ \input{\tikzitpathlem:linearity/E.tikz} 
		\\[5pt]
		\ddx{\theta} \left[ \substack{\input{\tikzitpathlem:linearity/D.tikz} \\[3pt] \input{\tikzitpathlem:linearity/E.tikz}} \right] = {\let\scriptstyle\textstyle \substack{\ddx{\theta} \left[ \input{\tikzitpathlem:linearity/D.tikz} \right] \\[3pt] \phantom{\ddx{\theta}} \input{\tikzitpathlem:linearity/E.tikz}}}
	\end{gather*}
	The equations also hold when switching the order of $D(\theta)$ and $E$.
\end{lemma} 
\begin{proof}
	Directly follows from the linearity of matrix differentiation for multiplication and tensor product, since for matrices $D(\theta), E$ we have $\ddx{\theta} \left[ D(\theta) E \right] = \left( \ddx{\theta} \left[ D(\theta) \right] \right) E$ and $\ddx{\theta} \left[ D(\theta) \otimes E \right] = \left( \ddx{\theta} \left[ D(\theta) \right] \right) \otimes E$.
\end{proof}

In the following, we will use brackets to denote the parts of the diagram we are differentiating.
Using the previous two lemmas, we can give the derivative of any green and red spider:
\begin{gather}
	\label{eqn:green-diff}
	\frac{\partial}{\partial\theta} \left[ \input{\tikzitpathex:green-diff/1.tikz} \right]
	\eqq{\sf,\text{Lem }\ref{lem:linearity}} \frac{\partial}{\partial\theta} ~ \input{\tikzitpathex:green-diff/2.tikz}
	\eqq{\ref{eqn:spider-diff}} if'(\theta)~ \input{\tikzitpathex:green-diff/3.tikz}
	\eqq{\sf} if'(\theta) ~ \input{\tikzitpathex:green-diff/4.tikz}
	\\
	\label{eqn:red-diff}
	\frac{\partial}{\partial\theta} \left[ \input{\tikzitpathex:red-diff/1.tikz} \right]
	\eqq{\sf,\cc,\text{Lem }\ref{lem:linearity}} \frac{\partial}{\partial\theta} ~ \input{\tikzitpathex:red-diff/2.tikz}
	%\eqq{\ref{eqn:spider-diff}} \sqrt 2 \cdot if'(\theta)~ \tikzfig{ex:red-diff/3}
	\eqq{\ref{eqn:spider-diff},\cc,\sf} \frac{if'(\theta)}{\sqrt 2} ~ \input{\tikzitpathex:red-diff/4.tikz}
\end{gather}

This allows us to differentiate all ZX diagrams with only a single parametrised spider.
In general, given some ZX diagram $D(\theta)$ in which $n$ spiders depend on $\theta$, we can always fuse out the parametrised spiders similar to first step in equations (\ref{eqn:green-diff}) and (\ref{eqn:red-diff}) and obtain a diagram of shape
\[ \input{\tikzitpathdiagram-fuse.tikz} \]
where $\theta$ does not occur in $D'$.
However, if we want to differentiate such diagrams using \Cref{lem:spider-diff}, we have to make use of the product rule:
\begin{lemma}[Product Rule] \label{lem:product-rule}
	Let $D(\theta)$ and $E(\theta)$ be parametrised ZX diagram depending on $\theta$. Then
	\begin{gather*}
		\ddx{\theta} \left[ \input{\tikzitpathlem:product-rule/DE.tikz} \right]
		=  \input{\tikzitpathlem:product-rule/E.tikz} \circ \ddx{\theta} \left[ \input{\tikzitpathlem:product-rule/D.tikz} \right] + \ddx{\theta} \left[ \input{\tikzitpathlem:product-rule/E.tikz} \right] \circ  \input{\tikzitpathlem:product-rule/D.tikz}
		\\[5pt]
		\ddx{\theta} \left[ \substack{\input{\tikzitpathlem:product-rule/D.tikz} \\[3pt] \input{\tikzitpathlem:product-rule/E.tikz}} \right] = {\let\scriptstyle\textstyle \substack{\ddx{\theta} \left[ \input{\tikzitpathlem:product-rule/D.tikz} \right] \\[3pt] \phantom{\ddx{\theta}} \input{\tikzitpathlem:product-rule/E.tikz}} ~+~ \substack{ \phantom{\ddx{\theta}}  \input{\tikzitpathlem:product-rule/D.tikz} \\[3pt] \ddx{\theta} \left[ \input{\tikzitpathlem:product-rule/E.tikz} \right]} }
	\end{gather*}
\end{lemma}
\begin{proof}
	Directly follows from the product rule for matrix differentiation.
\end{proof}

This allows us to differentiate diagrams with multiple occurrences of $\theta$, for example
\begin{align*}
	&\qquad\quad \ddx{\theta} \left[ \input{\tikzitpathex:product-rule/1.tikz} \right] \\[\linesep]
	\eqqa{\text{Lem }\ref{lem:product-rule}} \ddx{\theta} \input{\tikzitpathex:product-rule/2-1.tikz} ~+~ \ddx{\theta} \input{\tikzitpathex:product-rule/2-2.tikz} ~+~ \ddx{\theta} \input{\tikzitpathex:product-rule/2-3.tikz} \\[\linesep]
	\eqqa{\ref{eqn:spider-diff}} if_1'(\theta)~\input{\tikzitpathex:product-rule/3-1.tikz} ~+~ if_2'(\theta)~\input{\tikzitpathex:product-rule/3-2.tikz} ~+~ if_3'(\theta)~\input{\tikzitpathex:product-rule/3-3.tikz}
\end{align*}

Unfortunately, the ZX-calculus is not well-equipped to deal with such linear combinations of diagrams.
In particular, there are no rewrite rules that involve sums, which means that we would need to rewrite and simplify each term separately.
Clearly, it would be more convenient if we could express the derivative as a single diagram.
Luckily, Wang and Yeung~\cite{wang2022differentiating} developed a technique to achieve this in ZXW using the W spider:

\begin{theorem}[Wang and Yeung~\cite{wang2022differentiating}] \label{thm:zx-diff}
	Let $f_1,...,f_n$ be real differentiable functions and $D$ a ZX diagram. Then
	\[
	\ddx{\theta} \left[ \input{\tikzitpaththm:zx-diff/statement-1.tikz} \right]
	~=~ i ~ \input{\tikzitpaththm:zx-diff/statement-2.tikz}
	\]
\end{theorem}
\begin{proof}
	Follows from $(\w)$ and \Cref{lem:spider-diff}. The detailed proof can be found as Theorem 15 in~\cite{wang2022differentiating}.
\end{proof}

\section{Differentiating Quantum Circuits} \label{sec:diff-circ}

While \Cref{thm:zx-diff} can be used to obtain the derivative of any ZX diagram, for the purposes of this thesis, we are only interested in the special case of ZX diagrams representing parametrised quantum circuits.
We prove a novel, simplified version of \Cref{thm:zx-diff} for this special case that we will use for the remainder of this thesis.

To motivate the idea, recall that a two-legged green spider corresponds to the $R_Z$ gate up to a global phase:
\[ \def\arraystretch{.8}
R_Z(\theta) 
= \begin{pmatrix}
	e^{-i\frac{\theta}{2}} & 0 \\
	0 & e^{i\frac{\theta}{2}}
\end{pmatrix}
= e^{-i\frac{\theta}{2}} ~ \input{\tikzitpathex:RZ/RZ.tikz}
\]
Using Stone's theorem (see \Cref{thm:stone}), we can see that the derivative of the $R_Z$ gate is given by
\begin{gather*}
	\ddx{\theta} R_Z(\theta) 
	= \ddx{\theta} e^{-i\frac{\theta}{2}Z} 
	= -\frac{i}{2}Ze^{-i\frac{\theta}{2}Z} 
	= -\frac{i}{2}ZR_Z(\theta) \\
	= -\frac{i}{2}e^{-i\frac{\theta}{2}} ~ \input{\tikzitpathex:RZ/RZ-diff-1.tikz} 
	= -\frac{i}{2}e^{-i\frac{\theta}{2}} ~ \input{\tikzitpathex:RZ/RZ-diff-2.tikz}
\end{gather*}
Compared to Lemma \ref{lem:spider-diff}, we obtain the derivative by simply adding $\pi$ to the phase.
Similarly, we obtain an alternative version of \Cref{lem:spider-diff} by adding a global phase:

\begin{lemma} \label{lem:spider-diff-phase}
	Let $f$ be a differentiable real function. Then
	\begin{equation} \label{eqn:spider-diff-phase}
		\frac{\partial}{\partial\theta} \left[ e^{-i\frac{f(\theta)}{2}}~\input{\tikzitpathlem:spider-diff-phase/spider.tikz} \right] = -\frac{if'(\theta)}{2} e^{-i\frac{f(\theta)}{2}}~\input{\tikzitpathlem:spider-diff-phase/spider-pi.tikz}
	\end{equation}
\end{lemma}
\begin{proof}
	We have 
	\begin{align*}
	\frac{\partial}{\partial\theta} \left[ e^{-i\frac{f(\theta)}{2}}~\input{\tikzitpathlem:spider-diff-phase/spider.tikz} \right] 
	\eqqa{\ref{eqn:green-spider-def}} \frac{\partial}{\partial\theta} \left( e^{-i\frac{f(\theta)}{2}}\ket{0} + e^{i\frac{f(\theta)}{2}}\ket{1} \right) \\
	&= -\frac{if'(\theta)}{2} e^{-i\frac{f(\theta)}{2}}\ket{0} + \frac{if'(\theta)}{2} e^{i\frac{f(\theta)}{2}}\ket{1} \\
	%&= -\frac{if'(\theta)}{2} e^{-i\frac{f(\theta)}{2}} (\ket{0} - e^{i\theta}\ket{1}) \\
	&= -\frac{if'(\theta)}{2} e^{-i\frac{f(\theta)}{2}} (\ket{0} + e^{i(f(\theta)+\pi)}\ket{1}) \\
	\eqqa{\ref{eqn:green-spider-def}} -\frac{if'(\theta)}{2} e^{-i\frac{f(\theta)}{2}}~\input{\tikzitpathlem:spider-diff-phase/spider-pi.tikz} \qedhere
	\end{align*}
\end{proof}

Since we work with quantum circuits, we can ignore global phases as they cancel out because of the doubling construction (see \Cref{sec:zx-quantum}).
Therefore, we can use \Cref{lem:spider-diff-phase} instead of \Cref{lem:spider-diff} to differentiate all spiders in a circuit, yielding a simplified version of \Cref{thm:zx-diff}:

\begin{theorem} \label{thm:circ-diff}
	The derivative of a parametrised quantum circuit can be expressed as the following diagram:
	\[ \frac{\partial}{\partial\theta} \left[  \input{\tikzitpaththm:circ-diff/circ.tikz} \right] = -2^{n-1}i~ \input{\tikzitpaththm:circ-diff/diff.tikz} \]
	In other words, we can replace the triangles in \Cref{thm:zx-diff} with Hadamards.
\end{theorem}
\begin{proof}
	\renewcommand{\linesep}{20pt}
	\begin{align*}
	&\phantom{={}} \frac{\partial}{\partial\theta} \left[ \input{\tikzitpaththm:circ-diff/circ.tikz} \right] %\\[\linesep]
	= \frac{\partial}{\partial\theta} \left[ \input{\tikzitpaththm:circ-diff/proof-1.tikz} \right] \\[\linesep]
	\eqqa{\text{Lem }\ref{lem:product-rule},\ref{eqn:spider-diff-phase}}
	\begin{aligned}[t]
		& -\frac{if_1'(\theta)}{2}~\input{\tikzitpaththm:circ-diff/proof-2-1.tikz} \\[5pt]
		& +\frac{if_1'(\theta)}{2}~~~~~~\input{\tikzitpaththm:circ-diff/proof-2-2.tikz} \\[5pt]
		& - ~~~ ... \\
		& + ~~~ ... \\
		& -\frac{if_n'(\theta)}{2}~\input{\tikzitpaththm:circ-diff/proof-2-3.tikz} \\[5pt] & +\frac{if_n'(\theta)}{2}~~~~~~\input{\tikzitpaththm:circ-diff/proof-2-4.tikz}
	\end{aligned} \\[\linesep]
	&=
	\begin{aligned}[t]
		& -\frac{if_1'(\theta)}{2}~\input{\tikzitpaththm:circ-diff/proof-20-1.tikz} ~+~ \frac{if_1'(\theta)}{2}~\input{\tikzitpaththm:circ-diff/proof-20-2.tikz} \\
		& - ~~~ ... ~~~ + ~~~ ... \\
		& -\frac{if_n'(\theta)}{2}~\input{\tikzitpaththm:circ-diff/proof-20-3.tikz} ~+~ \frac{if_n'(\theta)}{2}\input{\tikzitpaththm:circ-diff/proof-20-4.tikz}
	\end{aligned} \\[\linesep]
	&= \begin{aligned}[t]
		& -\frac{if_1'(\theta)}{2} \left( \input{\tikzitpaththm:circ-diff/proof-3-1.tikz} ~-~ \input{\tikzitpaththm:circ-diff/proof-3-2.tikz}\right) \\
		& - ~~~ ... \\
		& -\frac{if_n'(\theta)}{2} \left( \input{\tikzitpaththm:circ-diff/proof-3-3.tikz} ~-~ \input{\tikzitpaththm:circ-diff/proof-3-4.tikz}\right)
	\end{aligned} \\[\linesep]
	\eqqa{\sf,\hh}
	\begin{aligned}[t]
		& -\frac{if_1'(\theta)}{2}2^n \left( \input{\tikzitpaththm:circ-diff/proof-4-1.tikz} ~-~ \input{\tikzitpaththm:circ-diff/proof-4-2.tikz}\right) \\
		& - ~~~ ... \\
		& -\frac{if_n'(\theta)}{2}2^n \left( \input{\tikzitpaththm:circ-diff/proof-4-3.tikz} ~-~ \input{\tikzitpaththm:circ-diff/proof-4-4.tikz}\right)
	\end{aligned} \\[\linesep]
	\eqqa{\cp} -2^{n-1}i \cdot 
	\begin{aligned}[t]
		& \left( \input{\tikzitpaththm:circ-diff/proof-5-1.tikz} \right. \\[5pt]
		& ~~~+~ \input{\tikzitpaththm:circ-diff/proof-5-2.tikz} \\[5pt]
		& ~~~+~ ... \\[5pt]
		& ~~~+~ \input{\tikzitpaththm:circ-diff/proof-5-3.tikz} \\[5pt]
		& ~~~+~ \left. \input{\tikzitpaththm:circ-diff/proof-5-4.tikz} \right)
	\end{aligned} \\[\linesep]
	\eqqa{\w} -2^{n-1}i ~~ \input{\tikzitpaththm:circ-diff/diff.tikz}
	\end{align*}
\end{proof}

Thus, in order to differentiate a circuit, we just have to connect the following \textit{differentiation gadget} to the parametrised spiders.
\begin{definition}
	The differentiation gadget is given by the following diagram:
	\[ \input{\tikzitpathdifferentiation-gadget.tikz} \]
\end{definition}

\section{Properties of the Differentiation Gadget} \label{sec:diff-facts}

We close this chapter by deriving some interesting properties of our differentiation gadget.
First, we consider the case where all spider have the same phase $f(\theta)$:
\begin{fact} \label{fact:chain-rule}
	Let $f$ be a differentiable real function. Then
	\[ \ddx{\theta} \left[ \left. \input{\tikzitpathfact:chain-rule/proof-1.tikz} \right\} {\scriptstyle{n}} \right] = -f'(\theta)  \cdot 2^{n-1}i ~ \input{\tikzitpathfact:chain-rule/proof-4.tikz} \]
\end{fact}
\begin{proof}
	By \Cref{thm:circ-diff}, the derivative is given by
	\begin{align*}
		&~ -2^{n-1}i \input{\tikzitpathfact:chain-rule/proof-2.tikz} \\[\linesep]
		\eqqa{\pcy,\sf} -2^{n-1}i \input{\tikzitpathfact:chain-rule/proof-3.tikz} \\[\linesep]
		\eqqa{\cp} -f'(\theta) \cdot 2^{n-1}i ~ \input{\tikzitpathfact:chain-rule/proof-4.tikz}
	\end{align*}
	In the last step we also used the fact that $\input{\tikzitpathfact:chain-rule/proof-5.tikz}$.
\end{proof}

Note that this fact is essentially a version of the chain rule since we have shown
\[ \ddx{\theta} C(f(\theta)) = f'(\theta) \cdot \frac{\partial C}{\partial\theta}(f(\theta)). \]
Another interesting question is how the derivative in \Cref{thm:circ-diff} behaves if one function $f_i$ is constant, i.e. one of the differentiated spiders does not actually depend on $\theta$.
We can graphically show that such a spider does not contribute to the derivative:

\begin{fact}
	Let $f_1,...,f_n$ be differentiable real functions where $f_i$ is a constant function. Then
	\begin{adjustwidth}{-2cm}{-2cm}
		\[ \input{\tikzitpathfact:constant-diff/statement-1.tikz} =~  \frac{1}{2} \input{\tikzitpathfact:constant-diff/statement-2.tikz} \]
	\end{adjustwidth}
\end{fact}
\begin{proof}
	Note that $f_i'(\theta) = 0$ since $f_i$ is constant. Thus
	\begin{align*}
		&\qquad \input{\tikzitpathfact:constant-diff/statement-1.tikz} \\[\linesep]
		\eqqa{\wf} \input{\tikzitpathfact:constant-diff/proof-1.tikz} \\[\linesep]
		\eqqa{\pcy} \input{\tikzitpathfact:constant-diff/proof-2.tikz} \\[\linesep]
		\eqqa{\ref{eqn:box-zero}} \input{\tikzitpathfact:constant-diff/proof-3.tikz} \\[\linesep]
		\eqqa{\w,\cc,\sf} \frac{1}{2} \input{\tikzitpathfact:constant-diff/proof-4.tikz} \\[\linesep]
		\eqqa{\ref{eqn:W-plug-leg}} \input{\tikzitpathfact:constant-diff/statement-2.tikz}
	\end{align*}
\end{proof}

Finally, we emphasize that the ZX diagram representing a given linear map is not unique.
In particular, diagrams representing the same parametrised circuit can differ in the number of parametrised spiders.
A trivial example of this is evidenced by the spider fusion rule:
\[ \input{\tikzitpathex:spider-fuse/1.tikz} \eqq{\sf} \input{\tikzitpathex:spider-fuse/2.tikz} \]
On the left-hand side we have 2 parametrised spiders, whereas we have 4 on the right-hand side.
This also means that the differentiation gadgets that we plug into either side need to have a different number of legs.
Of course, both representations still represent the same linear map.
We can verify this graphically by showing that the differentiation gadget respects spider fusion.
This requires the following auxiliary lemma:

\begin{lemma} \label{lem:W-fuse}
	For all $a,b \in \mathbb C$ we have
	\begin{equation*}
		\input{\tikzitpathlem:W-fuse/statement-1.tikz} ~=~ \input{\tikzitpathlem:W-fuse/statement-2.tikz}
	\end{equation*}
\end{lemma}
\begin{proof}
	If $a=0$, then
	\[
		\input{\tikzitpathlem:W-fuse/zero/proof-1.tikz}
		\eqq{\ref{eqn:box-zero},\sf} \input{\tikzitpathlem:W-fuse/zero/proof-2.tikz}
		\eqq{\ref{eqn:W-plug-leg},\id} \input{\tikzitpathlem:W-fuse/zero/proof-3.tikz}
	\]
	If $a \neq 0$, then
	\begin{gather*}
		\input{\tikzitpathlem:W-fuse/nonzero/proof-1.tikz}
		\eqq{\pcy} \input{\tikzitpathlem:W-fuse/nonzero/proof-2.tikz}
		\eqq{\wdc} \input{\tikzitpathlem:W-fuse/nonzero/proof-3.tikz}
		\eqq{\sf} \input{\tikzitpathlem:W-fuse/nonzero/proof-4.tikz} \\[\linesep]
		\eqq{\ho,\id} \input{\tikzitpathlem:W-fuse/nonzero/proof-5.tikz}
		\eqq{\suc,\sf} \input{\tikzitpathlem:W-fuse/nonzero/proof-6.tikz}
		\eqq{\sf} \input{\tikzitpathlem:W-fuse/nonzero/proof-7.tikz} \qedhere
	\end{gather*}
\end{proof}

Now, it easily follows that the differentiation gadget respects spider fusion:

\begin{fact} \label{fact:fuse-diff}
	Let $f_1,...,f_n,g,h$ be differentiable real functions. Then
	\begin{adjustwidth}{-2cm}{-2cm}
	\[ \input{\tikzitpathfact:fuse-diff/statement-1.tikz} =~ \frac{1}{2} \input{\tikzitpathfact:fuse-diff/statement-2.tikz} \]
	\end{adjustwidth}
\end{fact}
\begin{proof}
	\begin{align*}
		&\qquad\qquad \input{\tikzitpathfact:fuse-diff/statement-1.tikz} \\[\linesep]
		\eqqa{\wf} \input{\tikzitpathfact:fuse-diff/proof-1.tikz} \\[\linesep]
		\eqqa{\sf,\cc} \frac{1}{2} \input{\tikzitpathfact:fuse-diff/proof-2.tikz} \\[\linesep]
		\eqqa{\text{Lem. }\ref{lem:W-fuse},\wf} \frac{1}{2} \input{\tikzitpathfact:fuse-diff/statement-2.tikz}
	\end{align*}
\end{proof}

\chapter{Gradient Recipes} \label{chap:gradient-recipes}

\def\tikzitpath{chapter4/figs/}

This chapter deals with the problem of computing gradients of parametrised quantum circuits.
Given a PQC $C(\vec\theta)$, we are usually interested in the gradient of the expectation value w.r.t. a parameter $\theta_i$, i.e. $\ddx{\theta_i} \langle H \rangle$ for some Hamiltonian $H$.
While we can represent this gradient as a ZXW diagram (c.f. \Cref{thm:circ-diff}), computing it classically is very hard, akin to simulating the quantum system.
Therefore, the gradient computation should be ideally performed on quantum hardware.
We can use linearity (and the product rule if multiple gates depend on $\theta_i$) to break the gradient of the expectation value down to gradients of a single gate $U(\theta_i)$ that depends on $\theta_i$:
Suppose $C(\theta_i) = EU(\theta_i)D$, then
\begin{align*}
	\ddx{\theta_i} \langle H\rangle
	&= \ddx{\theta_i} \bra{0}D^\dagger U(\theta_i)^\dagger E^\dagger HEU(\theta_i)D\ket{0} \\
	&= \ddx{\theta_i} \left[ \input{\tikzitpathexpval-1.tikz} \right] \\
	\eqqa{\text{Lem. }\ref{lem:linearity}} \ddx{\theta_i} ~ \input{\tikzitpathexpval-2.tikz} \numberthis\label{eqn:expval-decompose}
\end{align*}

Ideally, we would like to replace this gate with a new sub-circuit that represents the matrix $\ddx{\theta_i} U(\theta_i)$.
Running this modified circuit would then yield the desired gradient $\ddx{\theta_i} \langle H \rangle$.
Unfortunately, the derivate of a parametrised unitary $U(\theta)$ is usually no longer unitary.
For a trivial example of this, consider the $R_Z(\theta)$ gates whose derivate
\[ 
\ddx{\theta} R_Z(\theta)
= \ddx{\theta} \begin{pmatrix}
	e^{-i\frac{\theta}{2}} & 0 \\
	0 & e^{i\frac{\theta}{2}}
\end{pmatrix}
= \begin{pmatrix}
	-i\frac{\theta}{2}e^{-i\frac{\theta}{2}} & 0 \\
	0 & i\frac{\theta}{2}e^{i\frac{\theta}{2}}
\end{pmatrix}
\]
is clearly not unitary.
One common approach to deal with this issue involves decomposing the gate into a linear combination of $k$ unitaries that can be run on quantum hardware.
Thus, computing the gradient in such a way involves $k$ circuit executions.
Such decompositions are called \textit{gradient recipes}.

An important detail to note here is that the gate $U(\theta_i)$ occurs doubled in (\ref{eqn:expval-decompose}), matching our previous discussion of quantum circuits in ZX.
Thus, we actually have to study decompositions of $\doubled(U(\theta_i))$ into a linear combination $\sum_{i=1}^k x_i\cdot \doubled(V_i)$ of doubled unitaries.
Thus, we can use the circuit differentiation machinery from \Cref{sec:diff-circ} to analyse this problem.

After discussing a custom ZX representation of parametrised unitaries in \Cref{sec:recipes-unitaries}, we focus on a popular class of gradient recipes in \Cref{sec:recipes-shift}, the so-called \textit{parameter-shift rules}.
We give new proofs of various shift rules based on our diagrammatic gradient representation.
Furthermore, we prove a conjecture by Anselmetti et al.~\cite{anselmetti2021local} establishing that their 4-term recipe is optimal.
For this, we prove a no-go theorem lower bounding the number of terms needed to compute gradients of a certain class of circuits in \Cref{sec:nogo}.
Finally, we remark on a gradient recipe using ancillae in \Cref{sec:ancilla-recipe}.

\section{Parametrised Unitaries as ZX Diagrams} \label{sec:recipes-unitaries}

In the literature, gradient recipes are usually derived based on properties of the matrices representing the gates.
For example, the validity of different parameter-shift rules for a unitary $e^{i\theta H}$ depends on the make-up of the eigenvalues of the Hermitian generator $H$.
The goal of this section is to bridge the gap between this eigenvalue-description and the higher-level ZX representation of parametrised unitaries.
This is necessary for our derived rules to be comparable with the results in the literature.

In order to do this, we have to determine the number of parametrised spiders needed to implement a parametrised unitary $U(\theta)$.
Parametrised spiders in this context refer to spiders whose phase is a (non-constant) function in $\theta$.
This number is important, because the cost of our recipes will depend on the number of legs that the differentiation gadget for the unitary has.
We have already seen that this can vary because of the spider-fusion rule (c.f. \Cref{fact:fuse-diff}).
However, there are also less trivial examples.
For instance, consider the $CU_1$ gate that has the following two representations:

\begin{equation} \label{eqn:CU1-def}
	CU_1(\theta) = \begin{pmatrix}
		1 & 0 & 0 & 0 \\
		0 & 1 & 0 & 0 \\
		0 & 0 & 1 & 0 \\
		0 & 0 & 0 & e^{i\theta}
	\end{pmatrix}
	~=~ \input{\tikzitpathex:represent/CU1/3.tikz} ~=~ \input{\tikzitpathex:represent/CU1/2.tikz}	
\end{equation}

where $\input{\tikzitpathex:represent/CU1/and.tikz}$ is the \textit{and-gate} acting like conjunction on the computational basis.

\subsection{Diagonalising Parametrised Unitaries}

First, we reduce the problem to constructing diagrams for diagonal matrices.
For this, let us consider an $n$-dimensional parametrised unitary $U(\theta)$.
By Stone's theorem (\ref{thm:stone}), we know that $U(\theta) = e^{i\theta H}$ for some self-adjoint $n$-dimensional matrix $H$.
By the spectral theorem for finite dimensional self-adjoint matrices, $H$ is diagonalisable.
This means, there is an orthonormal basis $\mathcal B = \{ \ket{v_1}, ..., \ket{v_n} \}$ called the \textit{eigenbasis}, satisfying $H \ket{v_j} = \lambda_j \ket{v_j}$ for all $j$.
Here, $\lambda_j$ are the eigenvalues of $H$ which must all be real since $H$ is self-adjoint.
We can now define a unitary $V := \sum_{j = 1}^n \ket{v_j} \bra{j}$ mapping each computational basis element $\ket{j}$ to the corresponding eigenvector $\ket{v_j}$ in the eigenbasis.
Furthermore, we define $D := \text{diag}(\lambda_1,...,\lambda_n)$ as the diagonal matrix consisting of the eigenvalues of $H$.
Noting that we can also write $D = \sum_{j=1}^n \lambda_j \ket{j}\bra{j}$, we have
\[ VDV^\dagger = \sum_{j_1,j_2,j_3 = 1}^n \lambda_{j_2} \ket{v_{j_1}} \braket{j_1}{j_2} \braket{j_2}{j_3} \bra{v_{j_3}} = \sum_{j=1}^n \lambda_j \ket{v_j} \bra{v_j} = H. \]
Therefore,
\[ U(\theta) = e^{i\theta VDV^\dagger} = V e^{i\theta D} V^\dagger = V \text{diag}(e^{i\theta \lambda_1},...,^{i\theta \lambda_n}) V^\dagger. \]
Note that only the diagonal matrix in the middle depends on the parameter $\theta$.
Hence, when trying to determine the number of parametrised spiders needed to implement $U(\theta)$ in the ZX-calculus, it suffices to look at $e^{i\theta D}$.

\subsection{General Construction}

%The goal of this section is to find a ZX representation for $e^{i\frac{\theta}{2}D}$ that requires a low number of parametrised spiders.
%Suppose the Hermitian generator $H$ has $m$ non-zero eigenvalues that make up the entries of $D$ and denote them by $\Lambda = \{\lambda_1, ..., \lambda_m \}$.
%We begin by describing a naive construction that realises $e^{i\frac{\theta}{2}D}$ using $m$ parametrised spiders.

Consider a $2^n$-dimensional parametrised unitary $U(\theta) = e^{i\theta H}$ whose Hermitian generator $H$ has $m$ non-zero eigenvalues $\lambda_1,...,\lambda_m \neq 0$ (it does not matter if $H$ has 0 as an additional eigenvalue).
In this section, we describe a construction to realise $U(\theta)$ in the ZX-calculus using $m$ parametrised spiders.
Following the previous section, we perform a diagonalization $U(\theta) = V^\dagger e^{i\theta D}V$ where $D$ is a diagonal matrix with non-zero entries $\lambda_1,...,\lambda_m$.
We define Boolean functions $f_{\lambda_1},...,f_{\lambda_m}$ that match on the eigenvectors corresponding to $\lambda_1,...\lambda_m$:
\[ f_{\lambda_j}(\vec x) = \begin{cases}
1 & \text{if } D\ket{\vec x} = \lambda_j \ket{\vec x} \\
0 & \text{otherwise.}
\end{cases} \]
This allows us to characterise the action of $e^{i\theta D}$ on computational basis states as follows:

%\begin{lemma} \label{lem:D-char}
%	For all $x \in \{0,1\}^n$, we have
%	\[ D \ket{\vec x} = \left( \sum_{j=1}^m \lambda_j \cdot f_{\lambda_j}(\vec x) \right) \ket{\vec x}.  \]
%\end{lemma}
%\begin{proof}
%	If $D\ket{\vec x} = \vec 0$, then also $f_{\lambda_j}(\vec x) = \vec 0$ for all $j$ such that
%	$ \left( \sum_{j=1}^m \lambda_j \cdot f_{\lambda_j}(\vec x) \right) \ket{\vec x} = \left( \sum_{j=1}^m \vec 0 \right) \ket{\vec x} = \vec 0 = D\ket{\vec x}$.
%	Otherwise, there must be $\lambda_j$ with $D\ket{\vec x} = \lambda_j\ket{\vec x}$.
%	This also means that $D\ket{\vec x} \neq \lambda_k\ket{\vec}$ for all $k \neq j$.
%	Therefore,
%	$ \left( \sum_{j=1}^m \lambda_j \cdot f_{\lambda_j}(\vec x) \right) \ket{\vec x} = \lambda_j \ket{\vec x} = D\ket{\vec x}$.
%\end{proof}

\begin{lemma} \label{lem:D-char}
	For all $x \in \{0,1\}^n$, we have
	\begin{equation} \label{eqn:D-char}
		e^{i\theta D} \ket{\vec x} = \left( \prod_{j=1}^m e^{i\theta\lambda_j \cdot f_{\lambda_j}(\vec x)}  \right) \ket{\vec x}.
	\end{equation}
\end{lemma}
\begin{proof}
	$D$ is a diagonal matrix whose entries are either zero or one of the non-zero eigenvalues $\lambda_j$.
	Thus, we either have $D\ket{\vec x} = \vec 0$, or $D\ket{\vec x} = \lambda_j\ket{\vec x}$ for some $j$.
	\begin{itemize}
		\item 
		Suppose $D\ket{\vec x} = \vec 0$, then $e^{i\theta D}\ket{\vec x} = \ket{\vec x}$.
		Furthermore, by definition $f_{\lambda_j}(\vec x) = 0$ for all $j$.
		Thus,
		$ \left( \prod_{j=1}^m e^{i\theta\lambda_j \cdot f_{\lambda_j}(\vec x)}  \right) \ket{\vec x} = \left( \prod_{j=1}^m 1 \right) \ket{\vec x} = \ket{\vec x} = e^{i\theta D}\ket{\vec x}$.
		
		\item
		Suppose $D\ket{\vec x} = \lambda_j \ket{\vec x}$ for some $j$, then $e^{i\theta D}\ket{\vec x} = e^{i\lambda_j\theta} \ket{\vec x}$.
		Furthermore, $f_{\lambda_j}(\vec x) = 1$ and $f_{\lambda_k}(\vec x) = 0$ for all $k \neq j$.
		Therefore,
		$ \left( \prod_{j=1}^m e^{i\theta\lambda_j \cdot f_{\lambda_j}(\vec x)}  \right) \ket{\vec x} = e^{i\lambda_j\theta} \ket{\vec x} = e^{i\theta D}\ket{\vec x}$. \qedhere
	\end{itemize}
\end{proof}

In order to realise this construction diagrammatically, we first need a result regarding the representably of our Boolean functions $f_{\lambda_1},...,f_{\lambda_m}$ in the ZX-calculus:

\begin{lemma}
	For every Boolean function $f : \{0,1\}^n \to \{0,1\}$ there is a ZX diagram such that for all $\vec x \in \mathbb \{0,1\}^n$ we have
	\[ \input{\tikzitpathlem:boolean-func/statement.tikz} \]
\end{lemma}
\begin{proof}
	We can express $f$ as a propositional formula in variables $x_1,...,x_n$ using only conjunction ($\land$) and negation ($\neg$) connectives.\footnote{This follows from the fact that conjunction and negation form a functionally complete set and can thus encode all possible truth tables~\cite{wernick1942complete}.}
	We can implement this formula as a diagram using gates that act like conjunction, negation, and copying on the computational basis.\footnote{The copying is necessary since inputs might be used multiple times.}
	In ZX calculus, those are given by
	\[ \input{\tikzitpathlem:boolean-func/and.tikz} \qquad\qquad \input{\tikzitpathlem:boolean-func/neg.tikz} \qquad\qquad \input{\tikzitpathlem:boolean-func/copy.tikz} \]
	By appropriately wiring those gate together, we get the desired ZX representation of $f$.
\end{proof}

%\begin{corollary}
%	Let $f : \{0,1\}^n \to \{0,1\}$ be a Boolean function, $\alpha \in \mathbb R$, and $\vec x \in \mathbb \{0,1\}^n$. Then
%	\begin{equation}
%		\tikzfig{corr:func-gadget/statement-1} =~ \tikzfig{corr:func-gadget/statement-2} \eqq{\cp} e^{i\alpha\cdot f(\vec x)}.
%	\end{equation}
%\end{corollary}

This yields the following construction in the ZX-calculus:

\begin{theorem} \label{thm:U-repr-naive}
	Let $U(\theta) = e^{i\theta H}$ be a parametrised unitary whose Hermitian generator $H$ has $m$ non-zero eigenvalues $\lambda_1,...,\lambda_m$ and admits the diagonalization $H = V^\dagger DV$.
	Then
	\[ U(\theta) =~ \input{\tikzitpaththm:U-repr-naive/statement.tikz} \]
\end{theorem}
\begin{proof}
	It suffices to show that the middle part is equal to $e^{i\theta D}$.
	We verify this by plugging in a computational basis state $\ket{\vec x}$:
	\begin{gather*}
		\input{\tikzitpaththm:U-repr-naive/proof-1.tikz}
		\eqq{\cp} \input{\tikzitpaththm:U-repr-naive/proof-2.tikz}
		~=~ \input{\tikzitpaththm:U-repr-naive/proof-3.tikz} \\[\linesep]
		\eqq{\cp} \left( \prod_{j=1}^m e^{i\theta\lambda_j \cdot f_{\lambda_j}(\vec x)} \right) ~ \input{\tikzitpaththm:U-repr-naive/proof-4.tikz}
		~=~ \left( \prod_{j=1}^m e^{i\theta\lambda_j \cdot f_{\lambda_j}(\vec x)} \right) \ket{\vec x}
		\eqq{\ref{eqn:D-char}} e^{i\theta D}\ket{\vec x}. \qedhere
	\end{gather*}
\end{proof}

As an example, consider the $CR_Z(\theta)$ and $CU_1(\theta)$ gate, whose Hermitian generators
\[
H_{CR_Z} = \begin{pmatrix}
	0 & 0 & 0 & 0 \\
	0 & 0 & 0 & 0 \\
	0 & 0 & -\frac{1}{2} & 0 \\
	0 & 0 & 0 & \frac{1}{2}
\end{pmatrix}
\qquad\qquad 
H_{CU_1} = \begin{pmatrix}
	0 & 0 & 0 & 0 \\
	0 & 0 & 0 & 0 \\
	0 & 0 & 0 & 0 \\
	0 & 0 & 0 & 1
\end{pmatrix}
\]
have non-zero eigenvalues $-\frac{1}{2}, \frac{1}{2}$, and $1$, respectively.
The functions matching on the eigenvectors are given by $f_{-1/2}(x_1,x_2) = x_1 \land \neg x_2$ and $f_{1/2}(x_1,x_2) = f_1(x_1,x_2) = x_1 \land x_2$.
Invoking \Cref{thm:U-repr-naive}, we get
\[ CR_Z(\theta) ~=~ \input{\tikzitpathex:represent/CRZ/1.tikz} ~=~ \input{\tikzitpathex:represent/CRZ/2.tikz} \eqq{\cp} \input{\tikzitpathex:represent/CRZ/3.tikz} \]
\[ CU_1(\theta) ~=~ \input{\tikzitpathex:represent/CU1/1.tikz} ~=~ \input{\tikzitpathex:represent/CU1/2.tikz} \]
To summarise, we constructed an alternate two-spider representation for $CR_Z(\theta)$ different from (\ref{eqn:CRZ-zx-def}) and recovered the one-spider representation of $CU_1(\theta)$ from (\ref{eqn:CU1-def}).

Note that phase gadgets are a special case of \Cref{thm:U-repr-naive} where the function computes the XOR of its inputs.
Also note that in general, the construction from \Cref{thm:U-repr-naive} is not optimal, in the sense that there might be representations that require less parametrised spiders.
For example, consider the parametrised unitary
\[ 
U(\theta) 
~=~ \input{\tikzitpathex:represent/RZ-RZ/1.tikz}
~=~ \begin{pmatrix}
	1 & 0 & 0 & 0 \\
	0 & e^{i\alpha} & 0 & 0 \\
	0 & 0 & e^{2i\alpha} & 0 \\
	0 & 0 & 0 & e^{3i\alpha}
\end{pmatrix} \]
whose Hermitian generator has three non-zero eigenvalues.
However, we can show that the construction in \Cref{thm:U-repr-naive} is in fact optimal if the eigenvectors are $-\lambda,0,\lambda$:

\begin{factE} \label{fact:U-repr-optimal}
	It is not possible to represent a parametrised unitary $e^{i\theta H}$ whose Hermitian generator has eigenvalue $-\lambda,0,\lambda$ with less than two parametrised spiders.
\end{factE}
\begin{proofE}
	\def\tikzitpath{chapter4/figs/}
	We first prove than we cannot represent the $CR_Z(\theta)$ gate using less than two parametrised spiders.	
	Suppose we had
	\[ CR_Z(\theta) = e^{ig(\theta)}~\input{\tikzitpathlem:CRZ-no-go/proof-1.tikz} \]
	But then 
	\begin{equation} \label{eqn:lem-CRZ-no-go/ddx}
	\begin{aligned}[b]
		\ddx{\theta} CR_Z(\theta) 
		\eqqa{\ref{eqn:spider-diff}} ig'(\theta)e^{ig(\theta)}~\input{\tikzitpathlem:CRZ-no-go/proof-1.tikz} ~+~ e^{ig(\theta)}if'(\theta) ~ \input{\tikzitpathlem:CRZ-no-go/proof-2.tikz} \\[\linesep]
		\eqqa{\cp} ig'(\theta)CR_Z(\theta) ~+~ e^{ig(\theta)}if'(\theta)e^{if(\theta)} ~ \input{\tikzitpathlem:CRZ-no-go/proof-3.tikz}
	\end{aligned}
	\end{equation}
	Note that the diagram on the right-hand side no longer depends on $\theta$.
	Since $\ddx{\theta} CR_Z(\theta) = \text{diag}(0, 0, -\frac{i\theta}{2}e^{-i\frac{\theta}{2}}, \frac{i\theta}{2}e^{i\frac{\theta}{2}})$ is diagonal, we must also have
	\[ \input{\tikzitpathlem:CRZ-no-go/proof-3.tikz} ~=~ \text{diag}(a,b,c,d) \]
	for some constants $a,b,c,d \in \mathbb C$.
	By comparing the diagonal entries to (\ref{eqn:lem-CRZ-no-go/ddx}), we get the equations
	\begin{align}
		g'(\theta)+ae^{i(f(\theta) + g(\theta))}f'(\theta) &= 0 \label{eqn:lem-CRZ-no-go/1} \\
		g'(\theta)+be^{i(f(\theta) + g(\theta))}f'(\theta) &= 0 \label{eqn:lem-CRZ-no-go/2} \\
		g'(\theta)e^{-i\frac{\theta}{2}}+ce^{i(f(\theta) + g(\theta))}f'(\theta) &= -\frac{\theta}{2}e^{-i\frac{\theta}{2}} \label{eqn:lem-CRZ-no-go/3} \\
		g'(\theta)e^{i\frac{\theta}{2}}+de^{i(f(\theta) + g(\theta))}f'(\theta) &= \frac{\theta}{2}e^{i\frac{\theta}{2}} \label{eqn:lem-CRZ-no-go/4}
	\end{align}
	Since (\ref{eqn:lem-CRZ-no-go/1}) implies $g'(\theta) = -ae^{i(f(\theta) + g(\theta))}f'(\theta)$, we can rewrite (\ref{eqn:lem-CRZ-no-go/3}) and (\ref{eqn:lem-CRZ-no-go/4}) to
	\begin{align*}
		e^{i(f(\theta) + g(\theta))}f'(\theta) \cdot (c-ae^{-i\frac{\theta}{2}}) &= -\frac{\theta}{2}e^{-i\frac{\theta}{2}} \\
		e^{i(f(\theta) + g(\theta))}f'(\theta) \cdot (d-ae^{i\frac{\theta}{2}}) &= \frac{\theta}{2}e^{i\frac{\theta}{2}}
	\end{align*}
	In particular, this implies that
	\begin{align*}
		c - ae^{-i\frac{\theta}{2}} &= -d + ae^{i\frac{\theta}{2}}
	\end{align*}
	for all $\theta$, which is clearly not possible for constant $a,b,c$.
	
	Thus, we can conclude that we cannot represent $CR_Z(\theta)$ with less than two parametrised spiders.
	Now, suppose there were some other unitary $U(\theta) = e^{i\theta H}$ whose Hermitian generator has eigenvalues $-\lambda, 0,\lambda$ that can be implemented with a single parametrised spider:
	\[ U(\theta) ~=~ \input{\tikzitpathfact:U-repr-optimal/proof-1.tikz} \]
	However, by \Cref{lem:CRZ-sim} this would immediately yield a one-spider representation on $CR_Z(\theta)$ via
	\[ CR_Z(\theta) ~=~ \input{\tikzitpathfact:U-repr-optimal/proof-2.tikz} \]
	which is not possible.
\end{proofE}

Furthermore, we discuss an improved optimal construction for unitaries with only two eigenvalues $\lambda_1,\lambda_2$ in the next section.

\subsection{Special Case for Two Eigenvalues}

In the special case where $H$ has only two eigenvalues $\lambda_1,\lambda_2$, it is possible to implement $e^{i\theta H}$ using only a single parametrised spider.
In particular, this is the case for all single-qubit unitaries.

\begin{theorem} \label{thm:U-repr-two}
	Let $U(\theta) = e^{i\theta H}$ be a parametrised unitary whose Hermitian generator $H$ has only eigenvalues $\lambda_1, \lambda_2$ and admits the diagonalization $H = V^\dagger DV$.
	Then
	\[ U(\theta) = e^{i\lambda_2} ~ \input{\tikzitpaththm:U-repr-two/statement.tikz} \]
\end{theorem}
\begin{proof}
	Follows from \Cref{thm:U-repr-naive} and the observation that $f_{\lambda_2}(\vec x) = \neg f_{\lambda_1}(\vec x)$ since either $D\ket{\vec x} = \lambda_1\ket{\vec x}$ or $D\ket{\vec x} = \lambda_2\ket{\vec x}$:
	\begin{gather*}
		U(\theta) 
		~=~ \input{\tikzitpaththm:U-repr-two/proof-1.tikz}
		~=~ \input{\tikzitpaththm:U-repr-two/proof-2.tikz} \\
		~=~ e^{i\lambda_2}~ \input{\tikzitpaththm:U-repr-two/proof-3.tikz}
		\eqq{*} e^{i\lambda_2} ~ \input{\tikzitpaththm:U-repr-two/statement.tikz}
	\end{gather*}
%	We can verify the last step $(*)$ by plugging in a computational basis state:
%	\begin{gather*}
%		\tikzfig{thm:U-repr-two/proof-4}
%		\eqq{\cp} \tikzfig{thm:U-repr-two/proof-5}
%		~=~ \tikzfig{thm:U-repr-two/proof-6} \\
%		\eqq{\cp} \tikzfig{thm:U-repr-two/proof-7}
%		\eqq{\sf} \tikzfig{thm:U-repr-two/proof-8}
%		~=~ \tikzfig{thm:U-repr-two/proof-9}
%		\eqq{\cp} \tikzfig{thm:U-repr-two/proof-10}
%	\end{gather*}
	The step $(*)$ holds since function boxes acting on the computational basis form a bialgebra with the green spider and thus
	\[ \input{\tikzitpaththm:U-repr-two/proof-11.tikz} ~=~ \input{\tikzitpaththm:U-repr-two/proof-12.tikz} \qedhere \]
	
\end{proof}

This construction is optimal since it is clearly not possible to implement parametrised unitaries using zero parametrised spiders.

\section{Parameter-Shift Rules} \label{sec:recipes-shift}

The first parameter-shift rule was discovered by Mitarai et al.~\cite{mitarai2018quantum} and extended by Schuld et al.~\cite{schuld2019evaluating}.
It says that the derivate of a gate $U(\theta) = e^{i\theta H}$ whose Hermitian generator $H$ has only two eigenvalues $\lambda_1,\lambda_2$ satisfies
\begin{equation} \label{eqn:shift-schuld}
	\ddx{\theta}U(\theta) = \frac{\lambda_1-\lambda_2}{2\sin((\lambda_1-\lambda_2)\alpha)} (U(\theta + \alpha) - U(\theta - \alpha))
\end{equation}
for an arbitrary shift angle $\alpha$ with $\sin((\lambda_1-\lambda_2)\alpha) \neq 0$.
Thus, computing the gradient requires two evaluations of the circuit on the quantum device with parameter values shifted by $\pm\alpha$.
Remarkably, equation (\ref{eqn:shift-schuld}) is an exact representation of the gradient and should not be mistaken for a numerical gradient approximation, which might look similar:
\[ \ddx{\theta} U(\theta) \approx \frac{1}{2h} (U(\theta + h) - U(\theta-h)) \]
Unlike this noisy approximation, parameter-shift rules provide an unbiased estimator for the gradient of the expectation value.
Hence, they are widely used in practice.\footnote{For example by the QML library \texttt{pennylane} \cite{bergholm2018pennylane}.}
There has also been a focus in recent years on finding shift rules for a wider class of gates going beyond two eigenvalues.
For example, there is the four-term rule by Anselmetti et al.~\cite{anselmetti2021local} for Hermitians with eigenvalues $-\lambda,0,\lambda$ in addition to further generalisations depending on the differences between eigenvalues by Wierichs et al.~\cite{wierichs2022general}.

In this section, we graphically derive the original rule by Schuld et al. and then move on to gates with more than two eigenvalues.
%In particular, we give a novel proof of Anselmetti et al.'s rule and derive a novel, generalised rule that we compare to the one given by Wierichs et al.

\subsection{Two-Term Shift Rule}

A diagrammatic proof for a simplified version of Schuld. et al.'s~\cite{schuld2019evaluating} parameter shift rule (\ref{eqn:shift-schuld}) has already been given in~\cite{toumi2021diagrammatic} and~\cite{wang2022differentiating}.
However, all previous ZX-based proofs only derived the special case $\alpha = \frac{\pi}{2}$.
Furthermore, they only consider simple rotation gates without generalising to arbitrary parametrised unitaries with two eigenvalues.
We extend the proof to derive the two-term shift rule in its most general form:

\begin{lemmaE}
	For all $\alpha \in \mathbb R$ with $\alpha \neq \pi n$ for all $n \in \mathbb Z$, we have
	\begin{equation} \label{eqn:pi-pi-decomp}
		\input{\tikzitpathlem:pi-pi-decomp/statement-1.tikz} ~=~ \frac{1}{2i\sin(\alpha)} \left( \input{\tikzitpathlem:pi-pi-decomp/statement-2.tikz} ~-~ \input{\tikzitpathlem:pi-pi-decomp/statement-3.tikz} \right).
	\end{equation}
\end{lemmaE}
\begin{proofE}
	\def\tikzitpath{chapter4/figs/}
%	First, note that
%	\begin{align*}
%		\tikzfig{lem:pi-pi-decomp/proof-1} ~
%		&= \ket{++} - \ket{--} \\
%		&= \frac{1}{2}(\ket{0} + \ket{1})(\ket{0} + \ket{1}) - \frac{1}{2}(\ket{0} - \ket{1})(\ket{0} - \ket{1}) \\
%		&= \ket{01} + \ket{10} \\
%		&= \tikzfig{lem:pi-pi-decomp/proof-2} ~+~ \tikzfig{lem:pi-pi-decomp/proof-3}
%	\end{align*}
%	such that
%	\begin{gather*}
%		\tikzfig{lem:pi-pi-decomp/statement-1}
%		~=~ \tikzfig{lem:pi-pi-decomp/proof-4} ~+~ \tikzfig{lem:pi-pi-decomp/proof-5}
%		\eqq{\cp} -~ \tikzfig{lem:pi-pi-decomp/proof-2} ~+~ \tikzfig{lem:pi-pi-decomp/proof-3}
%		~= \ket{10} - \ket{01}.
%	\end{gather*}
	First, note that
	\begin{gather*}
		\input{\tikzitpathlem:pi-pi-decomp/statement-1.tikz}
		\eqq{\ref{eqn:green-spider-def}} \input{\tikzitpathlem:pi-pi-decomp/proof-6.tikz} ~-~ \input{\tikzitpathlem:pi-pi-decomp/proof-7.tikz} \\[\linesep]
		\eqq{\sf} \input{\tikzitpathlem:pi-pi-decomp/proof-9.tikz} ~-~ \input{\tikzitpathlem:pi-pi-decomp/proof-8.tikz}
		\eqq{\ref{eqn:zx-states}} \ket{1}\bra{0} - \ket{0}\bra{1} \numberthis\label{eqn:pi-pi-decomp-help}
	\end{gather*}
	On the other hand, we have
	\begin{gather*}
		\input{\tikzitpathlem:pi-pi-decomp/statement-2.tikz} 
		\eqq{\ref{eqn:green-spider-def}} (\ket{0} + e^{-i\alpha}\ket{1}) (\bra{0} + e^{i\alpha}\bra{1})
		= \ket{0}\bra{0} + e^{i\alpha}\ket{0}\bra{1} + e^{-i\alpha}\ket{1}\bra{0} + \ket{1}\bra{1} \\
		\input{\tikzitpathlem:pi-pi-decomp/statement-3.tikz} 
		\eqq{\ref{eqn:green-spider-def}} (\ket{0} + e^{i\alpha}\ket{1}) (\bra{0} + e^{-i\alpha}\bra{1})
		= \ket{0}\bra{0} + e^{-i\alpha}\ket{0}\bra{1} + e^{i\alpha}\ket{1}\bra{0} + \ket{1}\bra{1}
	\end{gather*}
	such that
	\begin{align*}
		\input{\tikzitpathlem:pi-pi-decomp/statement-2.tikz} ~-~ \input{\tikzitpathlem:pi-pi-decomp/statement-3.tikz} ~
		&= (e^{i\alpha} - e^{-i\alpha})\ket{0}\bra{1} + (e^{-i\alpha} - e^{i\alpha})\ket{1}\bra{0} \\
		&= (e^{i\alpha} - e^{-i\alpha})(\ket{0}\bra{1} - \ket{1}\bra{0}) \\
		%&= 2i\sin(\alpha) (\ket{10} - \ket{01}) \\
		\eqqa{\ref{eqn:pi-pi-decomp-help}} 2i\sin(\alpha) ~ \input{\tikzitpathlem:pi-pi-decomp/statement-1.tikz} ~. \qedhere
	\end{align*}
\end{proofE}

This allows us to decompose a version of the two-legged differentiation gadget:

\begin{lemma} \label{lem:W-shift-2}
	For all $\alpha \in \mathbb R$ with $\alpha \neq \pi n$ for all $n \in \mathbb Z$, we have
	\begin{equation} \label{eqn:W-shift-2}
		-i ~ \input{\tikzitpathlem:W-shift-2/statement-1.tikz} ~=~ \frac{1}{2\sin(\alpha)} \left( \input{\tikzitpathlem:W-shift-2/statement-2.tikz} ~-~ \input{\tikzitpathlem:W-shift-2/statement-3.tikz} \right).
	\end{equation}
\end{lemma}
\begin{proof}
	We have
	\begin{gather*}
		-i ~ \input{\tikzitpathlem:W-shift-2/statement-1.tikz}
		\eqq{\ref{eqn:W2-act}} -i ~ \input{\tikzitpathlem:W-shift-2/proof-1.tikz}
		\eqq{\cc,\hh} -i ~ \input{\tikzitpathlem:W-shift-2/proof-2.tikz}
		\eqq{\p} i ~ \input{\tikzitpathlem:W-shift-2/proof-3.tikz} \\
		\eqq{\ref{eqn:pi-pi-decomp}} \frac{1}{2\sin(\alpha)} \left( \input{\tikzitpathlem:W-shift-2/statement-2.tikz} ~-~ \input{\tikzitpathlem:W-shift-2/statement-3.tikz} \right) \qedhere
	\end{gather*}
\end{proof}

Combining this with the results from \Cref{sec:recipes-unitaries}, we obtain the two-term shift rule:

\begin{theorem}[Schuld et al.~\cite{schuld2019evaluating}] \label{thm:shift-2}
	Every parametrised circuit $C(\theta)$ described by the unitary $e^{i\theta H}$ whose Hermitian generator $H$ has only two eigenvalues $\lambda_1,\lambda_2$ satisfies
	\[ \ddx{\theta} C(\theta) = \frac{\lambda_1-\lambda_2}{2\sin((\lambda_1-\lambda_2)\alpha)} \left( C(\theta + \alpha) - C(\theta - \alpha) \right) \]
	for all $\alpha \in \mathbb R$ with $(\lambda_1-\lambda_2)\alpha \neq \pi n$ for all $n \in \mathbb Z$.
\end{theorem}
\begin{proof}
	Using the construction from \Cref{thm:U-repr-two}, we can write $C(\theta)$ as
	\[ \input{\tikzitpaththm:shift-2/proof-1.tikz} \]
	Thus, we have
	\begin{align*}
		\ddx{\theta} C(\theta)
		~&=~ \ddx{\theta} \left[ \input{\tikzitpaththm:shift-2/proof-1.tikz} \right] 
		\\[\linesep]
		\eqqa{\text{Fact }\ref{fact:chain-rule}} -i(\lambda_1-\lambda_2) ~ \input{\tikzitpaththm:shift-2/proof-2.tikz} 
		\\[\linesep]
		\eqqa{\ref{eqn:W-shift-2}} \frac{\lambda_1 - \lambda_2}{2\sin((\lambda_1-\lambda_2)\alpha)}
		\begin{aligned}[t]
			& \left( \input{\tikzitpaththm:shift-2/proof-3.tikz} \right. \\ & ~~~-~ \left. \input{\tikzitpaththm:shift-2/proof-4.tikz} \right)
		\end{aligned}
		\\[\linesep]
		\eqqa{\sf} \frac{\lambda_1-\lambda_2}{2\sin((\lambda_1-\lambda_2)\alpha)} \left( C(\theta + \alpha) - C(\theta - \alpha) \right) 
		\qedhere
	\end{align*}
\end{proof}

\subsection{Shift Rules Beyond Two Terms}

One way to extend the result from \Cref{thm:shift-2} to a wider class of circuits is to invoke the product rule.
For example, this gives us
\begin{align*}
	&\qquad\quad \ddx{\theta} \left[ \input{\tikzitpathex:shift-product/1.tikz} \right] \\[\linesep]
	\eqqa{\text{Lem. }\ref{lem:product-rule}} \ddx{\theta} ~ \input{\tikzitpathex:shift-product/2-1.tikz} ~+~ \ddx{\theta} ~ \input{\tikzitpathex:shift-product/2-2.tikz} \\[\linesep]
	\eqqa{\text{Thm. }\ref{thm:circ-diff}, \ref{eqn:W-shift-2}} 
	\begin{aligned}[t]
		& \frac{1}{2\sin(\alpha)} \left( \input{\tikzitpathex:shift-product/3-1.tikz} ~-~ \input{\tikzitpathex:shift-product/3-2.tikz} \right) \\
		+ & \frac{1}{2\sin(\beta)} \left( \input{\tikzitpathex:shift-product/3-3.tikz} ~-~ \input{\tikzitpathex:shift-product/3-4.tikz} \right)
	\end{aligned}  \numberthis\label{eqn:shift-naive}
\end{align*}
The downside of this approach is that it requires the gate to be decomposed such that individual rotation angles can be shifted~\cite{crooks2019gradients}.
This introduces additional overhead if the considered gate is hardware-native.
One example of this studied in the literature is the $\mathit{fsim}$ gate native to Google's \texttt{gmon} architecture~\cite{foxen2020demonstrating}.
As argued in~\cite{kyriienko2021generalized}, it is more efficient to use a rule that shifts all parameter occurrences simultaneously, avoiding the depth increase invoked by decomposing $\mathit{fsim}$ into elementary gates.
Furthermore, it was proven in~\cite{wierichs2022general} that rules shifting all gates simultaneously sometimes require less shots to get accurate gradient estimates.

The natural way to extend our proof of \Cref{thm:shift-2} is to find decompositions of the differentiation gadget for more than two legs.
Unfortunately, the proof of \Cref{lem:W-shift-2} does not scale since the simple representation of the $W$-state as a red $\pi$-spider no longer holds if we add more legs.
Instead, we characterise the validity of all possible $m$-term shift rules for $n$-legged differentiation gadgets via a system of (complex) polynomial equations:

\begin{lemma} \label{lem:W-shift-eq}
	For $\vec \xi,\vec \alpha \in \mathbb R^m$, the diagram equation
	\[ -2^{n-1}i~ \input{\tikzitpathlem:W-shift-eq/statement-1.tikz} ~=~ \sum_{i=1}^m \xi_i ~ \input{\tikzitpathlem:W-shift-eq/statement-2.tikz} \]
	holds iff for all $k \in \{0, 1, ..., n\}$ we have
	\[ \sum_{j=1}^m \xi_j \cdot e^{ik\alpha_j} = ki. \]
\end{lemma}
\begin{proof}
	The diagram equation holds iff both sides are equal when plugging in computational basis states.
	First, consider the left-hand side:
	\begin{gather*}
		-2^{n-1}i ~ \input{\tikzitpathlem:W-shift-eq/left/1.tikz}
		\eqq{\cc,\sf} -\frac{1}{2}i ~ \input{\tikzitpathlem:W-shift-eq/left/2.tikz} \\[\linesep]
		%~=~ -\frac{1}{2}i ~ \tikzfig{lem:W-shift-eq/left/3}
		\eqq{\ref{eqn:W-add}} -\frac{1}{2}i ~ \input{\tikzitpathlem:W-shift-eq/left/4.tikz}
		~=~ -\frac{1}{2}i \sum_{j=1}^n (-1)^{x_j} - (-1)^{y_j}
	\end{gather*}
	For the right-hand side, we get
	\begin{gather*}
		\sum_{i=1}^m \xi_i \input{\tikzitpathlem:W-shift-eq/right/1-1.tikz}
		~=~ \sum_{i=1}^m \xi_i \cdot e^{i\alpha_i \sum_{j=1}^n x_j-y_j}.
	\end{gather*}
	Equating both sides yields
	\begin{align*}
		-\frac{1}{2}i \sum_{j=1}^n (-1)^{x_j} - (-1)^{y_j} &= \sum_{i=1}^m \xi_i \cdot e^{i\alpha_i \sum_{j=1}^n x_j-y_j} \\
		\Leftrightarrow ~ i\sum_{j=1}^n x_j-y_j &= \sum_{i=1}^m \xi_i \cdot e^{i\alpha_i \sum_{j=1}^n x_j-y_j}
	\end{align*}
	since $(-1)^{x_j} - (-1)^{y_j} = -2(x_j-y_j)$ for all $x_j,y_j\in\{0,1\}$.
	The equation above must hold for all choices of $\vec x,\vec y \in \{0,1\}^n$.
	Noting that $\sum_{j=1}^n x_j-y_j \in \{-n,...,n\}$, we can represent this more compactly as
	\[ \sum_{j=1}^m \xi_j \cdot e^{\pm ik\alpha_j} = \pm ki \]
	for $k\in\{0,...,n\}$.
	Finally, we can drop the $\pm$ sign since negating just corresponds to taking the complex conjugate on both sides.
\end{proof}

This general characterisation of shift rules will be useful for proving a no-go result in \Cref{sec:nogo}.
However, for the purposes of this section it suffices to look at the special case of symmetric shifts as in \Cref{lem:W-shift-2}.
This simplifies the system of equations:

\begin{corollary} \label{corr:W-shift-eq}
	For $\vec \xi,\vec \alpha \in \mathbb R^m$, the diagram equation
	\[ -2^{n-1}i~ \input{\tikzitpathlem:W-shift-eq/statement-1.tikz} ~=~ \sum_{i=1}^m \xi_i \left( \input{\tikzitpathlem:W-shift-eq/statement-2.tikz} ~-~ \input{\tikzitpathlem:W-shift-eq/statement-3.tikz} \right) \]
	holds iff for all $k \in \{1, ..., n\}$ we have
	\[ \sum_{j=1}^m \xi_j \cdot \sin(k\alpha_j) = \frac{1}{2}k. \]
	%We write this system of equations in matrix form as $\mathbf S \vec\xi = \vec b$.
\end{corollary}
\begin{proof}
	Invoking \Cref{lem:W-shift-eq}, we get the system
	\[
		\sum_{j=1}^m \xi_j \left( e^{ik\alpha_j} - e^{\mp ik\alpha_j} \right) = ki 
		\quad\Leftrightarrow\quad 
		2i \sum_{j=1}^m \xi_j \cdot \sin(k\alpha_j) = ki
	\]
	for $k \in \{0,1,...,n\}$. Notice that the case $k = 0$ is now trivially satisfied.
\end{proof}

To make the notation more concise, we will write this system of equations in matrix form as
\begin{equation} \label{eqn:W-shift-sys}
	\mathbf S_{\vec\alpha} \cdot \vec\xi = \frac{1}{2} \vec \tau
\end{equation}
where
\[
\mathbf S_{\vec\alpha} = \begin{pmatrix}
	\sin(\alpha_1) & ... & \sin(\alpha_m) \\
	\sin(2\alpha_1) & ... & \sin(2\alpha_m) \\
	\vdots & & \vdots \\
	\sin(n\alpha_1) & ... & \sin(n\alpha_m)
\end{pmatrix}
\qquad\qquad
\vec\tau = \begin{pmatrix}
	1 \\ 2 \\ \vdots \\ n
\end{pmatrix}
\]

If $m=n$, the system is square and solvable under some mild conditions on the $\alpha_i$.
For example, in the case $n = m = 1$, we get the single equation $\xi \sin(\alpha) = \frac{1}{2}$ whose solution $\xi = \frac{1}{2\sin(\alpha)}$ is exactly the shift rule from \Cref{lem:W-shift-2}.
In the case $n = m = 2$, we get the system
\[
\xi_1 \sin(\alpha_1) + \xi_2 \sin(\alpha_2) = \frac{1}{2}
\qquad\qquad
\xi_1 \sin(2\alpha_1) + \xi_2 \sin(2\alpha_2) = 1
\]
which for $\alpha_1 \neq \alpha_2$ and $\sin(2\alpha_1),\sin(2\alpha_2)\neq 0$ is solved by
\begin{gather*}
	\label{eqn:W-shift-4-xi1}
	\xi_1 = \frac{2\sin(\alpha_2) - \sin(2\alpha_2)}{2(\sin(2\alpha_1)\sin(\alpha_2) - \sin(\alpha_1)\sin(2\alpha_2))} \\[\linesep]
	\label{eqn:W-shift-4-xi2}
	\xi_2 = \frac{\sin(2\alpha_1) - 2\sin(\alpha_1)}{2(\sin(\alpha_1)\sin(2\alpha_2) - \sin(2\alpha_1)\sin(\alpha_2))}.
\end{gather*}

This allows us to immediately derive the four-term shift rule given by Anselmetti et al.~\cite{anselmetti2021local}:

\begin{theorem}[Anselmetti et al.~\cite{anselmetti2021local}] \label{thm:shift-4}
	Every parametrised circuit $C(\theta)$ described by the unitary $e^{i\theta H}$ whose Hermitian generator $H$ has eigenvalues $-\lambda,0,\lambda$ satisfies
	\[ \ddx{\theta} C(\theta) =  \xi_1 \left( C(\theta + \alpha_1) - C(\theta - \alpha_1) \right) + \xi_2 \left( C(\theta + \alpha_2) - C(\theta - \alpha_2) \right) \]
	for $\xi_1 = \frac{2\sin(\lambda\alpha_2) - \sin(2\lambda\alpha_2)}{2(\sin(2\lambda\alpha_1)\sin(\lambda\alpha_2) - \sin(\lambda\alpha_1)\sin(2\lambda\alpha_2))}$ and $\xi_2 = \frac{\sin(2\lambda\alpha_1) - 2\sin(\lambda\alpha_1)}{2(\sin(\lambda\alpha_1)\sin(2\lambda\alpha_2) - \sin(2\lambda\alpha_1)\sin(\lambda\alpha_2))}$.
	%for $\xi_1,\xi_2$ satisfying the system of equations $\mathbf S(\lambda\alpha_1,\lambda\alpha_2) \cdot \vec \xi = \frac{1}{2} \vec\tau$ (c.f. \Cref{corr:W-shift-eq} and equations (\ref{eqn:W-shift-4-xi1}), (\ref{eqn:W-shift-4-xi2})).
\end{theorem}
\begin{proof}
	Using the construction from \Cref{thm:U-repr-naive}, we can write $C(\theta)$ as
	\[ \input{\tikzitpaththm:shift-4/proof-1.tikz} \]
	Thus, we have
	\begin{align*}
		\ddx{\theta} C(\theta)
		&= \ddx{\theta} \left[ \input{\tikzitpaththm:shift-4/proof-1.tikz} \right] 
		\\[\linesep]
		\eqqa{\sym,\text{Fact \ref{fact:chain-rule}}} -2\lambda i ~ \input{\tikzitpaththm:shift-4/proof-2.tikz} 
		\\[\linesep]
		\eqqa{\sym,\text{Corr. } \ref{corr:W-shift-eq}} \lambda \sum_{i=1}^2 \xi_i \left( \input{\tikzitpaththm:shift-4/proof-3-1.tikz} - \input{\tikzitpaththm:shift-4/proof-3-2.tikz} \right) 
		\\[\linesep]
		\eqqa{\sf} \lambda\xi_1 \left( C(\theta + \alpha_1) - C(\theta - \alpha_1) \right) + \lambda\xi_2 \left( C(\theta + \alpha_2) - C(\theta - \alpha_2) \right) \qedhere
	\end{align*}
\end{proof}

Similarly, solving the system (\ref{eqn:W-shift-sys}) for $m=n=3,4,...$ yields $2n$-term shift rules for circuits with more than $2$ parametrised spiders.
Unfortunately, we are not aware of a closed-form solution for the coefficients $\vec\xi$ for arbitrary $n$ and $\vec\alpha$.
However, if we fix equidistant shift angles $\alpha_j = \frac{j\pi}{n+1}$, we can in fact derive a closed-form solution for $\vec\xi$:

\begin{lemma}
	If $\alpha_j = \frac{j\pi}{n+1}$, then $\mathbf S_{\vec\alpha} \cdot \vec\xi = \frac{1}{2}\vec\tau$ has the solution $\xi_j = \frac{1}{n+1} \sum_{k=1}^n k \cdot \sin\left(\frac{kj\pi}{n+1}\right)$.
\end{lemma}
\begin{proof}
	For equidistant angles, the equations correspond to a type-I discrete sine transform (DST-I)~\cite{britanak2010discrete,jain1979sinusoidal}
	\[ x_k = \sum_{j=1}^n \xi_j \cdot \sin\left(\frac{kj\pi}{n+1}\right) \]
	where $x_k = \frac{1}{2}k$.
	Since the inverse of the DST-I is again given by the DST-I scaled by $\frac{2}{n+1}$, we get
	\[ \xi_j = \frac{2}{n+1} \sum_{k=1}^n x_k \cdot \sin\left(\frac{kj\pi}{n+1}\right) = \frac{1}{n+1} \sum_{k=1}^n k \cdot \sin\left(\frac{kj\pi}{n+1}\right). \qedhere \]
\end{proof}

%Hypothesis: $\sum_{k=1}^n k \cdot \sin\left(\frac{kj\pi}{n+1}\right) = (-1)^j\frac{n+1}{2} \cot\left( \frac{j\pi}{2(n+1)} \right)$. TODO: Prove this! \todo{!!!}

This corresponds to the following generalised parameter-shift rule:

\begin{theorem} \label{thm:shift-general}
	Let $C(\theta)$ be a parametrised circuit that is represented by
	\[ \input{\tikzitpaththm:shift-n/statement.tikz} \]
	%where $x_1,...,x_n \in \{0,1\}$ and $\lambda\in\mathbb R$. 
	Then
	\[ \ddx{\theta} C(\theta) = \lambda \sum_{j=1}^n \xi_j \left( C(\theta+\alpha_j) - C(\theta-\alpha_j) \right) \]
	if $\vec\alpha$ and $\vec\xi$ satisfy the equations $\mathbf S_{\vec\alpha}\cdot\vec\xi = \frac{1}{2}\vec\tau$.
	One possible solution is given by $\alpha_j = \frac{j\pi}{n+1}$ and $\xi_j = \frac{1}{n+1} \sum_{k=1}^n k \cdot \sin\left(\frac{kj\pi}{n+1}\right)$.
\end{theorem}
\begin{proof}
	Similar to \Cref{thm:shift-4}, this follows from \Cref{fact:chain-rule} and \Cref{corr:W-shift-eq}.
\end{proof}

A similar generalised shift rule has been proven by Wierichs et al.~\cite{wierichs2022general}.
Their rule requires $2k$ terms where $k$ is the number of unique eigenvalue differences of the Hermitian generator, whereas the cost of our rule depends on the number of parametrised spiders needed to implement the gate.
We already established a connection between eigenvalues and ZX diagrams by upper-bounding the number of parametrised spiders by the number of non-zero eigenvalues (c.f. \Cref{thm:U-repr-naive}).
Furthermore, we lower-bounded the number of spiders for eigenvalues $0,\pm\lambda$ (c.f. \Cref{fact:U-repr-optimal}).
An interesting future research direction would be to explore whether there are tighter bounds for more general cases and any deeper relationships between eigenvalues and parametrised spiders in diagrams.
This could possibly lead to a diagrammatic proof of Wierichs et al.'s version of the generalised shift rule.
Note that this might also require decompositions of the differentiation gadget where the spider-phases do not all have the same absolute value as assumed in \Cref{thm:shift-general}.

%The downside of this shift rule is that it is only applicable if all parametrised spiders have phase $\pm\lambda\alpha$.
%If the phases are non-linear functions in $\theta$ 

%Concretely, we get the following shift rule:
%
%As a special case, we can derive a result similar to Wierichs et al.'s~\cite{wierichs2022general} general closed-form shift rule:
%Given a set $\Lambda$ of eigenvalues of a Hermitian generator $H$, they define the set $D_\Lambda = \{ |\lambda_i - \lambda_j| ~|~ \lambda_i,\lambda_j \in \Lambda \text{ with } \lambda_i \neq \lambda_j \}$ of unique eigenvalue differences.
%They make the assumption that the differences in $D_\Lambda$ are evenly spaced:
%Under the assumption that 
% and give a shift rule with closed-form coefficients requiring $2|D_\Lambda|$ terms.

\subsection{Proof of Anselmetti's No-Go Conjecture} \label{sec:nogo}

A general pattern in parameter shift rules seems to be that the number of terms required is the same as when using the naive approach of the product rule combined with two-term shifts (see equation (\ref{eqn:shift-naive})).
This holds true for our generalised rule (\Cref{thm:shift-general}), as well as for Wierichs et al.'s general rule~\cite{wierichs2022general} and Anselmetti et al.'s four-term rule~\cite{anselmetti2021local} (\Cref{thm:shift-4}).
An obvious question at this point is whether we can do any better than that.

As far as we are aware, no results regarding the optimality of shift rules in this sense have been proven in the literature.
In particular, Anselmetti et al.~\cite{anselmetti2021local} conjecture that their four-term rule is optimal, but do not give a proof.
In this section, we give the first proof (to our knowledge) of this conjecture.
We show that it is indeed impossible to compute gradients for gates whose generators have eigenvalue $-\lambda,0,\lambda$ using less than four shifts.

For this, we first look at an example.
One common gate whose generator has eigenvalues of this shape is $CR_Z(\theta)$.
Recall that when calculating derivatives, we always have to work with the doubling construction.
Thus, we define $U(\theta) := \doubled(CR_Z(2\theta)) = CR_Z(2\theta) \otimes \overline{CR_Z(2\theta)}$.
We multiply $\theta$ by $2$ to avoid fractions in the matrix:
\[
	U(\theta) = \text{diag}(1, 1, e^{i\theta}, e^{-i\theta}, 1, 1, e^{i\theta}, e^{-i\theta}, e^{-i\theta}, e^{-i\theta}, 1, e^{-i\theta}, e^{i\theta}, e^{i\theta}, e^{2i\theta}, 1)
\]
The corresponding derivative is thus given by
\begin{adjustwidth}{0cm}{-2cm}
\[
	\ddx{\theta}U(\theta) = \text{diag}(0, 0, ie^{i\theta}, -ie^{-i\theta}, 0, 0, ie^{i\theta}, -ie^{-i\theta}, -ie^{-i\theta}, -ie^{-i\theta}, 0, -ie^{-i\theta}, ie^{i\theta}, ie^{i\theta}, 2ie^{2i\theta}, 0)
\]
\end{adjustwidth}
Suppose we had a three-term shift rule for $U(\theta)$, i.e. $\xi_1,\xi_2,\xi_3,\alpha,\beta,\gamma \in \mathbb R$ such that $\ddx{\theta} U(\theta) = \xi_1 U(\theta+\alpha) + \xi_2 U(\theta+\beta) + \xi_3 U(\theta+\gamma)$.
By comparing the matrix elements, this rule would need to satisfy the following equations:
\begin{align}
	\xi_1 + \xi_2 + \xi_3 &= 0 \\
	\xi_1 e^{i(\theta+\alpha)} + \xi_2 e^{i(\theta+\beta)} + \xi_3 e^{i(\theta+\gamma)} &= ie^{i\theta} \label{eqn:nogo-1} \\
	\xi_1 e^{-i(\theta+\alpha)} + \xi_2 e^{-i(\theta+\beta)} + \xi_3 e^{-i(\theta+\gamma)} &= -ie^{-i\theta} \label{eqn:nogo-2} \\
	\xi_1 e^{2i(\theta+\alpha)} + \xi_2 e^{2i(\theta+\beta)} + \xi_3 e^{2i(\theta+\gamma)} &= 2ie^{2i\theta} \label{eqn:nogo-3}
\end{align}
Note that equation (\ref{eqn:nogo-1}) is redundant since it is the complex conjugate of equation (\ref{eqn:nogo-2}).
Furthermore, multiplying (\ref{eqn:nogo-2}) with $e^{-i\theta}$ and (\ref{eqn:nogo-3}) with $e^{-2i\theta}$ yields the following simplified system:
\begin{align*}
	\xi_1 + \xi_2 + \xi_3 &= 0 \\
	\xi_1 e^{i\alpha} + \xi_2 e^{i\beta} + \xi_3 e^{i\gamma} &= i \\
	\xi_1 e^{2i\alpha} + \xi_2 e^{2i\beta} + \xi_3 e^{2i\gamma} &= 2i
\end{align*}
Surprisingly, this is the exact same system we get in \Cref{lem:W-shift-eq} for decomposing the differentiation gadget.
We can show that this system is in fact not solvable:

\begin{lemma} \label{lem:nogo-system}
	This system of equations has no solution for $\xi_1,\xi_2,\xi_3,\alpha,\beta,\gamma \in \mathbb R$:
%	\begin{align*}
%		\xi_1 + \xi_2 + \xi_3 &= 0 \\
%		\xi_1 e^{i\alpha} + \xi_2 e^{i\beta} + \xi_3 e^{i\gamma} &= i \\
%		\xi_1 e^{2i\alpha} + \xi_2 e^{2i\beta} + \xi_3 e^{2i\gamma} &= 2i
%	\end{align*}
	\[ \begin{pmatrix}
			1 & 1 & 1 \\
			e^{i\alpha} & e^{i\beta} & e^{i\gamma} \\
			e^{2i\alpha} & e^{2i\beta} & e^{2i\gamma}
		\end{pmatrix} 
		\begin{pmatrix} \xi_1 \\ \xi_2 \\ \xi_3 \end{pmatrix} 
		= \begin{pmatrix} 0 \\ i \\ 2i \end{pmatrix} \]
\end{lemma}
\begin{proof}
	First, note that the system is given by a Vandermonde matrix.
	Thus, it has full rank if $\alpha,\beta,\gamma$ are pairwise distinct angles.
	In that case, the solution to the system is unique.
	Using the shorthand $a = e^{i\alpha}, b=e^{i\beta}, c=e^{i\gamma}$, Gaussian elimination yields
	\[ 
	\xi_1 = -i\frac{b+c-2}{(a-b)(a-c)}
	\qquad
	\xi_2 = -i\frac{a+c-2}{(b-a)(b-c)}
	\qquad
	\xi_3 = -i\frac{a+b-2}{(c-a)(c-b)}.
	\]
	Suppose that $\xi_1,\xi_2,\xi_3$ are real.
	This means that
	\begin{align*}
		\overline \xi_1 \cdot \xi_1 &= \xi_1^2 \\
		\Leftrightarrow -\frac{b^{-1}+c^{-1}-2}{(a^{-1}-b^{-1})(a^{-1}-c^{-1})} \cdot \frac{b+c-2}{(a-b)(a-c)} &= \frac{(b+c-2)^2}{(a-b)^2(a-c)^2} \\
		\Leftrightarrow \frac{a^2(b+c-2)(2bc-b-c)}{(a-b)^2(a-c)^2} &= \frac{(b+c-2)^2}{(a-b)^2(a-c)^2} \\
		\Leftrightarrow a^2 &= \frac{b+c-2}{2bc-b-c}
	\end{align*}
	Note that we have $2bc-b-c \neq 0$ since the equation $2e^{i(\beta+\gamma)} = e^{i\beta} + e^{i\gamma}$ has the only angle solution $\beta = \gamma = 0$ which is ruled out by the assumption that $\beta$ and $\gamma$ are distinct angles.
	
	Similarly, we get $b^2 = \frac{a+c-2}{2ac-a-c}$ and $c^2 = \frac{a+b-2}{2ab-a-b}$.
	Thus, we have a system of quadratic equations which we can solve using a computer algebra system.
	Using a Mathematica program, we find that $a=b=c=-\sqrt[3]{-1}$ and $a=b=c=(-1)^{\frac{2}{3}}$ are the only solutions.
	This violates the assumption that $\alpha$, $\beta$, and $\gamma$ are distinct angles. 

	Next, we consider the case where the angles are not distinct.
	W.l.o.g. assume that $\gamma = \alpha$.
	This means that the last column of the matrix becomes redundant and we can simplify the system to
	\[
	\xi_1 + \xi_2 = 0 \qquad\qquad
	\xi_1 e^{i\alpha} + \xi_2 e^{i\beta} = i \qquad\qquad
	\xi_1 e^{2i\alpha} + \xi_2 e^{2i\beta} = 2i \qquad\qquad
	\]
	Since $\xi_1$ and $\xi_2$ are real, we know that the conjugate equations $\xi_1 e^{-i\alpha} + \xi_2 e^{-i\beta} = -i$ and $\xi_1 e^{-2i\alpha} + \xi_2 e^{-2i\beta} = -2i$ also hold.
	Adding those conjugate equations to the original versions and using $\xi_2 = -\xi_1$ yields $\cos(\alpha) - \cos(\beta) = 0$ and $\cos(2\alpha) - \cos(2\beta) = 0$.
	This is only satisfied for $\alpha=\beta=\pi$.
	But then $\xi_1 e^{i\alpha} + \xi_2 e^{i\beta} = 0 \neq i$, which violates the second equation.
\end{proof}

With this lemma we have shown two things at once:
First, we cannot decompose the four-legged differentiation gadget into less than four shifts.
Secondly, $CR_Z(\theta)$ does not satisfy a shift rule with less than four terms.

Now the question is how to extend this result to arbitrary gates with generator eigenvalues $-\lambda,0,\lambda$?
The answer is surprisingly simple:
It relies on the fact that each such gate can be used to \enquote{simulate} $CR_Z(\theta)$:

\begin{lemma} \label{lem:CRZ-sim}
	Let $U(\theta) = e^{i\theta H}$ be an $n$-qubit unitary whose Hermitian generator $H$ has eigenvalue $-\lambda,0,\lambda$ with corresponding eigenvectors $\ket{\vec x_{-\lambda}}$, $\ket{\vec x_0}$, and $\ket{\vec x_{\lambda}}$.
	Define a Boolean function $f : \{0,1\}^2 \to \{0,1\}^n$ by
	\[ f(0,0) = \vec x_0 \qquad\qquad f(0,1) = \vec x_0 \qquad\qquad f(1,0) = \vec x_{-\lambda} \qquad\qquad f(1,1) = \vec x_\lambda. \]
	Then, we have
	\[ CR_Z(\theta) ~=~ \input{\tikzitpathlem:CRZ-sim/statement.tikz} \]
\end{lemma}
\begin{proof}
	We check how the diagram acts on computational basis states:
	\begin{gather*}
		\input{\tikzitpathlem:CRZ-sim/proof-1.tikz}
		\eqq{\cp} \input{\tikzitpathlem:CRZ-sim/proof-2.tikz} \\[\linesep]
		~=~ \input{\tikzitpathlem:CRZ-sim/proof-3.tikz}
		\eqq{\cp} \input{\tikzitpathlem:CRZ-sim/proof-4.tikz}
	\end{gather*}
	\begin{itemize}
		\item For $\ket{00}$ we get $\bra{\vec x_0}U(\theta)\ket{\vec x_0} \ket{00} = e^{i\cdot 0\cdot \frac{\theta}{2\lambda}}\ket{00} = \ket{00} = CR_Z(\theta)\ket{00}$.
		
		\item For $\ket{01}$ we get $\bra{\vec x_0}U(\theta)\ket{\vec x_0} \ket{01} = e^{i\cdot 0\cdot \frac{\theta}{2\lambda}} = \ket{01} = CR_Z(\theta)\ket{01}$.
		
		\item For $\ket{10}$ we get $\bra{\vec x_{-\lambda}}U(\theta)\ket{\vec x_{-\lambda}} \ket{10} = e^{-i\lambda\frac{\theta}{2\lambda}}\ket{10} = e^{-i\frac{\theta}{2}}\ket{10} = CR_Z(\theta)\ket{10}$.
		
		\item For $\ket{11}$ we get $\bra{\vec x_\lambda}U(\theta)\ket{\vec x_\lambda} \ket{11} = e^{i\lambda\frac{\theta}{2\lambda}}\ket{11} = e^{i\frac{\theta}{2}}\ket{11} = CR_Z(\theta)\ket{11}$. \qedhere
	\end{itemize}
\end{proof}

This allows us to immediately conclude Anselmetti et al.'s conjecture:

\begin{theorem}[Anselmetti's No-Go Conjecture] \label{thm:nogo}
	The shift rule in \Cref{thm:shift-4} is optimal, i.e. it is not possible to compute the gradient of gates with generator eigenvalues $-\lambda,0,\lambda$ using less than four shifts.
\end{theorem}
\begin{proof}
	Suppose there is such a gate $U(\theta)$ that admits a 3-term shift rule.
	But by \Cref{lem:CRZ-sim} this would also yield a 3-term rule for $CR_Z(\theta)$ which we have shown is not possible (\Cref{lem:nogo-system}).
\end{proof}

\begin{remark}
	This \enquote{proof by example} technique also generalises to the optimality of other shift rules.
	As soon as we can prove that a shift rule is optimal for an example gate, this immediately implies that the rule is optimal for all gates with the same eigenvalues.
	This could be used to generalise this no-go theorem to capture the cost of shift rules for all gates.
	The main difficulty lies in characterising for which combinations of $n$ and $m$ the system in \Cref{lem:W-shift-eq} is solvable.
\end{remark}

\section{Ancilla Recipes} \label{sec:ancilla-recipe}

One of the initial motivations for using the ZX calculus to study gradient recipes was the hope that the graphical representation of derivatives might make it easier to discover new recipes that possibly go beyond parameter shift rules.
Generally, this requires decomposing the differentiation gadget into doubled maps.
We have found the following promising decomposition:
\begin{gather*}
	\input{\tikzitpathfact:W-decomp/1.tikz}
	\eqq{\wf} \input{\tikzitpathfact:W-decomp/2.tikz} \\[\linesep]
	\eqq{\ref{eqn:W2-act}} \input{\tikzitpathfact:W-decomp/3.tikz}
	\eqq{\sf,\pcy} \input{\tikzitpathfact:W-decomp/4.tikz} \\[\linesep]
	\eqq{\ref{eqn:pi-pi-decomp}} \frac{1}{2\sin(\alpha)} \left( \input{\tikzitpathfact:W-decomp/5-1.tikz} ~-~ \input{\tikzitpathfact:W-decomp/5-2.tikz} \right) \\[\linesep]
	~=~ \frac{1}{2\sin(\alpha)} \left( \doubled\left(\input{\tikzitpathfact:W-decomp/6-1.tikz}\right) ~-~ \doubled\left(\input{\tikzitpathfact:W-decomp/6-2.tikz}\right) \right) \numberthis\label{eqn:W-decomp}
\end{gather*}
Unfortunately, applying this decomposition to gates yields non-unitary terms in general.
The same holds true for other compositions of the gadget we investigated.
However, (\ref{eqn:W-decomp}) suggests a general algorithm to compute gradients using \textit{ancillae}.
We illustrate this on the example of two $R_Z$ gates. Note that
\[ \input{\tikzitpathancilla/alg-1.tikz} \eqq{\id,\sf} \input{\tikzitpathancilla/alg-2.tikz} \]
Thus, we can perform this computation on a quantum computer by preparing two ancillae, i.e. extra qubits, in the state $\input{\tikzitpathancilla/state.tikz}$,\footnote{We discuss how to construct this state in \Cref{chap:appendix-W}} connecting them to the original qubits via CZs, and then performing \textit{post-selection}.
Post-selection means that we measure and ignore all executions where the outcome is not the one specified in the circuit.
This allows us to perform non-unitary operations like the gradient above.

This yields a 2-term recipe for the gate above which would have required four terms using the regular shit rules.
Generally, any gate with $n$ parametrised spider can be differentiated this way using 2 terms and $n$ ancilla qubits.
However, note that each term is more expensive to execute, requiring more shots to get accurate estimates of the expectation value because of the post-selection.
Furthermore, the linear requirement of ancillae is a significant limitation of this rule since qubits are a very scarce recourse on current quantum devices.
Thus, while theoretically interesting, the practical applicability of this rule is limited and shift rules should probably be preferred.

\chapter{Barren Plateaus} \label{chap:barren-plateaus}

\def\tikzitpath{chapter5/figs/}

\newcommand{\var}[1]{\text{Var}\left(#1\right)}
\newcommand{\Sim}[1]{\hyperref[fig:sim-circuits]{\text{Sim}_{#1}}}
\newcommand{\IQP}[1]{\hyperref[fig:IQP-circuits]{\text{IQP}_{#1}}}

After studying how gradients can be computed, we now turn to the question of how the gradient landscape of quantum circuits looks like.
A common challenge when training PQCs using gradient-based methods is the so-called \textit{barren plateau phenomenon}.
Roughly, it describes the problem that the gradient landscape of many quantum circuits, unlike classical neural networks~\cite{choromanska2015loss}, flattens exponentially with increasing circuit sizes.
In other words, the probability that the gradient $\frac{\partial \langle H\rangle}{\partial \theta_i}$ is non-zero to some fixed precision is exponentially small with regards to the number of qubits~\cite{mcclean2018barren}.
As a result of this, gradient-based optimisation becomes increasingly difficult or even numerically impossible.
Thus, identifying and studying which circuits exhibit this undesirable behaviour has been a major focus of QML research~\cite{mcclean2018barren,holmes2022connecting}.

Recently, Wang and Yeung proposed a new method to detect barren plateaus using the ZXW-calculus~\cite{wang2022differentiating}.
However, they only demonstrate their method on a trivial example circuit.
In this chapter we apply their diagrammatic approach to ansätze actually used in QML research (see \Cref{fig:sim-circuits}).
After formally defining the barren plateau phenomenon in \Cref{sec:barren-backgound} and the circuits we study in \Cref{sec:circuits}, we introduce a technique to empirically detect barren plateaus in \Cref{sec:barren-numeric}.
\begin{itemize}
	\item
	We develop a tool using the QuiZX library~\cite{kissinger2022simulating} in Rust that automatically computes the variance of the expectation value gradients.
	
	\item
	We use the tool to numerically analyse 7 circuits used by Sim et al.~\cite{sim2019expressibility} and conclude that they likely all already have barren plateaus for a single layer (Figures \ref{fig:sim-single-layer} and \ref{fig:sim-single-layer-parameters}).
	
	\item
	We also study 3 IQP circuits, concluding that two of them likely have barren plateaus while one does not (Figures \ref{fig:iqp-single-layer}, \ref{fig:iqp1-multi-layer}, and \ref{fig:iqp1-multi-layer-converge}).
\end{itemize}
We verify our empirical hypotheses in \Cref{sec:barren-formal} by diagrammatically proving the existence of barren plateaus:
\begin{itemize}
	\item 
	We prove that the first Sim ansatz has barren plateaus even when only using a single layer if we measure on $\Theta(n)$ qubits (\Cref{fact:sim1-barren}) and derive similar conditions for the second Sim ansatz (\Cref{fact:sim2-barren}).
	
	\item
	We derive a general result that can be used to analyse any single-layer IQP circuit for barren plateaus (\Cref{thm:IQP-variance}) and apply it to prove the existence of barren plateaus for 3 single-layer IQP ansätze (\Cref{fact:iqp123-barren}), including the main ansatz used by the quantum natural language processing library \texttt{lambeq}~\cite{kartsaklis2021lambeq} (\Cref{thm:iqp4-barren}).
	
	\item
	We also prove that one of the IQP ansätze does not have barren plateaus for any number of layers, with the variance converging to a constant in the limit (\Cref{corr:iqp1-barren}).
\end{itemize}

Finally, we give a brief overview of barren plateau mitigation techniques presented in the literature in \Cref{sec:barren-mitigation}.

\section{Background} \label{sec:barren-backgound}

Consider a parametrised quantum circuit $U(\vec\theta)$ on $n$ qubits and a Hamiltonian $H$.
We assume that the parameters of $U$ are independently and uniformly distributed over the interval $[-\pi,\pi]$ since this is a common initialisation strategy.
One can show that the mean gradient of $U$'s expectation value with regards to $H$ is zero in that case, i.e. $\mathbf E\left(\frac{\partial \langle H\rangle}{\partial \theta_i}\right) = 0$~\cite{zhao2021analyzing}.
Now, if furthermore $\var{\frac{\partial \langle H\rangle}{\partial \theta_i}} \approx 0$ then it is likely that the training starts in a barren plateau where the gradient $\frac{\partial \langle H\rangle}{\partial \theta_i} \approx 0$.
Formally, we say that barren plateaus are present if $\var{\frac{\partial \langle H\rangle}{\partial \theta_i}} \in O\left(\frac{1}{2^{\text{poly}(n)}}\right)$, i.e. the variance vanishes exponentially as a function of the number of qubits $n$.
Then, Chebyshev’s inequality implies that 
$\text{Pr}\left(\left|\frac{\partial \langle H\rangle}{\partial \theta_i}\right| \geq \varepsilon \right) \leq \var{\frac{\partial \langle H\rangle}{\partial \theta_i}} / \varepsilon^2$.
In other words, the probability that the gradient $\frac{\partial \langle H\rangle}{\partial \theta_i}$ is non-zero up to some precision $\varepsilon$ is exponentially small in $n$.

The barren plateau phenomenon was first studied by McLean et al.~\cite{mcclean2018barren} who proved that barren plateaus appear if an ansatz is sufficiently random such that its parametrisations match the uniform distribution of unitaries (the so-called \textit{Haar-measure}) up to the second moment, i.e. they form a \textit{unitary 2-design}.
The distance between the distribution of unitaries generated by an ansatz and the Haar distribution can be seen as a measure for ansatz expressivity since it captures how uniformly an ansatz explores the unitary space~\cite{holmes2022connecting,sim2019expressibility}.
Sim et al.~\cite{sim2019expressibility} studied the expressiveness of several commonly used ansätze.
We will analyse a selection of these in \Cref{sec:barren-numeric} and \Cref{sec:barren-formal}.
Holmes et al. relate the existence of barren plateaus to the expressiveness of an ansatz~\cite{holmes2022connecting}, showing that more expressive ansätze have flatter gradient landscapes.
Concretely, they upper-bound the variance of the gradient in terms of how far an ansatz is from a 2-design, implying a trade-off between ansatz expressiveness and trainability.
Interestingly, the existence of barren plateaus also depends on the Hamiltonian $H$:
If we only measure a subset of qubits (i.e. we use a so-called \textit{local cost function}), then some ansätze can avoid barren plateaus up to logarithmic circuit depth in $n$~\cite{cerezo2021cost}.

Zhao and Gao~\cite{zhao2021analyzing} were the first to employ the ZX-calculus to analyse barren plateaus.
They express $\var{\frac{\partial \langle H\rangle}{\partial \theta_i}}$ as a linear combination of diagrams with an exponential number of terms, which they handle using tensor networks.
Wang and Yeung~\cite{wang2022differentiating} improve on this by expressing the variance in a single diagram, allowing the analysis of barren plateaus to be carried out entirely within the framework of ZX.
They consider ansätze $U(\vec\theta)$ where each parameter only occurs a single time and introduce the following notation for the expectation value:
\[
\langle H \rangle 
= \bra{0} U^\dagger(\vec\theta) H U(\vec\theta) \ket{0} 
=~ \input{\tikzitpathbackground/expval-1.tikz}
~=:~ \input{\tikzitpathbackground/expval-2.tikz}
\]
Since $\mathbf E\left(\frac{\partial \langle H\rangle}{\partial \theta_i}\right) = 0$ (a diagrammatic proof of this is given as Lemma 25 in~\cite{wang2022differentiating}), we have
\[ 
\var{\frac{\partial \langle H\rangle}{\partial \theta_i}} 
%= \mathbf E\left(\left(\frac{\partial \langle H\rangle}{\partial \theta_i}\right)^2\right) - \mathbf E\left(\frac{\partial \langle H\rangle}{\partial \theta_i}\right)^2
= \mathbf E\left(\left(\frac{\partial \langle H\rangle}{\partial \theta_i}\right)^2\right)
= \frac{1}{(2\pi)^m} \int_{-\pi}^\pi ... \int_{-\pi}^\pi \left(\frac{\partial \langle H\rangle}{\partial \theta_i}\right)^2 d\theta_1 ... d\theta_m.
\]

Using their graphical integration approach, Wang and Yeung express those nested integrals as the following diagram:

\bigskip
\begin{theorem}[Wang and Yeung~\cite{wang2022differentiating}] \label{thm:variance}
	\[ \var{\frac{\partial \langle H\rangle}{\partial \theta_i}} ~=~ \input{\tikzitpaththm:variance/statement.tikz} \]
\end{theorem}
\begin{proof}
	See Theorem 28 in~\cite{wang2022differentiating}.
\end{proof}

However, Wang and Yeung only apply \Cref{thm:variance} to a small toy circuit with two qubits and four parameters.
In particular, they perform no actual barren plateau analysis which would require computing the variance for an arbitrary number of qubits $n$.
The goal of this chapter is to apply \Cref{thm:variance} to ansätze that are used in practice and characterise when barren plateaus show up.

\section{Studied Ansätze} \label{sec:circuits}

\begin{figure}[t]
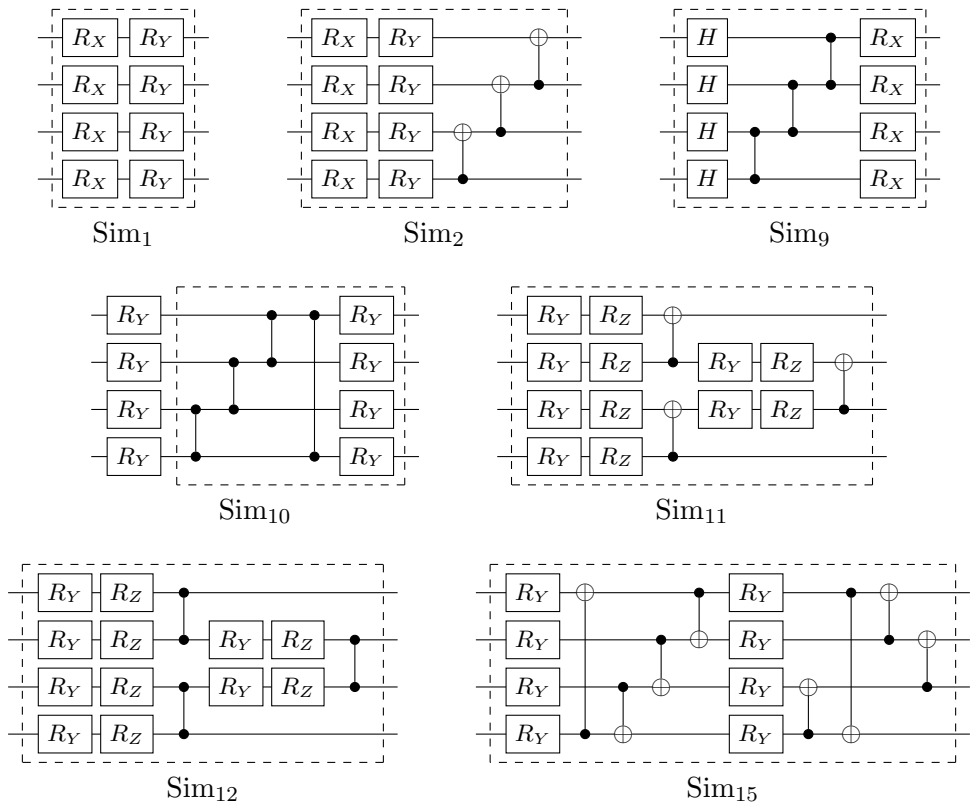

	\setlength{\tabcolsep}{10pt}
	\renewcommand{\arraystretch}{1.25}
	\begin{adjustwidth}{-2cm}{-2cm}
		\begin{center}
			\begin{tabular}{ccc}
				\input{\tikzitpathsim/sim1.tikz} & \input{\tikzitpathsim/sim2.tikz} & \input{\tikzitpathsim/sim9.tikz} \\
				$\text{Sim}_1$ & $\text{Sim}_2$ & $\text{Sim}_9$
			\end{tabular}
			\\[10pt]
			\begin{tabular}{cc}
				\input{\tikzitpathsim/sim10.tikz} & \input{\tikzitpathsim/sim11.tikz} \\
				$\text{Sim}_{10}$ & $\text{Sim}_{11}$
			\end{tabular}
			\\[10pt]
			\begin{tabular}{cc}
				\input{\tikzitpathsim/sim12.tikz} & \input{\tikzitpathsim/sim15.tikz} \\
				$\text{Sim}_{12}$ & $\text{Sim}_{15}$
			\end{tabular}
		\end{center}
	\end{adjustwidth}
	\caption{Circuits from Sim et al.~\cite{sim2019expressibility} we study in this chapter. The numbering follows Figure 2 in~\cite{sim2019expressibility}. The dashed box indicates a single layer that can be repeated multiple times where each layer has unique parameters. We omitted the parameters in the $R_X$, $R_Y$, and $R_Z$ gates for brevity.} 
	\label{fig:sim-circuits}
\end{figure}

In this chapter, we perform barren plateau analyses for two different classes of ansätze.
First, we consider a selection of circuits studied by Sim et al.~\cite{sim2019expressibility}.
Concretely, we analyse all ansätze for which Wang and Yeung's~\cite{wang2022differentiating} ZX-based variance computation from \Cref{thm:variance} is applicable.
They are depicted in \Cref{fig:sim-circuits}.
The remaining circuits in~\cite{sim2019expressibility} use controlled rotation gates which (as we have proven in \Cref{fact:U-repr-optimal}) need at least two parametrised spiders to be implemented.
Thus, \Cref{thm:variance} does not apply there.
\enlargethispage{\baselineskip}
Sim et al. \cite{sim2019expressibility} calculated the expressiveness of their ansätze, which makes them interesting cases to study as they are good candidates to empirically test the expressiveness vs. trainability trade-off described by Holmes et al \cite{holmes2022connecting}.

\begin{figure}[t]
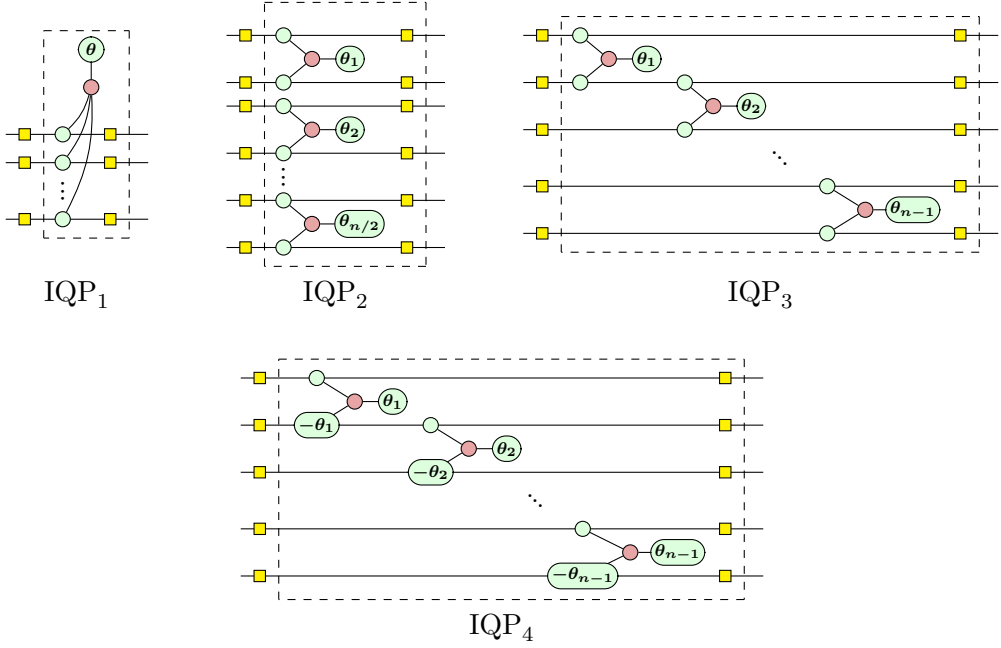

	\setlength{\tabcolsep}{10pt}
	\renewcommand{\arraystretch}{1.25}
	\begin{adjustwidth}{-2cm}{-2cm}
		\begin{center}
			\begin{tabular}{ccc}
				\input{\tikzitpathIQP/IQP1.tikz} & \input{\tikzitpathIQP/IQP2.tikz} & \input{\tikzitpathIQP/IQP3.tikz} \\
				$\text{IQP}_1$ & $\text{IQP}_2$ & $\text{IQP}_3$
			\end{tabular}
			\\[15pt]
			\begin{tabular}{c}
				\input{\tikzitpathIQP/IQP4.tikz} \\
				$\text{IQP}_4$
			\end{tabular}
		\end{center}
	\end{adjustwidth}
	\caption{Various $n$-qubit IQP ansätze we study in this chapter given in ZX notation. Similar to \Cref{fig:sim-circuits}, the dashed boxes can be repeated multiple times, where each layer has unique parameters. Note that $\text{IQP}_2$ is only defined for even $n$.}
	\label{fig:IQP-circuits}
\end{figure}

Secondly, we study \textit{instantaneous quantum polynomial} (IQP) circuits.
First introduced in~\cite{shepherd2009temporally}, IQPs consist of layers made up of diagonal gates, separated by columns of Hadamards, i.e.
\[ U(\vec\theta_1,...,\vec{\theta_\ell}) ~=~ \input{\tikzitpathIQP/IQP_gates.tikz} \]
where the blocks $D(\vec\theta_i)$ only contain gates with diagonal matrices.
Thus, all gates that make up $D(\vec\theta_i)$ commute with each other.
Therefore, it does not matter in which order they are executed which is the reason why this type of ansatz is called \textit{instantaneous}.
Remarkably, classical weak simulation of IQP circuits has been shown to be $\#P$-hard~\cite{bremner2016average,lund2017quantum} and an efficient simulation algorithm would collapse the polynomial hierarchy to the third level~\cite{bremner2011classical}.
Thus, the simple structure of IQP circuits already captures a quantum advantage, which makes them an interesting class of circuits to study.

For the purpose of this chapter, we use phase gadgets for the diagonal gates that make up the blocks.
This is motivated by the fact that they have an elegant representation in ZX an can be nicely reasoned about.
Concretely, \Cref{fig:IQP-circuits} shows the IQP ansätze we study in this chapter.
$\IQP{1}$ and $\IQP{2}$ are of more theoretical interest and will serve as demonstrations for our analytical techniques.
On the other hand, $\IQP{3}$ has been suggested in~\cite{meichanetzidis2020quantum} and $\IQP{4}$ is the default ansatz for the quantum natural language processing (QNLP) library \texttt{lambeq}~\cite{kartsaklis2021lambeq}.
Thus, the barren plateau analysis for this ansatz is of great practical interest.
Recalling (\ref{eqn:CRZ-zx-def}), $\IQP{4}$ can be seen as a ladder of $CR_Z$ gates.
Also note that each parameter occurs twice in $\IQP{4}$ which means that \Cref{thm:variance} is not directly applicable.
However, we can still compute the variance in some special cases which we discuss in \Cref{sec:barren-multiple-parameters}.

\section{Numerical Barren Plateau Detection} \label{sec:barren-numeric}

In this section, we develop a method to empirically test ansätze for barren plateaus by computing $\text{Var}(\frac{\partial\langle H\rangle}{\partial\theta_i})$.

\subsection{Method}

Our numerical barren plateau detection method relies on the following representation of the triangle in ZX calculus:

\bigskip
\begin{lemmaE}
	\[ \input{\tikzitpathlem:tri-zx/statement-1.tikz} ~=~ 2~ \input{\tikzitpathlem:tri-zx/statement-2.tikz} \]
\end{lemmaE}
\begin{proofE}
	\def\tikzitpath{chapter5/figs/}
	By comparing the action on the computational basis:
	\begin{gather*}
		\input{\tikzitpathlem:tri-zx/proof-1.tikz}
		\eqq{\cp} \input{\tikzitpathlem:tri-zx/proof-2.tikz}
		\eqq{\id,\sf} \input{\tikzitpathlem:tri-zx/proof-3.tikz}
		\eqq{*} 2~ \input{\tikzitpathlem:tri-zx/proof-4.tikz}
		\eqq{\tri} 2~ \input{\tikzitpathlem:tri-zx/proof-5.tikz} 
		\\[\linesep]
		\input{\tikzitpathlem:tri-zx/proof-6.tikz}
		\eqq{\cp} \input{\tikzitpathlem:tri-zx/proof-7.tikz}
		\eqq{\p,\sf} \input{\tikzitpathlem:tri-zx/proof-8.tikz}
		\eqq{\cp,\sf} \input{\tikzitpathlem:tri-zx/proof-9.tikz}
		\eqq{\tri} 2~ \input{\tikzitpathlem:tri-zx/proof-10.tikz} 
	\end{gather*}
	where the step $(*)$ follows again from plugging in the computational basis:
	\[
	\input{\tikzitpathlem:tri-zx/proof-11.tikz}
	\eqq{\sf,\id} \input{\tikzitpathlem:tri-zx/proof-12.tikz}
	\eqq{\sf} 2~ \input{\tikzitpathlem:tri-zx/proof-13.tikz}
	\qquad\qquad
	\input{\tikzitpathlem:tri-zx/proof-14.tikz}
	\eqq{\sf,\p} \input{\tikzitpathlem:tri-zx/proof-15.tikz}
	~=~ 0
	\eqq{\sf} 2~ \input{\tikzitpathlem:tri-zx/proof-16.tikz}
	\]
\end{proofE}

This allows us to represent the variance from \Cref{thm:variance} as a Clifford+T diagram\footnote{A ZX diagram is Clifford+T if all spider phases are multiples of $\frac{\pi}{4}$. The variance diagram is of course only Clifford+T if the ansatz is (not considering parametrised spiders), but this is the case for almost all ansätze used in practice.} which in turn allows us to use the ZX contraction techniques from Kissinger et al.~\cite{kissinger2022classical} to compute the scalar represented by the diagram.
Originally developed for classical simulation of quantum circuits, they employ decompositions of so-called \textit{magic states} and \textit{cat states} to successively simplify ZX diagrams leading to a runtime of $O(2^{\alpha t})$ where $\alpha \approx 0.396$ and $t$ is the number spiders with phase $\pm\frac{\pi}{4}$ or $\pm\frac{3\pi}{4}$.

Thus, contracting the variance diagram from \Cref{thm:variance} using this method has complexity $O(2^{4\alpha(p-1)})$ where $p$ is the number of parameters in the ansatz.
While the runtime scales exponentially, in practice the method is fast enough to handle a wide range of ansätze.
Concretely, all experiments in this section combined take roughly two hours to run using a single core on a standard desktop computer equipped with an Intel Core i7-8700k and 16Gb of RAM.
Furthermore, the execution speeds up linearly by utilising multiple CPU cores.

We implement the ansätze from \Cref{fig:sim-circuits} and \ref{fig:IQP-circuits}, and the variance diagram from \Cref{thm:variance} in the Rust programming language using the QuiZX library~\cite{kissinger2022simulating}.
Note that QuiZX uses cyclotomic rational numbers~\cite{coates2006cyclotomic} and has a special treatment for powers of $\sqrt 2$.
Therefore, all scalars that occur during the ZX contraction can be represented exactly, thus avoiding the imprecisions of floating point arithmetic.
%This allows us to represent the raw variance data in a human readable format (see ...)

Finally, we want to point out that one could also compute $\var{\frac{\partial\langle H\rangle}{\partial\theta_i}}$ by computing the gradient $\frac{\partial\langle H\rangle}{\partial\theta_i}$ for many random parameter samples using the shift rules discussed in \Cref{chap:gradient-recipes} and then compute the numerical variance.
However, this would require actually running the circuit on a quantum device or simulator for a large number of shots.
Furthermore, this method only yields noisy estimates of the variance (in particular when using a NISQ device) whereas our tool computes exact values for $\var{\frac{\partial\langle H\rangle}{\partial\theta_i}}$.

\subsection{Note on Zero Variance}

Before discussing our numerical results, we remark that we sometimes observe $\var{\frac{\partial\langle H\rangle}{\partial\theta_i}} = 0$, meaning that the gradient is constant. 
We show in \Cref{sec:sim1} that in those cases the gradient is actually zero, meaning that varying the parameter $\theta_i$ does not change $\langle H\rangle$. 
A trivial example of this is using the Hamiltonian $H=I^{\otimes n}$, i.e. performing no measurement.
However, there are also non-trivial cases where some parameters do not influence the expectation value.
Training such parameters is of course pointless.
Therefore, we can exclude them from the barren plateau analysis.
See \Cref{rem:var-zero} for more details on this.

\subsection{Results}

\subsubsection{Sim Ansätze}

\begin{figure}
	\begin{center}
		\includegraphics[draft=false]{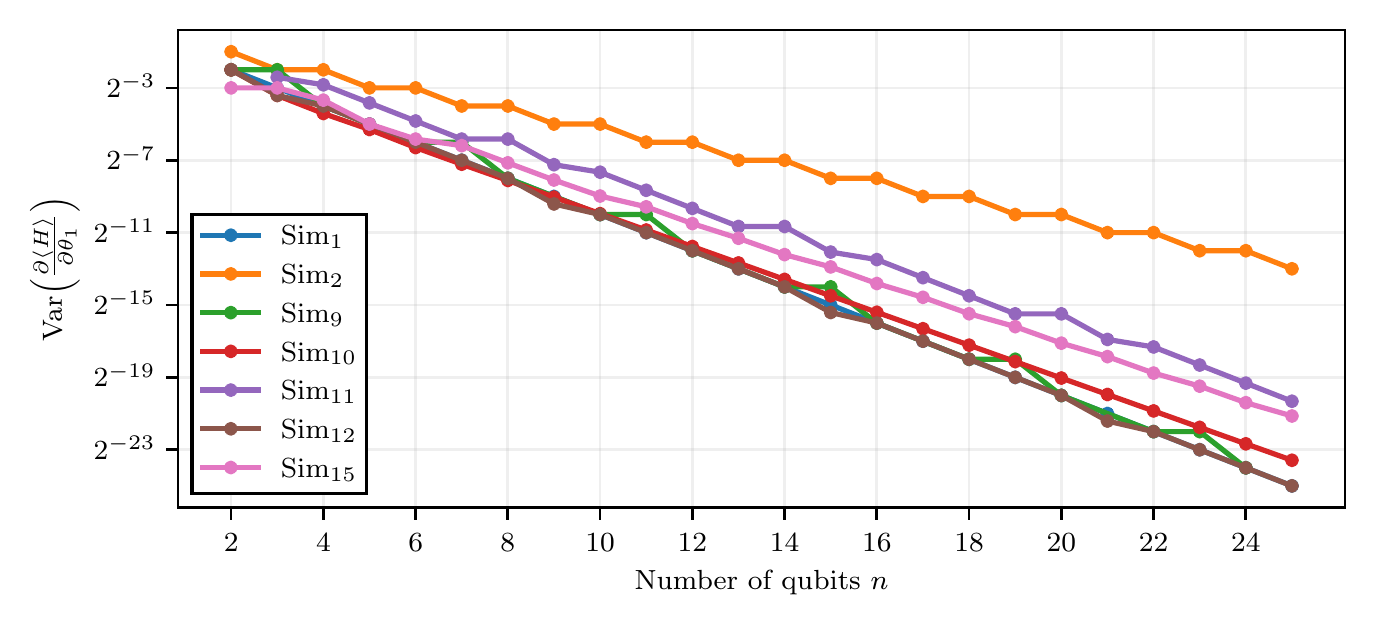}
	\end{center}
	\caption{Gradient variance as function of qubits for a single layer of the different Sim ansätze for the Hamiltonian $H = Z^{\otimes n}$.
		Concretely, the variance $\text{Var}( \frac{\partial\langle H\rangle}{\partial\theta_1})$ for the first parameter $\theta_1$ (i.e. the top-left rotations in \Cref{fig:sim-circuits}) is plotted.
		Note that the y-axis has a logarithmic scale.}
	\label{fig:sim-single-layer}
\end{figure}

We begin by analysing the Sim ansätze from \Cref{fig:sim-circuits}.
\Cref{fig:sim-single-layer} shows the gradient variance for a single layer of the circuits when measuring with the Hamiltonian $H = Z^{\otimes n}$.
As we can see, the gradient variance of all ansätze seems to vanish exponentially with increasing $n$.
This suggests that the Sim ansätze have barren plateaus for $H = Z^{\otimes n}$, even when using only a single layer.

\begin{figure}
	\begin{adjustwidth*}{-2cm}{-2cm}
		\begin{center}
			\includegraphics[draft=false]{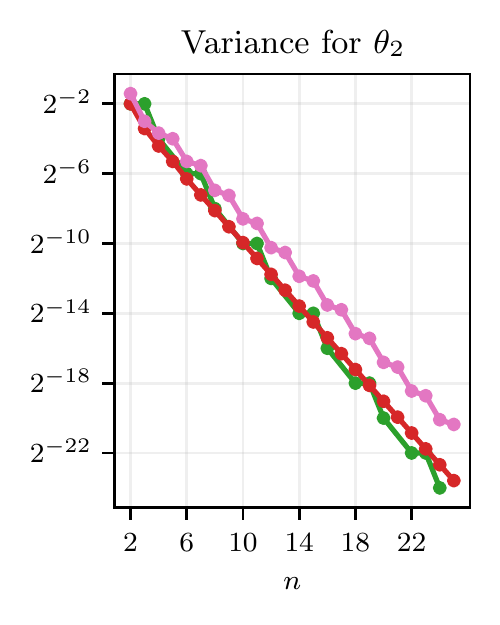}
			\includegraphics[draft=false]{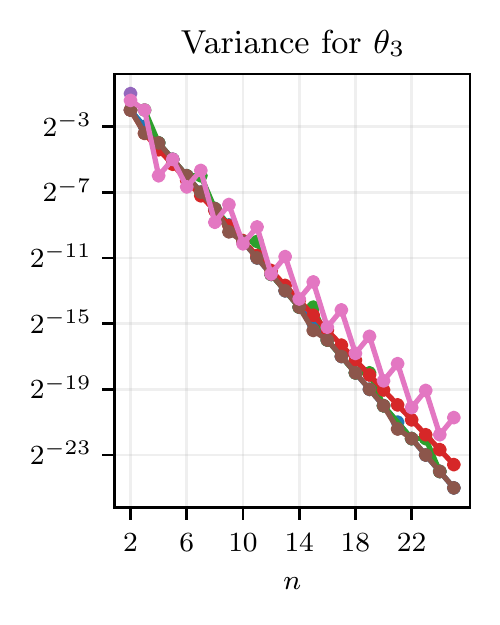}
			\includegraphics[draft=false]{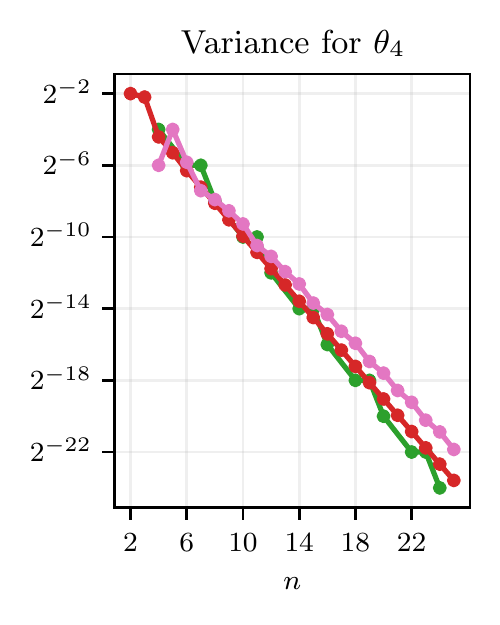}
			\includegraphics[draft=false]{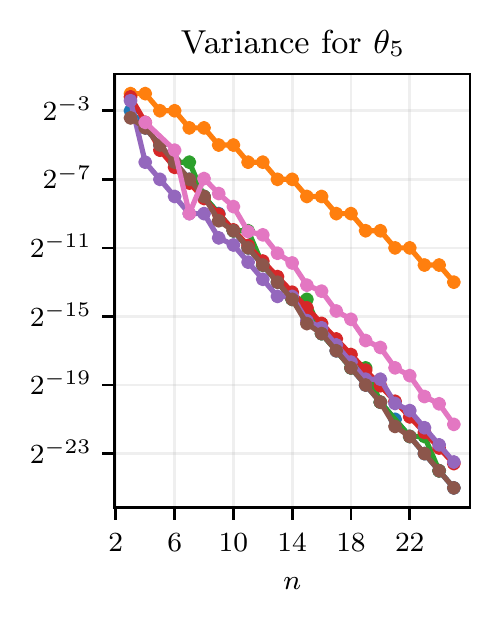}
			\includegraphics[draft=false]{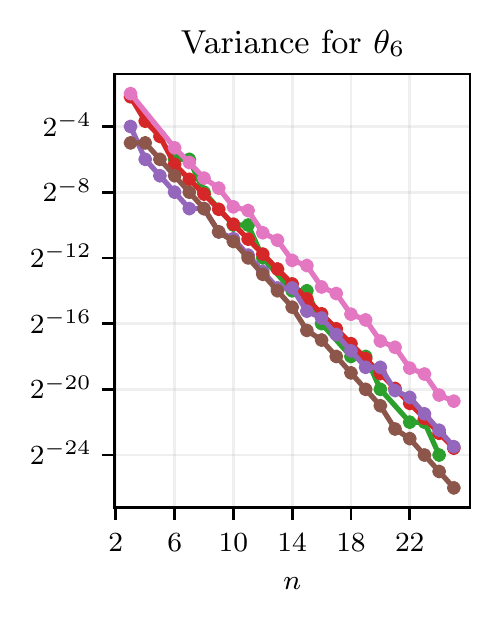}
			\includegraphics[draft=false]{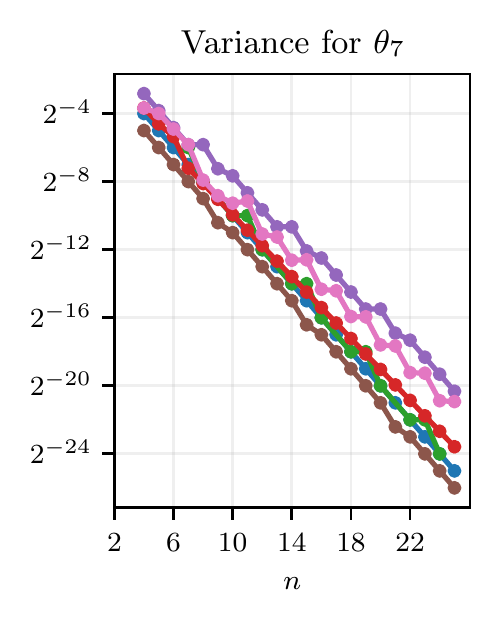}
			\includegraphics[draft=false]{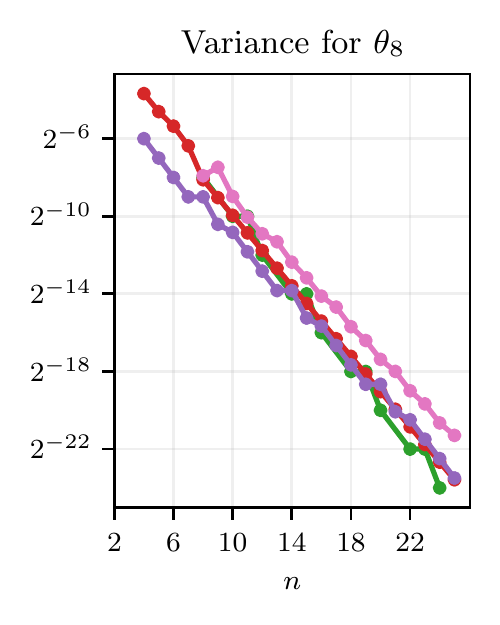}
			\includegraphics[draft=false]{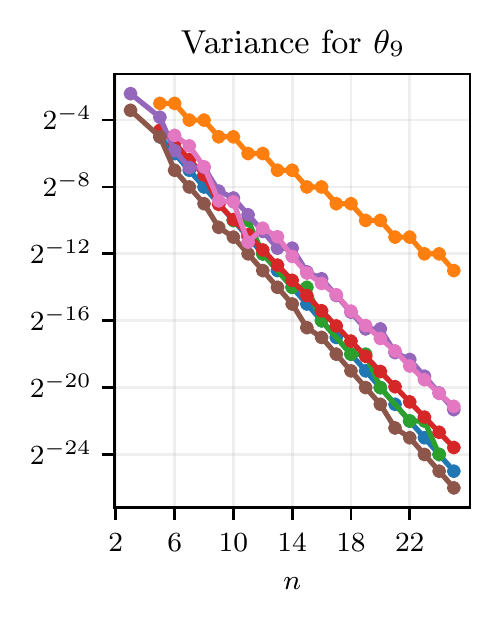}
			\includegraphics[draft=false]{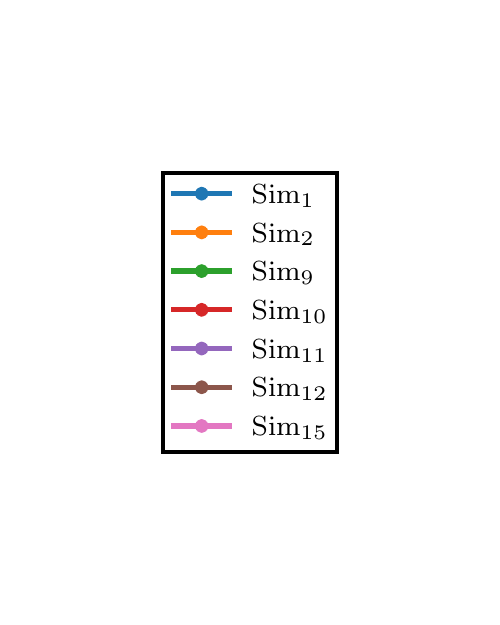}
		\end{center}
	\end{adjustwidth*}
	\caption{Gradient variance for different parameters $\theta_i$ as a function of qubits for a single layer of the different Sim ansätze for the Hamiltonian $H = Z^{\otimes n}$.
		We do not plot points if the parameter does not exist or $\text{Var}( \frac{\partial\langle H\rangle}{\partial\theta_i}) = 0$.}
	\label{fig:sim-single-layer-parameters}
\end{figure}

However, note that \Cref{fig:sim-single-layer} only plots the gradient variance w.r.t. the first parameter $\theta_1$.
It might be the case that other parameters do not vanish exponentially which would make learning possible.
To investigate this, we run the same experiment for different parameters $\theta_i$.
The results are shown in \Cref{fig:sim-single-layer-parameters}.
As we can see, as long as $\var{\frac{\partial\langle H\rangle}{\partial\theta_i}} \neq 0$ we get exponentially vanishing variances for all cases.
This leads us to stating the following hypothesis:

\begin{hypothesis} \label{hyp:sim-barren}
	For $H = Z^{\otimes n}$, all single-layer Sim ansätze from \Cref{fig:sim-circuits} have barren plateaus on all parameters.
\end{hypothesis}

This matches with the expressiveness results computed by Sim et al. \cite{sim2019expressibility}:
All ansätze in \Cref{fig:sim-circuits} have a similar expressiveness for a single layer.
Differences in expressiveness only show up when additional layers are added, with some circuits gaining more expressiveness by this than others.
However, our experiments suggest that even a single layer already suffices to generate barren plateaus.

We want to stress that just looking at graphs is of course not a proof for the existence of a barren plateau.
It could for example be the case that the curve in \Cref{fig:sim-single-layer} starts to flatten after some point $n_0$ outside of the range we investigated.
However, the number of qubits used in QML experiments today is limited.
Thus, in practical terms, the variance behaviour for small $n$ is most relevant to make statements about the trainability of ansätze.
Here, our experiments suggest that the Sim ansätze scale badly and might benefit from using barren plateau mitigation techniques (see \Cref{sec:barren-mitigation}).

While our empirical results are of practical use, there is also significant value in formal statements regarding the existence of barren plateaus.
We turn to this question in \Cref{sec:barren-formal} where we formally analyse $\Sim{1}$, $\Sim{2}$, and $\Sim{9}$ and prove \Cref{hyp:sim-barren} for those three ansätze.
Furthermore, we generalise to arbitrary Hamiltonians, moving beyond the case $H=Z^{\otimes n}$ considered here.

\subsubsection{IQP Ansätze}

\begin{figure}
	\begin{center}
		\includegraphics[draft=false]{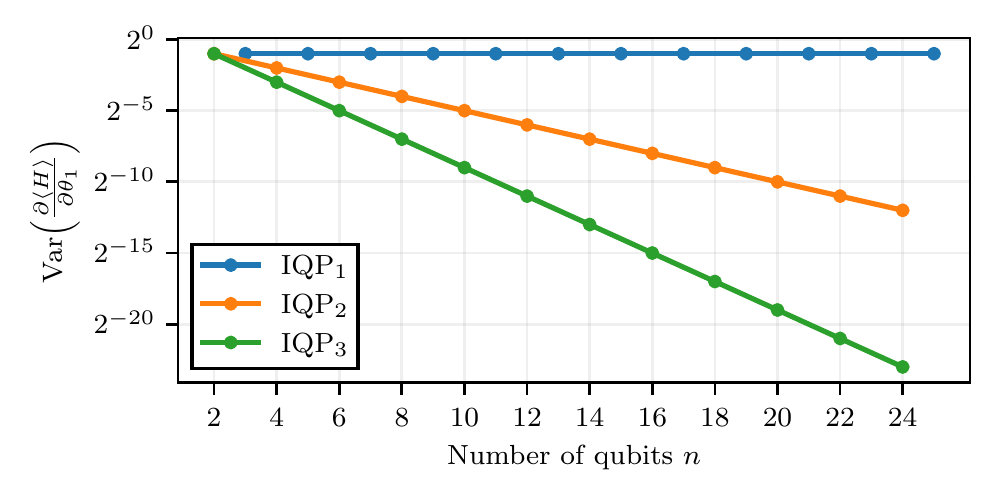}
	\end{center}
	\caption{Gradient variance as a function of qubits for a single layer of the different IQP ansätze. We do not plot points if $\text{Var}( \frac{\partial\langle H\rangle}{\partial\theta_i}) = 0$.}
	\label{fig:iqp-single-layer}
\end{figure}

We run similar experiments for a single layer of $\IQP{1}$, $\IQP{2}$, and $\IQP{3}$.\footnote{We cannot apply our method to $\IQP{4}$ since multiple spiders share the same parameter, however we will derive some theoretical results in \Cref{sec:barren-multiple-parameters}.}
However, while we still use the Hamiltonian $H=Z^{\otimes n}$ for $\IQP{1}$, we use an alternating Hamiltonian $H=Y\otimes X\otimes Y\otimes X...$ for $\IQP{2}$ and $\IQP{3}$.
This is because we get $\text{Var}(\frac{\partial\langle H\rangle}{\partial\theta_i}) = 0$ otherwise.\footnote{For a theoretical explanation of this see the proof of \Cref{fact:iqp123-barren}.}
The results are shown in \Cref{fig:iqp-single-layer}.
Similar to the Sim ansätze, $\IQP{2}$ and $\IQP{3}$ appear to have exponentially vanishing gradient variances.
Surprisingly, rerunning the experiment for different parameters $\theta_i$ yields the exact same numerical variance values.\footnote{We will prove later that this is in fact true for \textit{all} single layer IQP ansätze (see \Cref{thm:IQP-variance} and \Cref{rem:iqp-same-var}).}
Thus, we make the following hypothesis:

\begin{hypothesis} \label{hyp:iqb-barren}
	A single layer of $\IQP{2}$ and $\IQP{3}$ has barren plateaus on all parameter for $H = Y\otimes X\otimes Y\otimes X...$.
\end{hypothesis}

\begin{figure}
	\begin{adjustwidth*}{-2cm}{-2cm}
		\begin{center}
			\includegraphics[draft=false]{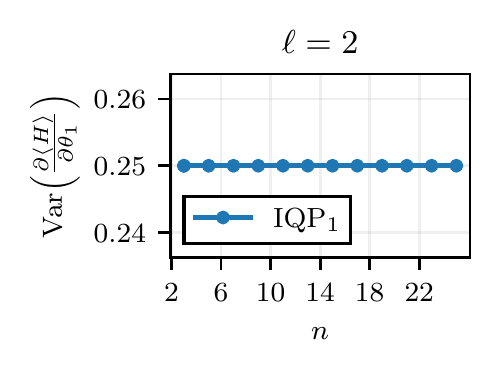}
			\includegraphics[draft=false]{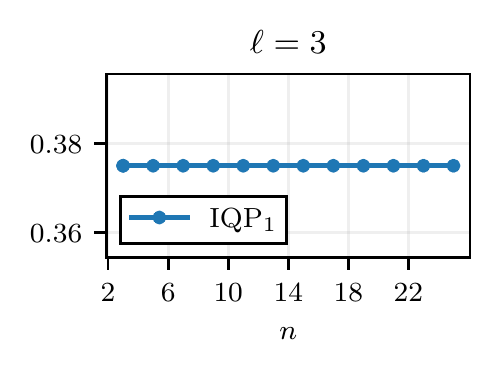}
			\includegraphics[draft=false]{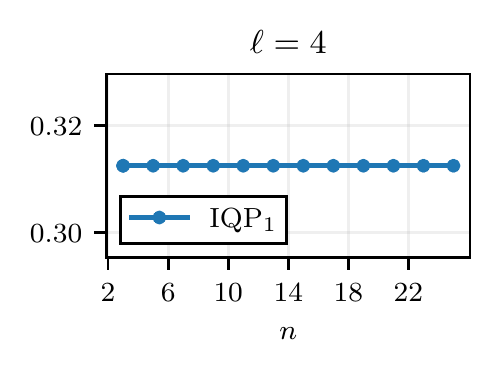}
		\end{center}
	\end{adjustwidth*}
	\caption{Gradient variance of $\IQP{1}$ as a function of qubits for different number of layers $\ell$. We do not plot points if $\text{Var}( \frac{\partial\langle H\rangle}{\partial\theta_i}) = 0$.}
	\label{fig:iqp1-multi-layer}
\end{figure}

\begin{figure}
	\begin{center}
		\includegraphics[draft=false]{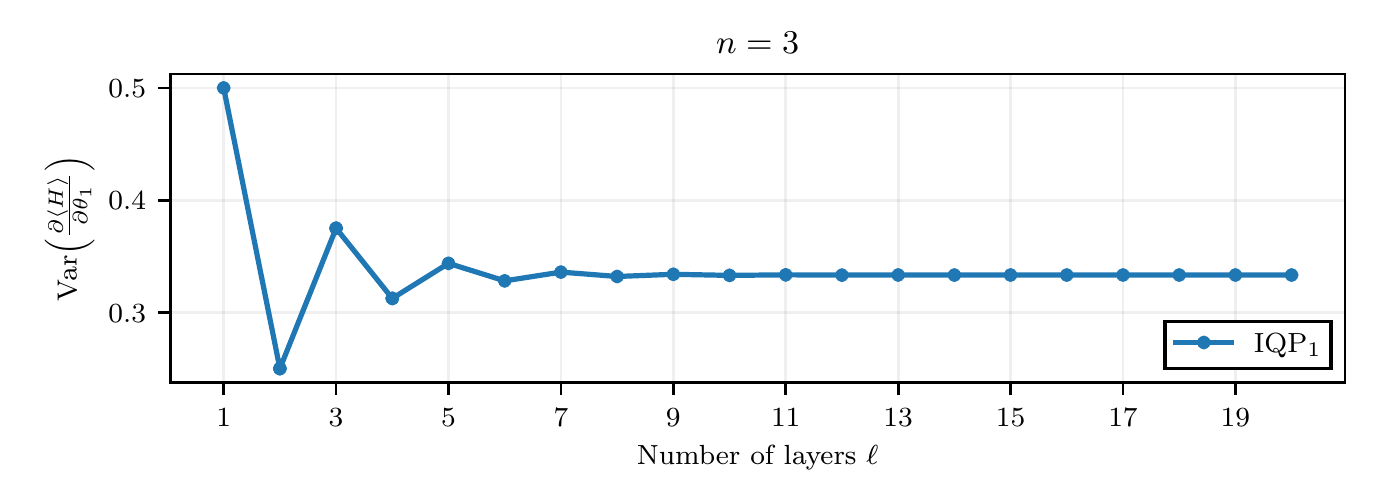}
	\end{center}
	\caption{Gradient variance of $\IQP{1}$ as a function of layers for $n = 3$.}
	\label{fig:iqp1-multi-layer-converge}
\end{figure}

However, the more interesting observation from \Cref{fig:iqp-single-layer} is that the gradient variance of $\IQP{1}$ does not vanish.
To investigate whether using more than one layer makes a barren plateau appear, we rerun the $\IQP{1}$ experiment for increasing numbers of layers.
But the results in \Cref{fig:iqp1-multi-layer} show that this is not the case.
While adding more layers changes the variance, it stays constant with increasing $n$.
Finally, we plot the variance for $n=3$ as a function of $\ell$ in \Cref{fig:iqp1-multi-layer-converge}.
As we can see, the variance seems to converge with increasing $\ell$.
To summarise, we can make the following hypothesis:

\begin{hypothesis} \label{hyp:iqb1-barren}
	$\IQP{1}$ does not have barren plateaus for $H = Z^{\otimes n}$.
	More specifically, the variance for $\theta_1$ is constant in $n$ and converges for $\ell \to \infty$.
\end{hypothesis}

Again following the trade-off described by Holmes et al. \cite{holmes2022connecting}, this observation might be explained by the fact that $\IQP{1}$ is a very simple ansatz with limited expressiveness.

We will prove both \Cref{hyp:iqb-barren} and \Cref{hyp:iqb1-barren} in in the next section (see \Cref{fact:iqp123-barren} and \Cref{corr:iqp1-barren}).
%Furthermore, we generalise to arbitrary Hamiltonians moving beyond the example considered here.

\section{Analytical Barren Plateau Detection} \label{sec:barren-formal}

After investigating the gradient landscape of ansätze numerically, we now turn to the formal analysis of barren plateaus using \Cref{thm:variance}.
For this, we introduce a bit of terminology to refer to the structure of the variance diagram from \Cref{thm:variance}:
Note that it is made up of two main building blocks which we call \textit{cycles}:
\[ \input{\tikzitpathbackground/cycle-1.tikz} \qquad\qquad \input{\tikzitpathbackground/cycle-2.tikz} \]
The left cycle is plugged into the positions corresponding to the variance parameter $\theta_i$.
The right cycle with the triangle is plugged into every other position, corresponding to parameters $\theta_j$ with $j \neq i$.
The remainder of this section will largely be concerned with simplifying those kinds of cycles for different expectation value diagrams plugged in the middle.
This will allow us to contract the diagram and obtain a numerical value for the variance as a function of $n$ or $\ell$.

We make use of the following two lemmas throughout this section:

\begin{lemmaE}
	For all $x,y \in \{0,1\}$ we have
	\begin{equation}\label{eqn:pi-cycle}
		\input{\tikzitpathlem:pi-cycle/statement-1.tikz} ~=~ \input{\tikzitpathlem:pi-cycle/statement-2.tikz}
		\qquad\qquad
		\input{\tikzitpathlem:pi-cycle/statement-3.tikz} ~=~ \input{\tikzitpathlem:pi-cycle/statement-4.tikz}
	\end{equation}
\end{lemmaE}
\begin{proofE}
	\def\tikzitpath{chapter5/figs/}
	\begin{gather*}
	 	\input{\tikzitpathlem:pi-cycle/statement-1.tikz}
	 	\eqq{\sf} \input{\tikzitpathlem:pi-cycle/proof-1.tikz}
	 	\eqq{\p,\sf} \input{\tikzitpathlem:pi-cycle/proof-2.tikz} \\[\linesep]
	 	\eqq{\sf} \input{\tikzitpathlem:pi-cycle/proof-3.tikz}
	 	\eqq{\ho} \input{\tikzitpathlem:pi-cycle/statement-2.tikz}
	\end{gather*}
	\begin{gather*}
		\input{\tikzitpathlem:pi-cycle/statement-3.tikz}
		\eqq{\sf} \input{\tikzitpathlem:pi-cycle/proof-4.tikz}
		\eqq{\p,\sf} \input{\tikzitpathlem:pi-cycle/proof-5.tikz} \\[\linesep]
		\eqq{\sf} \input{\tikzitpathlem:pi-cycle/proof-6.tikz}
		\eqq{\ho} \input{\tikzitpathlem:pi-cycle/statement-4.tikz}
	\end{gather*}
\end{proofE}

\bigskip
\begin{lemmaE}
	\begin{equation}\label{eqn:pi-connect}
		\input{\tikzitpathlem:pi-connect/statement-1.tikz} ~=~ e^{i\alpha}~ \input{\tikzitpathlem:pi-connect/statement-2.tikz}
		\qquad\qquad
		\input{\tikzitpathlem:pi-connect/statement-3.tikz} ~=~ 2^{n-1}~ \input{\tikzitpathlem:pi-connect/statement-4.tikz}
	\end{equation}
\end{lemmaE}
\begin{proofE}
	\def\tikzitpath{chapter5/figs/}
	\begin{gather*}
		\input{\tikzitpathlem:pi-connect/statement-1.tikz}
		\eqq{\p,\sf,\id} e^{i\alpha}~ \input{\tikzitpathlem:pi-connect/proof-1.tikz}
		\eqq{\sf} e^{i\alpha}~ \input{\tikzitpathlem:pi-connect/statement-2.tikz} 
		\\[\linesep]
		\input{\tikzitpathlem:pi-connect/statement-3.tikz}
		\eqq{\p,\sf,\id} \input{\tikzitpathlem:pi-connect/proof-2.tikz}
		\eqq{\sf} 2^{n-1}~ \input{\tikzitpathlem:pi-connect/statement-4.tikz} 
	\end{gather*}
	Note that the scalar $2^{n-1}$ appears in the last step according to \Cref{lem:sf-pink}.
\end{proofE}

\subsection{Introductory Example} \label{sec:barren-example}

Before discussing the Sim and IQP ansätze, we first show how to diagrammatically compute the variance of a smaller example ansatz:
\[ U(\theta_1,\theta_2) ~=~ \input{\tikzitpathintro/ansatz.tikz} \]
For the Hamiltonian $H = X^{\otimes n}$, we get the following expectation value:
\begin{gather*}
	\langle H \rangle
	~=~ \input{\tikzitpathintro/expval-1.tikz}  \\[\linesep]
	~=~ \input{\tikzitpathintro/expval-2.tikz} \\[\linesep]
	\eqq{\cc,\sf,\id} \frac{1}{2^n} ~ \input{\tikzitpathintro/expval-3.tikz} 
	\eqq{\cc} \frac{1}{4^n}~ \input{\tikzitpathintro/expval-4-2.tikz} \\[\linesep]
	\eqq{\cc} \frac{2^{n-1}}{4^n} ~ \input{\tikzitpathintro/expval-4.tikz} 
	\eqq{\sf} \frac{1}{2^{n+1}} ~ \input{\tikzitpathintro/expval-5.tikz} \\[\linesep]
	\eqq{\ref{eqn:pi-cycle}} \frac{1}{4} ~ \input{\tikzitpathintro/expval-6.tikz} 
	\eqq{\ho} \begin{cases}
		\frac{1}{4} ~ \input{\tikzitpathintro/expval-7-1.tikz} & \text{if $n$ is even} \\[20pt]
		\frac{1}{4} ~ \input{\tikzitpathintro/expval-7-2.tikz} & \text{if $n$ is odd}
	\end{cases} \\[\linesep]
	\eqq{\sf} \begin{cases}
		0 & \text{if $n$ is even} \\[10pt]
		\frac{1}{4} ~ \input{\tikzitpathintro/expval-7-2.tikz} & \text{if $n$ is odd}
	\end{cases}
\end{gather*}

If $n$ is even, we have $\langle H \rangle = 0$ and thus $\frac{\partial \langle H\rangle}{\partial \theta_1} = \frac{\partial \langle H\rangle}{\partial \theta_2} = 0$ such that $\text{Var}\left(\frac{\partial \langle H\rangle}{\partial \theta_1}\right) = \text{Var}\left(\frac{\partial \langle H\rangle}{\partial \theta_2}\right) = 0$.
If $n$ is odd, we can calculate the variance of $\frac{\partial \langle H\rangle}{\partial \theta_i}$ diagrammatically using \Cref{thm:variance}:
\begin{gather*}
	\text{Var}\left(\frac{\partial \langle H\rangle}{\partial \theta_1}\right)
	~=~ \frac{1}{16} ~ \input{\tikzitpathintro/odd/1-1.tikz}
	\eqq{\sf,\ref{eqn:pi-cycle}} \frac{1}{16} ~ \input{\tikzitpathintro/odd/1-2.tikz} \\[\linesep]
	\eqq{\id,\p,\sf} -\frac{1}{16} ~ \input{\tikzitpathintro/odd/1-3.tikz}
	\eqq{\sf,\id} -\frac{1}{8} ~ \input{\tikzitpathintro/odd/1-4.tikz} \\[\linesep]
	\eqq{\cc} -\frac{1}{8} ~ \input{\tikzitpathintro/odd/1-5.tikz} 
	\eqq{\p,\sf} -\frac{1}{8} ~ \input{\tikzitpathintro/odd/1-6.tikz}
	\eqq{\sf,\ho} -\frac{1}{8} ~ \input{\tikzitpathintro/odd/1-7.tikz} \\[\linesep]
	\eqq{\tri} -\frac{1}{8} ~ \input{\tikzitpathintro/odd/1-8.tikz}
	~=~ -\frac{1}{8} \cdot (-1) \cdot 2
	~=~ \frac{1}{4}
\end{gather*}
Similarly, for $\frac{\partial \langle H\rangle}{\partial \theta_2}$ we get
\begin{gather*}
	\text{Var}\left(\frac{\partial \langle H\rangle}{\partial \theta_2}\right)
	~=~ \frac{1}{16} ~ \input{\tikzitpathintro/odd/2-1.tikz}
	\eqq{\id,\cc} \frac{1}{32} ~ \input{\tikzitpathintro/odd/2-2.tikz} \\[\linesep]
	\eqq{\cc} \frac{1}{32} ~ \input{\tikzitpathintro/odd/2-3.tikz}
	\eqq{\sf} \frac{1}{32} ~ \input{\tikzitpathintro/odd/2-4.tikz} \\[\linesep]
	\eqq{\ref{eqn:pi-connect}} \frac{1}{16} ~ \input{\tikzitpathintro/odd/2-5.tikz}
	\eqq{\ref{eqn:pi-connect}} \frac{1}{8} ~ \input{\tikzitpathintro/odd/2-6.tikz}
	\eqq{\sf,\ho} \frac{1}{8} ~ \input{\tikzitpathintro/odd/2-7.tikz} \\[\linesep]
	\eqq{\ref{eqn:pi-cycle},\cp,\sf} \frac{1}{8} ~ \input{\tikzitpathintro/odd/2-8.tikz}
	~=~ \frac{1}{4} ~ \input{\tikzitpathintro/odd/2-9.tikz}
	~=~ \frac{1}{4} \cdot 1 \cdot 1
	~=~ \frac{1}{4}
\end{gather*}

In both cases, the variance of the gradient does not vanishes exponentially.
%We have also verified this result numerically using our tool from \Cref{sec:barren-numeric}.
Thus, we can conclude that the barren plateau phenomenon does not appear in this ansatz when measuring using the Hamiltonian $H=X^{\otimes n}$.
The diagrammatic calculation in this example was relatively straightforward since the ansatz $U(\theta_1,\theta_2)$ has a fixed number of parameters, independent of the number of qubits $n$.
Next, we will consider ansätze where the number of parameters increases when increasing $n$.

\subsection{Sim 1} \label{sec:sim1}

A single layer of $\Sim{1}$ can be represented in the ZX-calculus as 
\[ \Sim{1}(\vec\theta) ~=~ \input{\tikzitpathsim/1/circ-1.tikz} \]
Given some Hamiltonian $H$, the corresponding expectation value is given by
\begin{gather*}
	\langle H \rangle 
	~=~ \input{\tikzitpathsim/1/expval-1.tikz} 
	\eqq{\sf,\cc} \frac{1}{2^n}~ \input{\tikzitpathsim/1/expval-2.tikz} \\[\linesep]
	\eqq{\sf,\id} \frac{1}{2^n}~ \input{\tikzitpathsim/1/expval-3.tikz}
\end{gather*}
Recall that each Hamiltonian can be written as a sum of Pauli strings $\{X,Y,Z,I\}^{\otimes n}$.
Thus, it suffices to compute $\var{\frac{\partial\langle H\rangle}{\partial\theta_i}}$ for $H = P_1\otimes ...\otimes P_n$ where $P_j \in \{X,Y,Z,I\}$.
To make the following derivations more concise, we represent all three cases in a single diagram
\[ \input{\tikzitpathsim/1/H-1.tikz} ~=~ i^{a_jb_j} ~ \input{\tikzitpathsim/1/H-2.tikz} \]
where
\[
a_j = \begin{cases}
	1 & \text{if } H_j = Y, Z \\
	0 & \text{if } H_j = X, I
\end{cases}
\qquad\qquad
b_j = \begin{cases}
	1 & \text{if } H_j = X, Y \\
	0 & \text{if } H_j = Z, I.
\end{cases}
\]
Thus, we can write the expectation value as
\[ \langle H \rangle ~=~ \frac{i^{\sum a_jb_j}}{2^n}~ \input{\tikzitpathsim/1/expval-4.tikz} \]

Next, we consider the different types of cycles that show up in the variance diagram:

\begin{lemmaE} \label{lem:sim1-cycles}
	We have
	\begin{gather*}
		(-1)^{a_ib_i}~ \input{\tikzitpathsim/1/lem-1/1.tikz}
		~=~ \begin{cases}
		0 & \text{if } H_i=I \\
		1 & \text{if } H_i=Y,X \\
		2 & \text{if } H_i=Z
		\end{cases}
		\\[\linesep]
		(-1)^{a_ib_i}~ \input{\tikzitpathsim/1/lem-2/1.tikz}
		~=~ \begin{cases}
			0 & \text{if } H_i=Z,I \\
			1 & \text{if } H_i=X,Y
		\end{cases}
		\\[\linesep]
		(-1)^{a_ib_i}~ \input{\tikzitpathsim/1/lem-3/1.tikz}
		~=~ \begin{cases}
			1 & \text{if } H_i = X \\
			2 & \text{if } H_i = Z,Y \\
			4 & \text{if } H_i = I
		\end{cases}
	\end{gather*}
\end{lemmaE}
\begin{proofE}
	\def\tikzitpath{chapter5/figs/}
	We prove the three equations separately:
	\begin{itemize}
		\item We have
		\begin{gather*}
			(-1)^{a_ib_i}~ \input{\tikzitpathsim/1/lem-1/1.tikz}
			\eqq{\p,\cc,\sf} (-1)^{a_ib_i}(-1)^{b_i}~ \input{\tikzitpathsim/1/lem-1/2.tikz} \\[\linesep]
			\eqq{\sf} (-1)^{a_ib_i+b_i}~ \input{\tikzitpathsim/1/lem-1/3.tikz}
			\eqq{\ho,\cc} \frac{1}{2}(-1)^{a_ib_i+b_i}~ \input{\tikzitpathsim/1/lem-1/4.tikz} \\[\linesep]
			\eqq{\ref{eqn:pi-cycle}} \frac{1}{2}(-1)^{a_ib_i+b_i}~ \input{\tikzitpathsim/1/lem-1/5.tikz} 
			\eqq{\id,\sf} \frac{1}{2}(-1)^{a_ib_i+b_i}~ \input{\tikzitpathsim/1/lem-1/6.tikz} \\[\linesep]
			\eqq{\cc,\sf} \frac{1}{2}(-1)^{a_ib_i+b_i}~ \input{\tikzitpathsim/1/lem-1/7.tikz}
		\end{gather*}
		If $b_i=0$, we get
		\[  
		\frac{1}{2}~ \input{\tikzitpathsim/1/lem-1/even/8-0.tikz}
		\eqq{\tri,\sf} \frac{1}{2}~ \input{\tikzitpathsim/1/lem-1/even/8.tikz}
		\eqq{\id,\sf,\ho} \frac{1}{2}~ \input{\tikzitpathsim/1/lem-1/even/9.tikz}
		~=~ 2a_i.
		\]
		If $b_i=1$, we get
		\begin{gather*}
		\frac{1}{2}(-1)^{a_i+1}~ \input{\tikzitpathsim/1/lem-1/odd/8-0.tikz}
		\eqq{\tri,\cp,\sf} \frac{1}{2}(-1)^{a_i+1}~ \input{\tikzitpathsim/1/lem-1/odd/8.tikz} \\[\linesep]
		\eqq{\p,\sf,\id} \frac{1}{2}(-1)^{a_i+1}(-1)^{a_i+1}~ \input{\tikzitpathsim/1/lem-1/odd/9.tikz}
		~=~ 1.
		\end{gather*}
		
		\item We have
		\begin{gather*}
			(-1)^{a_ib_i}~ \input{\tikzitpathsim/1/lem-2/1.tikz}
			\eqq{\p,\cc,\sf} (-1)^{a_ib_i}~ \input{\tikzitpathsim/1/lem-2/2.tikz} \\[\linesep]
			\eqq{\sf} (-1)^{a_ib_i}~ \input{\tikzitpathsim/1/lem-2/3.tikz}
			\eqq{\ho,\cp,\cc,\sf} (-1)^{a_ib_i}(-1)^{a_i}~ \input{\tikzitpathsim/1/lem-2/4.tikz} \\[\linesep]
			\eqq{\cc} \frac{1}{2}(-1)^{a_ib_i+a_i}~ \input{\tikzitpathsim/1/lem-2/5.tikz}
			\eqq{\sf,\ho} \frac{1}{2}(-1)^{a_ib_i+a_i}~ \input{\tikzitpathsim/1/lem-2/6.tikz} \\[\linesep]
			\eqq{\p,\sf} \frac{1}{2}(-1)^{a_ib_i+a_i}~ \input{\tikzitpathsim/1/lem-2/7.tikz}
			~=~ \begin{cases}
			0 &\text{if } b_i = 0 \\
			1 &\text{if } b_i = 1
			\end{cases}
		\end{gather*}	
		
		\item We have
		\begin{gather*}
			(-1)^{a_ib_i}~ \input{\tikzitpathsim/1/lem-3/1.tikz} \\[\linesep]
			\eqq{\p,\cc,\sf} (-1)^{a_ib_i}~ \input{\tikzitpathsim/1/lem-3/2.tikz}
			\eqq{\sf,\ho} (-1)^{a_ib_i}~ \input{\tikzitpathsim/1/lem-3/3.tikz}
		\end{gather*}
		If $b_i = 0$, we get
		\begin{gather*}
			\input{\tikzitpathsim/1/lem-3/even/4-0.tikz}
			\eqq{\tri,\sf} \input{\tikzitpathsim/1/lem-3/even/4.tikz}
			\eqq{\id,\sf,\ho} \input{\tikzitpathsim/1/lem-3/even/5.tikz} \\[\linesep]
			\eqq{\cc} \input{\tikzitpathsim/1/lem-3/even/6.tikz}
			\eqq{\sf,\ho} \input{\tikzitpathsim/1/lem-3/even/7.tikz}
			\eqq{\cp,\tri,\sf} \input{\tikzitpathsim/1/lem-3/even/8.tikz}
			~=~ \begin{cases}
				4 & \text{if } a_i = 0 \\
				2 & \text{if } a_i = 1
			\end{cases}
		\end{gather*}
		If $b_i = 1$, we get
			\begin{gather*}
			(-1)^{a_i}~ \input{\tikzitpathsim/1/lem-3/odd/4-0.tikz}
			\eqq{\tri,\cp,\sf} \input{\tikzitpathsim/1/lem-3/odd/4.tikz} \\[\linesep]
			\eqq{\cp,\cc,\sf} (-1)^{a_i}(-1)^{a_i}~ \input{\tikzitpathsim/1/lem-3/odd/5.tikz} 
			\eqq{\cc} \frac{1}{2}~ \input{\tikzitpathsim/1/lem-3/odd/6.tikz} \\[\linesep]
			\eqq{\sf,\ho} \frac{1}{2}~ \input{\tikzitpathsim/1/lem-3/odd/7.tikz}
			\eqq{\cp,\sf} \frac{1}{2}~ \input{\tikzitpathsim/1/lem-3/odd/8.tikz}
			~=~ 1.
		\end{gather*}
	\end{itemize}
\end{proofE}

This leads to the following result regarding the variance of the gradients:

\begin{fact} \label{fact:sim1-variance}
	Let $h_P = |\{ H_j ~|~ H_j=P, j\in\{1,...,n\}\setminus\{i\} \}|$ be the number of times the Pauli $P\in\{X,Y,Z,I\}$ occurs in $H$, excluding the position $H_i$. Then
	\begin{gather*}
		\var{\frac{\partial\langle H\rangle}{\partial\theta_i^1}} 
		= \begin{cases}
			0 & \text{if } H_i = I \\
			\frac{1}{4^n} \cdot 2^{h_Z+h_Y} \cdot 4^{h_I} & \text{if } H_i = X,Y \\
			\frac{2}{4^n} \cdot 2^{h_Z+h_Y} \cdot 4^{h_I} & \text{if } H_i = Z \\
		\end{cases} 
		\\[\linesep]
		\var{\frac{\partial\langle H\rangle}{\partial\theta_i^2}} 
		= \begin{cases}
			0 & \text{if } H_i = Z,I \\
			\frac{1}{4^n} \cdot 2^{h_Z+h_Y} \cdot 4^{h_I} & \text{if } H_i = X,Y
			\end{cases}
	\end{gather*}
\end{fact}
\begin{proof}
	We start with the first equation where the gradient is w.r.t. $\theta_i^1$.
	By \Cref{thm:variance}, $\var{\frac{\partial\langle H\rangle}{\partial\theta_i^1}} $ is given by
	\[ (-1)^{\sum a_jb_j}~ \scalebox{0.7}{\input{\tikzitpathsim/1/fact:var/1.tikz}} \]
	Note that the different cycles are not connected with each other, which means we can arrange them as follows:
	\begin{adjustwidth}{-2cm}{-2cm}
		\[ (-1)^{\sum a_jb_j}~ \input{\tikzitpathsim/1/fact:var/2.tikz} \]
	\end{adjustwidth}
	By \Cref{lem:sim1-cycles}, we have
	\begin{equation*}
		A = \begin{cases}
			0 & \text{if } H_i=I \\
			1 & \text{if } H_i=Y,X \\
			2 & \text{if } H_i=Z
		\end{cases}
		\qquad\qquad
		\begin{aligned}[t]
		B &= \prod_{\substack{j=1\\j\neq i}}^n \begin{cases}
			1 & \text{if } H_j = X \\
			2 & \text{if } H_j = Z,Y \\
			4 & \text{if } H_j = I
		\end{cases} \\[\linesep]
		&= 1^{h_X} \cdot 2^{h_Z+h_X} \cdot 4^{h_I}
		\end{aligned}
	\end{equation*}
	such that $A \cdot B$ corresponds to desired equation.
	The proof for $\var{\frac{\partial\langle H\rangle}{\partial\theta_i^2}}$ is analogous.
\end{proof}

\begin{remark} \label{rem:var-zero}
	One might wonder why the variance is zero in some of the cases.
	If $H_i = I$, this corresponds to performing no measurement on qubit $i$.
	In this case, the value $\langle H\rangle$ actually does not depend on $\theta_i^1$ and $\theta_i^2$ since
	\[ \input{\tikzitpathsim/1/ex:zero-var/I-1.tikz} ~=~ \input{\tikzitpathsim/1/ex:zero-var/I-2.tikz} \eqq{\sf} \input{\tikzitpathsim/1/ex:zero-var/I-3.tikz}. \]
	Therefore, we have $\frac{\partial\langle H\rangle}{\partial\theta_i^1} = \frac{\partial\langle H\rangle}{\partial\theta_i^2} = 0$ and thus $\var{\frac{\partial\langle H\rangle}{\partial\theta_i^1}} = \var{\frac{\partial\langle H\rangle}{\partial\theta_i^2}} = 0$.
	Similarly, if $H_i=Z$ we have
	\[ \input{\tikzitpathsim/1/ex:zero-var/Z-1.tikz} ~=~ \input{\tikzitpathsim/1/ex:zero-var/Z-2.tikz} \eqq{\sf} \input{\tikzitpathsim/1/ex:zero-var/Z-3.tikz} \]
	such that $\theta_i^2$ does not contribute to the expectation value and thus $\var{\frac{\partial\langle H\rangle}{\partial\theta_i^2}} = 0$.
\end{remark}

Finally, \Cref{fact:sim1-variance} immediately yields the condition for $\Sim{1}$ to have a barren plateau:

\begin{theorem} \label{fact:sim1-barren}
	The barren plateau phenomenon appears in $\Sim{1}$ if we measure on $\Theta(n)$ qubits.
	In particular, this implies \Cref{hyp:sim-barren} for $\Sim{1}$.
\end{theorem}
\begin{proof}
	By \Cref{fact:sim1-variance}, the variance for all parameters (ignoring the scalar 2 in the case $\theta_i^1$) is either 0 or
	\[
	\frac{2^{h_Z+h_Y} \cdot 4^{h_I}}{4^n}
	= \frac{1}{2^{2n-h_Z-h_Y-2h_I}}
	= \frac{1}{2^{2h_X+h_Y+h_Z}}
	\]
	Since we measure on $\Theta(n)$ qubits, we must have $2h_X+h_Y+h_Z = \Theta(n)$ such that the variance vanishes exponentially.
\end{proof}

\subsection{Sim 2}

A single layer of $\Sim{2}$ can be represented in the ZX-calculus as
\[ \Sim{2}(\vec\theta) = \input{\tikzitpathsim/2/circ.tikz} \]
Using the same representation for a Hamiltonian $H$ as in \Cref{sec:sim1}, we can write the expectation value as
\begin{align*}
	\langle H \rangle
	~&=~ i^{\sum a_jb_j}~ \input{\tikzitpathsim/2/expval-1.tikz} \\[\linesep]
	\eqqa{\p,\sf} i^{\sum a_jb_j}~ \input{\tikzitpathsim/2/expval-2.tikz} \\[\linesep]
	\eqqa{\sf,\ho,\id} i^{\sum a_jb_j}~ \input{\tikzitpathsim/2/expval-3.tikz}
\end{align*}
Notice that this diagram has the same shape as the expectation value for $\Sim{1}$.
In fact, the only difference is the Hamiltonian in the middle, which in this case is given by
\[ H_i' = \left( \sum_{j=i}^n b_j \right)X \cdot \left(\sum_{j=1}^i a_j \right)Z. \]
In other words, $\langle H \rangle_{\Sim 2} = \langle H' \rangle_{\Sim 1}$.
Thus, we can use \Cref{fact:sim1-barren} to characterise the barren plateaus in $\Sim{2}$:

\begin{theorem} \label{fact:sim2-barren}
	The barren plateau phenomenon appears in $\Sim{2}$ if $H_i' \neq I$ at $\Theta(n)$ positions. In particular, this implies \Cref{hyp:sim-barren} for $\Sim{2}$.
\end{theorem}
\begin{proof}
	Follows from \Cref{fact:sim1-barren}.
	Note that for $H = Z^{\otimes n}$ we have $H' = Z\otimes I\otimes Z\otimes I ...$ such that the theorem applies and \Cref{hyp:sim-barren} is true.
\end{proof}

\subsection{Sim 9}

$\Sim{9}$ can be represented in the ZX-calculus as
\[ \Sim{9}(\vec\theta) = \sqrt{2}^{n-1}~ \input{\tikzitpathsim/9/circ.tikz} \]
yielding the expectation value
\begin{align*}
	\langle H\rangle
	~&=~ 2^{n-1}i^{\sum a_jb_j}~ \input{\tikzitpathsim/9/expval-1.tikz} \\[\linesep]
	\eqqa{\cc,\sf} \frac{i^{\sum a_jb_j}}{2}~ \input{\tikzitpathsim/9/expval-2.tikz} \\[\linesep]
	\eqqa{\sf,\cc} \frac{i^{\sum a_jb_j}}{2^{n+1}}~ \input{\tikzitpathsim/9/expval-3.tikz}
\end{align*}

As before, we consider how the cycles simplify:

\begin{lemmaE} \label{lem:sim9-cycles}
	We have
	\begin{gather*}
		(-1)^{a_ib_i}~ \input{\tikzitpathsim/9/lem-1/1.tikz} ~=~ \begin{cases}
			0 & \text{if } H_i=I,X \\[5pt]
			\input{\tikzitpathsim/9/lem-1/7.tikz} & \text{if } H_i=Y,Z
		\end{cases}
		\\[\linesep]
		(-1)^{a_ib_i}~ \input{\tikzitpathsim/9/lem-2/1.tikz} ~=~ \begin{cases}
			4~ \input{\tikzitpathsim/9/lem-2/even/7.tikz} & \text{if } H_i=I,X \\
			\\
			2~ \input{\tikzitpathsim/9/lem-2/odd/7.tikz} & \text{if } H_i=Y,Z
		\end{cases} 
	\end{gather*}
\end{lemmaE}
\begin{proofE}
	\def\tikzitpath{chapter5/figs/}
	We prove both cases separately:
	\begin{itemize}
		\item We have
		\begin{gather*}
			(-1)^{a_ib_i}~ \input{\tikzitpathsim/9/lem-1/1.tikz}
			\eqq{\cc} \frac{1}{2}(-1)^{a_ib_i}~ \input{\tikzitpathsim/9/lem-1/2.tikz} \\[\linesep]
			\eqq{\cc} \frac{1}{2}(-1)^{a_ib_i}~ \input{\tikzitpathsim/9/lem-1/3.tikz} 
			\eqq{\sf,\cp} \frac{1}{2}(-1)^{a_ib_i}(-1)^{b_i}~ \input{\tikzitpathsim/9/lem-1/4.tikz} \\[\linesep]
			\eqq{\ref{eqn:pi-cycle}} \frac{1}{2}(-1)^{a_ib_i+b_i}~ \input{\tikzitpathsim/9/lem-1/5.tikz} 
		\end{gather*}
		If $a_i = 0$, the diagram becomes 0. If $a_i=1$, we get
		\[ \input{\tikzitpathsim/9/lem-1/6.tikz} \eqq{\p,\sf} \input{\tikzitpathsim/9/lem-1/7.tikz} \]
		
		\item We have
		\begin{gather*}
			(-1)^{a_ib_i}~ \input{\tikzitpathsim/9/lem-2/1.tikz}
			\eqq{\cc,\hh} (-1)^{a_ib_i}~ \input{\tikzitpathsim/9/lem-2/2.tikz} \\[\linesep]
			\eqq{\sf,\cc,\ref{eqn:pi-cycle}} \sqrt{2}(-1)^{a_ib_i}~ \input{\tikzitpathsim/9/lem-2/3.tikz} \\[\linesep]
			\eqq{\cp,\cc,\sf} 2(-1)^{a_ib_i}~ \input{\tikzitpathsim/9/lem-2/3-1.tikz}
		\end{gather*}
		If $a_i = 0$, we get
		\begin{gather*}
			2~ \input{\tikzitpathsim/9/lem-2/even/4.tikz}
			\eqq{\tri,\sf,\cc} 2~ \input{\tikzitpathsim/9/lem-2/even/5.tikz} \\[\linesep]
			\eqq{\cp,\sf} 2~ \input{\tikzitpathsim/9/lem-2/even/6.tikz} 
			\eqq{\sf,\ho} 4~ \input{\tikzitpathsim/9/lem-2/even/7.tikz}
		\end{gather*}
		If $a_i = 1$, we get
		\begin{gather*}
			2(-1)^{b_i}~ \input{\tikzitpathsim/9/lem-2/odd/4.tikz}
			\eqq{\tri,\cp} 2~ \input{\tikzitpathsim/9/lem-2/odd/5.tikz} \\[\linesep]
			\eqq{\cc,\sf} \input{\tikzitpathsim/9/lem-2/odd/6.tikz} 
			\eqq{\ref{eqn:pi-connect}} 2 \input{\tikzitpathsim/9/lem-2/odd/7.tikz}
		\end{gather*}
	\end{itemize}
\end{proofE}

Unfortunately, this makes it difficult to give a closed-form expression of the gradient variance in terms of a general Hamiltonian $H$ as we have done in \Cref{fact:sim1-variance}.
However, we can easily investigate concrete instances.
For example we can verify Hypothesis \Cref{hyp:iqb-barren} for $\Sim{9}$:

\begin{theorem} \label{thm:sim9-barren}
	\Cref{hyp:sim-barren} holds for $\Sim{9}$, i.e. $\Sim{9}$ has barren plateaus for $H = Z^{\otimes n}$.
\end{theorem}
\begin{proof}
	By \Cref{thm:variance}, we have
	\[ \var{\frac{\partial\langle H\rangle}{\partial\theta_i}} ~=~ \frac{1}{2^{2n+2}}~ \input{\tikzitpathsim/9/var-1.tikz}  \]
	Using \Cref{lem:sim9-cycles}, we can simplify this to 
	\begin{gather*}
		\frac{2^{n-1}}{2^{2n+2}}~ \input{\tikzitpathsim/9/var-2.tikz}
		\eqq{\ref{eqn:pink-decompose}} \frac{2^{n-1}}{2^{3n+2}} \sum_{\vec x\in\{0,1\}^n} \input{\tikzitpathsim/9/var-3.tikz}
	\end{gather*}
	Each of those \enquote{lines} can only represent the scalars $0$, $\pm 1$, and $\pm 2$.
	Thus
	\[
	\var{\frac{\partial\langle H\rangle}{\partial\theta_i}}
	\leq \frac{2^{n-1}}{2^{3n+2}} \sum_{\vec x\in\{0,1\}^n} 16
	= \frac{2^{n-1}}{2^{3n+2}} \cdot 2^n \cdot 16
	= \frac{1}{2^{n - 1}}
	\]
	such that we have a barren plateau.
\end{proof}

\subsection{Single-Layer IQP Ansätze} \label{sec:IQP}

After discussing some of the Sim ansätze, we now move to IQPs.
In this section, we prove a general result that allows us to compute the gradient variance of any single-layer IQP circuit with single parameter occurrences.
To motivate our approach, we first look at an example IQP circuit:
\begin{equation} \label{eqn:IQP-single-layer-ex}
	U(\vec\theta) ~=~ \input{\tikzitpathIQP/single-layer/ex/circ.tikz}
\end{equation}
The corresponding diagram for the expectation value is given by
\begin{align*}
	\langle H\rangle 
	&=~ i^{\sum a_jb_j}~ \input{\tikzitpathIQP/single-layer/ex/expval-1.tikz} \\[\linesep]
	\eqqa{\cc,\hh} \frac{i^{\sum a_jb_j}}{2^4}~ \input{\tikzitpathIQP/single-layer/ex/expval-2.tikz} \\[\linesep]
	\eqqa{\p,\sf} \frac{i^{\sum a_jb_j}}{2^4}~ \input{\tikzitpathIQP/single-layer/ex/expval-3.tikz} \\[\linesep]
	\eqqa{\sf} \frac{i^{\sum a_jb_j}}{2^4}~ \input{\tikzitpathIQP/single-layer/ex/expval-4.tikz}
\end{align*}

The main insight is that the cycles for diagrams of this shape simplify nicely:

\begin{lemmaE} \label{lem:IQP-cycles}
	The cycles from single-layer IQP circuits simplify as follows:
	\begin{align*}
		\input{\tikzitpathIQP/single-layer/lem:cycle-1/statement-1.tikz} ~&=~ 0
		&\input{\tikzitpathIQP/single-layer/lem:cycle-1/statement-2.tikz} ~&=~ \input{\tikzitpathIQP/single-layer/lem:cycle-1/statement-3.tikz} \\[\linesep]
		\input{\tikzitpathIQP/single-layer/lem:cycle-2/statement-1.tikz} ~&=~ 	\input{\tikzitpathIQP/single-layer/lem:cycle-2/statement-3.tikz}
		&\input{\tikzitpathIQP/single-layer/lem:cycle-2/statement-2.tikz} ~&=~
		\input{\tikzitpathIQP/single-layer/lem:cycle-2/statement-4.tikz}
	\end{align*}
\end{lemmaE}
\begin{proofE}
	\def\tikzitpath{chapter5/figs/}
	\begin{gather*}
		\input{\tikzitpathIQP/single-layer/lem:cycle-1/proof-0.tikz}
		\eqq{\sf} \input{\tikzitpathIQP/single-layer/lem:cycle-1/proof-1.tikz}
		\eqq{\sc} \input{\tikzitpathIQP/single-layer/lem:cycle-1/proof-2.tikz}
		\eqq{\cp,\sf} \input{\tikzitpathIQP/single-layer/lem:cycle-1/proof-3.tikz} \\[\linesep]
		\eqq{\sf,\ho} \input{\tikzitpathIQP/single-layer/lem:cycle-1/proof-4.tikz}
		\eqq{\cp,\sf} \input{\tikzitpathIQP/single-layer/lem:cycle-1/proof-5.tikz}
		\eqq{\sf,\id} \input{\tikzitpathIQP/single-layer/lem:cycle-1/proof-6.tikz}
	\end{gather*}
	
	\begin{gather*}
		\input{\tikzitpathIQP/single-layer/lem:cycle-2/proof-1.tikz}
		\eqq{\sc} \input{\tikzitpathIQP/single-layer/lem:cycle-2/proof-2.tikz}
		\eqq{\cp,\sf} \input{\tikzitpathIQP/single-layer/lem:cycle-2/proof-3.tikz}
		\eqq{\sf\ho} \input{\tikzitpathIQP/single-layer/lem:cycle-2/proof-4.tikz}
	\end{gather*}
	If $k=0$, we get
	\begin{gather*}
		\input{\tikzitpathIQP/single-layer/lem:cycle-2/even/proof-5.tikz}
		\eqq{\tri,\sf} \input{\tikzitpathIQP/single-layer/lem:cycle-2/even/proof-6.tikz}
		\eqq{\sf,\ho} \input{\tikzitpathIQP/single-layer/lem:cycle-2/even/proof-7.tikz}
		\eqq{\cp,\sf} \input{\tikzitpathIQP/single-layer/lem:cycle-2/even/proof-8.tikz}
	\end{gather*}
	If $k=1$, we get
	\begin{gather*}
	\input{\tikzitpathIQP/single-layer/lem:cycle-2/odd/proof-5.tikz}
	\eqq{\tri,\cp,\sf} \input{\tikzitpathIQP/single-layer/lem:cycle-2/odd/proof-6.tikz}
	\eqq{\cp,\sf} \input{\tikzitpathIQP/single-layer/lem:cycle-2/odd/proof-7.tikz}
	\eqq{\sf,\id} \input{\tikzitpathIQP/single-layer/lem:cycle-2/odd/proof-8.tikz} \\
	\end{gather*}
\end{proofE}

To illustrate the application of \Cref{lem:IQP-cycles}, we show how to compute $\var{\frac{\partial\langle H\rangle}{\partial\theta_2}}$ for the example circuit (\ref{eqn:IQP-single-layer-ex}) for the Hamiltonian $H = X\otimes Z\otimes X\otimes Y$ which
corresponds to $a_1 = b_2 = a_3 = 0$ and $b_1 = a_2 = b_3 = b_4 = a_4 = 1$ yielding the expectation value
\begin{equation} \label{eqn:iqp-expval-ex´} 
	\langle H \rangle ~=~ \frac{i}{2^4}~ \input{\tikzitpathIQP/single-layer/ex/expval-5.tikz}
\end{equation}
Invoking \Cref{thm:variance} we get the following diagram for $\var{\frac{\partial\langle H\rangle}{\partial\theta_2}}$:
\begin{gather*}
	-\frac{1}{2^8}~ \input{\tikzitpathIQP/single-layer/ex/calc-1.tikz} 
	\eqq{\text{Lem. }\ref{lem:IQP-cycles}} -\frac{1}{2^8}~ \input{\tikzitpathIQP/single-layer/ex/calc-2.tikz} \\[\linesep]
	\eqq{\text{Lem. }\ref{lem:IQP-cycles}} -\frac{1}{2^8}~ \input{\tikzitpathIQP/single-layer/ex/calc-3.tikz}
	\eqq{\text{Lem. }\ref{lem:IQP-cycles}} -\frac{1}{2^8}~ \input{\tikzitpathIQP/single-layer/ex/calc-4.tikz}
	\eqq{\cp,\sf} -\frac{1}{2^8}~ \input{\tikzitpathIQP/single-layer/ex/calc-5.tikz} \\[\linesep]
	\eqq{\cp,\sf} \frac{1}{2^8}~ \input{\tikzitpathIQP/single-layer/ex/calc-6.tikz}
	~=~ \frac{1}{4}
\end{gather*}

\subsubsection{Generalising to arbitrary IQPs}

Following the technique from the example circuit, we can compute the variance for arbitrary single-layer IQP circuits.
In the general case, our ansatz $U(\vec\theta)$ consists of $m$ phase gadgets, given by exponentials of Pauli strings $P_1,...,P_m \in \{I,Z\}^{\otimes n}$:
\[ U(\vec\theta) ~=~ \input{\tikzitpathIQP/single-layer/circ.tikz} \]
Furthermore fix a Hamiltonian $H$ and define
\[ k_i = \sum_{\substack{1 \leq j \leq n \\ P_i^j = Z}} a_j \]
such that
\begin{align*}
	\langle H\rangle
	~&=~ i^{\sum a_jb_j}~ \input{\tikzitpathIQP/single-layer/expval-1.tikz} \\[\linesep]
	\eqqa{\cc,\hh} \frac{i^{\sum a_jb_j}}{2^n}~ \input{\tikzitpathIQP/single-layer/expval-2.tikz} \\[\linesep]
	\eqqa{\p,\sf} \frac{i^{\sum a_jb_j}}{2^n}~ \input{\tikzitpathIQP/single-layer/expval-3.tikz} \\[\linesep]
	%\eqqa{\ref{}} \frac{i^{\sum a_jb_j}}{2^n}~ \tikzfig{IQP/single-layer/expval-4} \\[\linesep]
	\eqqa{\sf} \frac{i^{\sum a_jb_j}}{2^n}~ \input{\tikzitpathIQP/single-layer/expval-6.tikz}
\end{align*}

Compare this with (\ref{eqn:iqp-expval-ex´}):
The green spiders in the middle represent one qubit each.
Furthermore, we get pink spiders on the left and right side for each parameter $\theta_j$.
Those pink spiders are connected to all the qubits where the gadget associated with $\theta_j$ has legs.

This leads to the following characterisation of the variance:

\begin{theorem} \label{thm:IQP-variance}
	If $k_i=0$, then $\var{\frac{\partial\langle H\rangle}{\partial\theta_i}} = 0$.
	If $k_i=1$, then
	\[ \var{\frac{\partial\langle H\rangle}{\partial\theta_i}} ~=~ \frac{(-1)^{\sum_{j=1}^n a_jb_j}}{4^n}~ \input{\tikzitpathIQP/single-layer/thm:variance/statement.tikz} \]
\end{theorem}
\begin{proof}
	Follows by simplifying the cycles in the variance diagram according to \Cref{lem:IQP-cycles}.
\end{proof}

\begin{remark} \label{rem:iqp-same-var}
	A remarkable consequence of \Cref{thm:IQP-variance} is that the variance for every parameter is either zero, or the same as all other parameters with non-zero variance.
	Thus, when determining whether an ansatz exhibits the barren plateau phenomenon, it suffices to look at a single parameter whose gradient has non-zero variance.
\end{remark}

Using \Cref{thm:IQP-variance}, we can analyse the IQP ansätze for barren plateaus:

\begin{theorem} \label{fact:iqp123-barren}
	\Cref{hyp:iqb-barren} is true, i.e. we get the following results for single-layer IQPs:
	\begin{itemize}
		\item 
		$\IQP{1}$ does not have barren plateaus.
		
		\item
		$\IQP{2}$ has barren plateaus for $H = (Y \otimes X)^{\otimes n/2}$.
		%However, there are other Hamiltonians like $H = (Y\otimes X)^{\otimes c} \otimes Z^{\otimes n-2c}$ for constant $c$ for which the first $c$ parameters have a non-vanishing variance.
		\item
		$\IQP{3}$ has barren plateaus for $H = (Y \otimes X)^{\otimes n/2}$.
	\end{itemize}
\end{theorem}
\begin{proof} $ $
	\begin{itemize}
		\item $\IQP{1}$:
		If $k = \sum a_j = 0$, we get $\var{\frac{\partial\langle H\rangle}{\partial\theta}} = 0$.
		If $k = 1$, \Cref{thm:IQP-variance} gives us
		\begin{gather*}
			\var{\frac{\partial\langle H\rangle}{\partial\theta}}
			~=~ \frac{(-1)^{\sum a_jb_j}}{4^n}~ \input{\tikzitpathIQP/IQP1/var-1.tikz} \\[\linesep]
			\eqq{\cp,\sf} \frac{(-1)^{\sum a_jb_j}}{4^n}(-1)^{b_1}~ \input{\tikzitpathIQP/IQP1/var-2.tikz}
		\end{gather*}
		This is either 0, or $\frac{1}{2}$.
		Thus, the variance does not vanish exponentially.
		
		\item $\IQP{2}$:
		For $\IQP{2}$, we have $k_1 = a_1+a_2$, $k_2=a_3+a_4$, ..., $k_{n/2} = a_{n-1}+a_n$.
		By \Cref{thm:IQP-variance} we get $\var{\frac{\partial\langle H\rangle}{\partial\theta_i}} = 0$ if $k_i$ is even. Otherwise
		\[ \var{\frac{\partial\langle H\rangle}{\partial\theta_i}} ~=~ \frac{(-1)^{\sum a_jb_j}}{4^n}~ \input{\tikzitpathIQP/IQP2/var-1.tikz} \]
		where the pink spiders only exist if the annotated condition is met.
		Concretely, for $H = (Y \otimes X)^{\otimes n/2}$ we have $a_1=a_3=...=a_{n-1} = 1$, $a_2=a_4=...=a_n = 0$, $b_j=1$ for all $j$ and hence
		\begin{gather*}
		\var{\frac{\partial\langle H\rangle}{\partial\theta_i}}
		~=~ \frac{(-1)^{n/2}}{4^n} \left( \input{\tikzitpathIQP/IQP2/var-2.tikz} \right)^{n/2}
		\eqq{\cp} \frac{(-1)^{n/2}}{4^n} \left( -~ \input{\tikzitpathIQP/IQP2/var-3.tikz} \right)^{n/2} \\
		= \frac{8^{n/2}}{4^n}
		= \frac{1}{2^{n/2}}.
		\end{gather*}
		Note that this exactly matches with the numerical data from \Cref{fig:iqp-single-layer}.
		We conclude that we have a barren plateau.
%		Concretely, for $H = (Y\otimes X)^{\otimes c} \otimes Z^{\otimes n-2c}$ we get
%		\begin{align*}
%			\var{\frac{\partial\langle H\rangle}{\partial\theta_1}}
%			~&=~ \frac{(-1)^c}{4^n} \left( \tikzfig{IQP/IQP2/var-2} \right)^c \cdot \left( \tikzfig{IQP/IQP2/var-4} \right)^{n/2-c} \\
%			\eqqa{\cp} \frac{(-1)^c}{4^n} \left(-~ \tikzfig{IQP/IQP2/var-3} \right)^c \cdot \left( \tikzfig{IQP/IQP2/var-4} \right)^{n/2-c} \\
%			&=~ \frac{8^c \cdot 16^{n/2-c}}{4^n}
%			= \frac{1}{2^c}.
%		\end{align*}
%		However, we get $\var{\frac{\partial\langle H\rangle}{\partial\theta_i}} = 0$ for all $i > c$.

		\item $\IQP{3}$:
		Again, \Cref{thm:IQP-variance} yields
		\begin{gather*}
			\var{\frac{\partial\langle H\rangle}{\partial\theta_i}}
			~=~ \frac{(-1)^{\lfloor n/2\rfloor}}{4^n}~ \input{\tikzitpathIQP/IQP3/var-1.tikz}
			\eqq{\cp,\sf} -\frac{(-1)^{\lfloor n/2\rfloor}}{4^n}~ \input{\tikzitpathIQP/IQP3/var-2.tikz} \\[\linesep]
			\eqq{\cp,\sf} -\frac{(-1)^{\lfloor n/2\rfloor}}{4^n}~ \input{\tikzitpathIQP/IQP3/var-3.tikz} 
			~=~ ...
			~=~ \frac{1}{4^n}~ \input{\tikzitpathIQP/IQP3/var-4.tikz}
		\end{gather*}
		Thus, if $n$ is odd we get variance 0.
		If $n$ is even we get variance $\frac{2^n}{4^n} = \frac{1}{2^n}$.
		This exactly matches with the numerical data from \Cref{fig:iqp-single-layer}.
		Thus, we have a barren plateau. \qedhere
	\end{itemize}
\end{proof}

\subsection{Dealing with multiple parameter occurrences} \label{sec:barren-multiple-parameters}

The biggest limitation to the current ZX based analysis of barren plateaus is the fact that \Cref{thm:variance} only applies if each parameter occurs once in the diagram.
However, many circuits of interest (for example $\IQP{4}$) require multiple spiders with the same parameter.
Ideally, one would want alternate versions of \Cref{thm:variance} that support all possible combinations of parameter occurences.
This would require extending and generalising the integration results by Wang an Yeung~\cite{wang2022differentiating}.
While this is principally possible using the summing technique from~\cite{shaikh2022sum}, the main challenge is finding a representation that is amenable to rewriting and offering a way to break up  cycles.
%This has proven to be difficult and still 

In this section, we describe a trick that can sometimes be used instead to compute variances using \Cref{thm:variance}, even if parameters occur multiple times.
The idea is that in some special cases, one can choose a Hamiltonian for which the extra parameter occurrences cancel out.
We demonstrate this using the $\IQP{4}$ ansatz with the Hamiltonian $H = Z^{\otimes n}$.
In this case, we can rewrite the expectation value as follows:
\begin{gather*}
	\langle H\rangle
	~=~ \input{\tikzitpathIQP/IQP4/expval-1.tikz} \\[\linesep]
	\eqq{\cc} \input{\tikzitpathIQP/IQP4/expval-2.tikz} \\[\linesep]
	\eqq{\sf,\p} \input{\tikzitpathIQP/IQP4/expval-3.tikz}
	\eqq{\text{Lem. }\ref{eqn:pauli-exp-fuse}} \input{\tikzitpathIQP/IQP4/expval-4.tikz} \\[\linesep]
	\eqq{\cp,\sf} \input{\tikzitpathIQP/IQP4/expval-5.tikz} 
	~=~ \input{\tikzitpathIQP/IQP4/expval-6.tikz}
\end{gather*}
Thus, we got rid of all two-legged phase gadgets.
This is now amenable for barren plateau analysis using \Cref{thm:variance} and \Cref{lem:IQP-cycles}:

\begin{theorem} \label{thm:iqp4-barren}
	The barren plateau phenomenon appears in $\IQP{4}$ for $H = Z^{\otimes n}$.
\end{theorem}
\begin{proof}
	By \Cref{thm:variance} and \Cref{lem:IQP-cycles}, we have
	\[
		\var{\frac{\partial\langle H\rangle}{\partial\theta_i}}
		= \frac{1}{4^{n-1}}~ \input{\tikzitpathIQP/IQP4/var-1.tikz}
		\eqq{\cp,\sf} \frac{1}{4^{n-1}}~ \input{\tikzitpathIQP/IQP4/var-2.tikz}
		~= \frac{2^{n-1}}{4^{n-1}}
		= \frac{1}{2^{n-1}}. \qedhere
	\]
\end{proof}

\subsection{Commuting Multi-Layer IQP Ansätze}

So far, we only studied single-layered circuits.
In this section we analyse a special case where the multi-layers analysis of IQPs is straightforward.
Note that since the layers are separated by Hadamards, we can view multi-layer IQPs as alternating layers of $Z$- and $X$-Pauli exponentials.
For example, consider $\IQP{2}$ for an even number of layers:
\begin{align*}
	\IQP{2}(\vec\theta)
	~&=~ \input{\tikzitpathIQP/IQP6/ansatz-small.tikz} \\[\linesep]
	~&=~ \input{\tikzitpathIQP/IQP6/ansatz-boxes.tikz}
\end{align*}
Recalling the commutation properties of Pauli boxes (see \Cref{lem:pauli-exp-comm}), we see that $X$- and $Z$-layers commute with each other for this ansatz.
In this special case, computing the variance is actually not difficult since we can fuse all odd and even layers together via \Cref{lem:pauli-exp-fuse}:
\[ \IQP{2}(\vec\theta) ~=~ \input{\tikzitpathIQP/IQP6/ansatz-fuse.tikz} \]
As a result, an $\ell$-layer $\IQP{2}$ circuit with even $\ell$\footnote{We focus on the case where $\ell$ is even. For odd $\ell$, the derivations are orthogonal, noting that the fused second layer will not have Hadamards at the end.} is equivalent to a $2$-layer $\IQP{2}$ circuit.
%In fact, this generalises to all multi-layer IQP ansätze where the $Z$- and $X$-layers commute.
The barren plateau analysis in this case is straightforward:

\begin{theorem} \label{thm:iqp2-barren-multilayer}
	The barren plateau phenomenon appears in $\IQP{2}$ for $H = (Z\otimes Y)^{\otimes n/2}$ for any number of layers.
\end{theorem}
\begin{proof}
	As discussed before, it suffices to consider the case $\ell = 2$.
	The expectation value is given by
	\begin{gather*}
		\langle H\rangle
		~=~ \input{\tikzitpathIQP/IQP6/expval-1.tikz} \\[\linesep]
		\eqq{\cc,\p,\sf} \frac{1}{4^n}~ \input{\tikzitpathIQP/IQP6/expval-2.tikz} \\[\linesep]
		\eqq{\ref{eqn:pauli-exp-fuse}} \frac{1}{4^n}~ \input{\tikzitpathIQP/IQP6/expval-3.tikz}
		\eqq{\cp,\sf,\id,\hh} \frac{1}{4^n}~ \input{\tikzitpathIQP/IQP6/expval-4.tikz}
	\end{gather*}
	This allows us to use \Cref{lem:IQP-cycles} to remove the cycles showing up during the variance calculation via \Cref{thm:IQP-variance}.
	Concretely, we get
	\[ 
		\var{\frac{\partial\langle H\rangle}{\partial\theta_i^1}}
		= \frac{(-1)^{n/2}}{4^n}~ \input{\tikzitpathIQP/IQP6/var-1.tikz}
		\eqq{\cp,\sf} \frac{1}{4^n}~ \input{\tikzitpathIQP/IQP6/var-2.tikz}
		~= \frac{8^{n/2}}{4^n}
		= \frac{1}{2^{n/2}}.
	\]
	Thus, we have a barren plateau.
\end{proof}

This result is not surprising since we showed that the single-layer version of $\IQP{3}$ already has barren plateaus (see \Cref{fact:iqp123-barren}).
The more interesting question is how $\IQP{1}$ behaves for multiple layers since it does not have barren plateaus for a single layer.
Unfortunately, the layers of $\IQP{1}$ only commute for an even number of qubits.
Hence, the technique discussed in this section is not applicable if $n$ is odd.
In that case, the necessary calculations become significantly more involved, which we explore in the next section.

\subsection{Non-Commuting Multi-Layer IQP Ansätze}

Apart from specifically designed examples like $\IQP{2}$, it is uncommon that IQP layers fully commute.
For example, the QNLP ansätze $\IQP{3}$ and $\IQP{4}$ do not form commuting layers in a multi-layer configuration.
In that case, the variance computation becomes significantly more difficult.
Since we have already shown that the QNLP ansätze have barren plateaus even for a single layer, we will not consider them in this section.
Instead, we focus on $\IQP{1}$ to demonstrate our variance computation technique for non-commuting layers.
$\IQP{1}$'s layers do not commute for odd $n$ and its single-layer version does not have barren plateaus which makes it an interesting case to study.

%\subsubsection{Step 1: Drawing the Expectation Value}

To make the diagrams more concise, we introduce the following notation to denote layers of $\IQP{1}$:
\[ \input{\tikzitpathIQP/IQP5/layer-def.tikz} \]

The diagram for the expectation value is then given by
\begin{align*}
	\langle H \rangle
	&=~ i^{\sum a_jb_j} ~ \input{\tikzitpathIQP/IQP5/expval-1.tikz} \\[\linesep]
	\eqqa{\cp,\sf,\cc} i^{\sum a_jb_j} \cdot (-1)^d ~ \input{\tikzitpathIQP/IQP5/expval-2.tikz}
\end{align*}
where
\begin{displaymath}
	\input{\tikzitpathIQP/IQP5/layer-prime-def.tikz}
	\qquad\qquad
	k_i = \begin{cases}
		a_1+...+a_n & \text{if $i$ has same parity as $\ell$} \\
		b_1+...+b_n & \text{otherwise}
	\end{cases}
\end{displaymath}
\begin{displaymath}
	c_i = \begin{cases}
	a_i & \text{if $\ell$ is even} \\
	b_i & \text{if $\ell$ is odd}
	\end{cases}
	\qquad\qquad
	d = \begin{cases}
		0 & \text{if $\ell$ is even} \\
		\sum_{i=1}^n a_ib_i & \text{if $\ell$ is odd} 
	\end{cases}
\end{displaymath}

Similar to the calculations we did before, we have pushed the Hamiltonian through the layers on the right-hand side, occasionally adding a phase of $\pi$ to the phase gadgets.
The factor $(-1)^d$ is introduced because if $\ell$ is odd, we get the following situation after pushing the Hamiltonian through:
\[ \input{\tikzitpathIQP/IQP5/expval-expl-1.tikz} \eqq{\cp} (-1)^{\sum a_jb_j} ~\input{\tikzitpathIQP/IQP5/expval-expl-2.tikz} ~=~ (-1)^d ~ \input{\tikzitpathIQP/IQP5/expval-expl-3.tikz} \]
However, the factor $(-1)^d$ does not really matter since it cancels out when computing the variance.
In order to draw the variance diagram, we add a wire coming out of each layer that replaces the parametrised spider:
\[ \input{\tikzitpathIQP/IQP5/layer-wire-def.tikz} \qquad\qquad \input{\tikzitpathIQP/IQP5/layer-prime-wire-def.tikz} \]
Note that in the following variance diagram, we only explicitly draw the cycle connecting $L_\ell$ and $L_\ell'$.
We only hint at remaining cycles using dots:
\begin{align*}
	\text{Var}\left(\frac{\partial \langle H\rangle}{\partial \theta_i}\right)
	~&=~ (-1)^{\sum a_jb_j}~ \input{\tikzitpathIQP/IQP5/var-0.tikz} \\[\linesep]
	\eqqa{\sf} (-1)^{\sum a_jb_j}~ \input{\tikzitpathIQP/IQP5/var-0-1.tikz}
\end{align*}
We can cut this cycle using our existing simplification strategy from \Cref{lem:IQP-cycles}.
Concretely, if $k_\ell = 0$, we get
\[ (-1)^{\sum a_jb_j}~ \input{\tikzitpathIQP/IQP5/var-1.tikz} \]
In this case, $L_{\ell-1}$ and $L_{\ell-1}'$ are now directly next to each other and we can continue the same argument recursively.

However, if $k_\ell = 1$ is odd, we get a red $\pi$-spider:
\[ (-1)^{\sum a_jb_j}~ \input{\tikzitpathIQP/IQP5/var-2.tikz} \]
Assuming that $n$ is odd, commuting $L_{\ell-1}$ past this will add some extra Hadamard wires according to \Cref{lem:pauli-exp-comm}:
\begin{gather*}
	\frac{(-1)^{\sum a_jb_j}}{2}~ \input{\tikzitpathIQP/IQP5/var-3.tikz} \\[\linesep]
	\eqq{\sf} \frac{(-1)^{\sum a_jb_j}}{2}~ \input{\tikzitpathIQP/IQP5/var-3-1.tikz}
\end{gather*}
Note that the two new Hadamard wires connected to the red spiders in $L_{\ell-1}'$ come with a scalar of $\frac{1}{\sqrt 2}$ each.
In order to proceed from here, we need a new cycle cutting lemma that applies when the right side is connected to shared pink spider(s):

\enlargethispage{\baselineskip}
\begin{lemmaE} \label{lem:IQP-cycle-multi}
	We have
	\begin{gather*}
		\input{\tikzitpathIQP/IQP5/cycle-1/1.tikz} ~=~ \sqrt 2^m ~ \input{\tikzitpathIQP/IQP5/cycle-1/6.tikz} \\[\linesep]
		\input{\tikzitpathIQP/IQP5/cycle-2/1.tikz} ~=~ \sqrt 2^m ~ \input{\tikzitpathIQP/IQP5/cycle-2/6.tikz}
	\end{gather*}
\end{lemmaE}
\begin{proofE}
	\def\tikzitpath{chapter5/figs/}
	We have
	\begin{gather*}
		\input{\tikzitpathIQP/IQP5/cycle-1/1.tikz}
		\eqq{\ref{eq:pi-copy}} \frac{1}{\sqrt 2^m}~ \input{\tikzitpathIQP/IQP5/cycle-1/2.tikz} \\[\linesep]
		\eqq{\ref{eqn:hopf-had}} \sqrt 2^m ~ \input{\tikzitpathIQP/IQP5/cycle-1/3.tikz} 
		\eqq{\sc} \sqrt 2^m ~ \input{\tikzitpathIQP/IQP5/cycle-1/4.tikz} \\[\linesep]
		\eqq{\sc} \sqrt 2^m ~ \input{\tikzitpathIQP/IQP5/cycle-1/5.tikz}
		\eqq{\sf,\ho,\id} \sqrt 2^m ~ \input{\tikzitpathIQP/IQP5/cycle-1/6.tikz}
	\end{gather*}
	For the second cycle, we have
	\begin{gather*}
		\input{\tikzitpathIQP/IQP5/cycle-2/1.tikz}
		\eqq{\ref{eq:pi-copy}} \frac{1}{\sqrt 2^m}~ \input{\tikzitpathIQP/IQP5/cycle-2/2.tikz} \\[\linesep]
		\eqq{\ref{eqn:hopf-had}} \sqrt 2^m ~ \input{\tikzitpathIQP/IQP5/cycle-2/3.tikz} 
		\eqq{\sc} \sqrt 2^m ~ \input{\tikzitpathIQP/IQP5/cycle-2/4.tikz} \\[\linesep]
		\eqq{\sf,\ho} \sqrt 2^m ~ \input{\tikzitpathIQP/IQP5/cycle-2/5.tikz}
		\eqq{\p,\sf} \sqrt 2^m ~ \input{\tikzitpathIQP/IQP5/cycle-2/6.tikz} \\
	\end{gather*}
\end{proofE}

\begin{figure}[t]
	\begin{adjustwidth}{-2cm}{-2cm}
		\begin{center}
			$\frac{(-1)^{\sum a_jb_j}}{\sqrt 2^h}~ \input{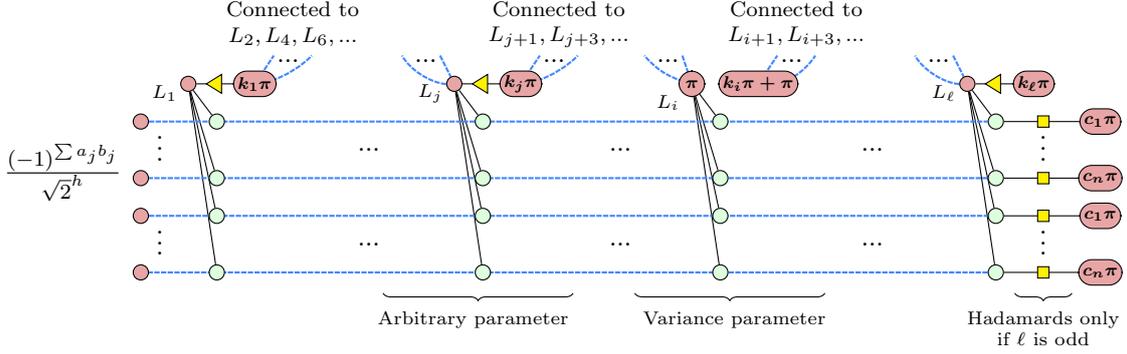}$
		\end{center}
	\end{adjustwidth}
	\caption{Diagram for $\text{Var}\left(\frac{\partial \langle H\rangle}{\partial \theta_i}\right)$ where $h$ is the number of Hadamard wires connecting the red spiders.} \label{fig:IQP-var-multi}
\end{figure}

Note that in the special case $m = 0$, \Cref{lem:IQP-cycle-multi} exactly corresponds to \Cref{lem:IQP-cycles}.
We can now use \Cref{lem:IQP-cycle-multi} to simplify the cycle in the variance computation above, also replacing the previous application of \Cref{lem:IQP-cycles} with the more general \Cref{lem:IQP-cycle-multi}:
\begin{adjustwidth}{-2cm}{-2cm}
\begin{gather*}
	\frac{(-1)^{\sum a_jb_j}}{\sqrt 2}~ \input{\tikzitpathIQP/IQP5/var-4.tikz} \\[\linesep]
	=~ \frac{(-1)^{\sum a_jb_j}}{\sqrt 2^2}~ \input{\tikzitpathIQP/IQP5/var-5.tikz} \\[\linesep]
	=~ \frac{(-1)^{\sum a_jb_j}}{\sqrt 2^4}~ \input{\tikzitpathIQP/IQP5/var-6.tikz}
\end{gather*}
\end{adjustwidth}
Iterating this process for all layers yields the diagram shown in \Cref{fig:IQP-var-multi} which we evaluate using a recursive strategy.
However, we only sketch the proof here, fixing $i = 1$ and skipping over some details.
We refer to \Cref{chap:appendix-contract} for the full derivation.

One can show that $\text{Var}\left(\frac{\partial \langle H\rangle}{\partial \theta_1}\right)$ is only non-zero if $c_1 = c_2 = ... = c_n$.
Furthermore, it turns out that the value of the $c$'s only effects the sign of the scalar represented by the diagram.
Therefore we will ignore them here.
Also note that by definition we have $k_i = k_j$ if $i$ and $j$ have the same parity.
Thus, we define the following shorthands:
\[ k_o := k_1=k_3=k_5=... \qquad\qquad k_e := k_2=k_4=k_6=... \]
This means that the diagram in \Cref{fig:IQP-var-multi} only depends on the numbers $k_e$, $k_o$, and $\ell$.
We can show that it satisfies the following recurrence relation.

\begin{lemma}
	Let $V_\ell(k_e,k_o,c)$ denote the diagram in \Cref{fig:IQP-var-multi} and let $\ell>1$ be odd. Then
	\begin{align*}
		V_1(k_e,0) &= 0 &
		V_1(k_e,1) &= \frac{1}{2} \\
		V_\ell(0,0) &= 0 &
		V_\ell(0,1) &= \frac{1}{4} (3V_{\ell-2}(0,1) - V_{\ell-2}(1,0)) \\
		V_\ell(1,1) &= V_{\ell}(0,1) &
		V_\ell(1,0) &= \frac{1}{2} (V_{\ell-2}(1,0) - V_{\ell-2}(0, 1))
	\end{align*}
\end{lemma}
\begin{proof}
	See equations (\ref{eqn:VO-00}), (\ref{eqn:VO-11=VO-01}), (\ref{eqn:VO-01'}), and (\ref{eqn:VO-10'}) in \Cref{chap:appendix-contract}.
\end{proof}

Similarly, we can obtain recursive equations for even $\ell$.
This yields a recursive algorithm for computing $\text{Var}\left(\frac{\partial \langle H\rangle}{\partial \theta_1}\right)$.
However, it is also possible to derive a closed-form solution:

\begin{lemma} \label{lem:IQP-closed}
	For odd $\ell$ we have
	\[ 
	V_{\ell}(1,1) = V_{\ell}(0,1) = \frac{2 \cdot 4^{\lfloor \ell/2\rfloor} + 1}{6 \cdot 4^{\lfloor \ell/2 \rfloor}}
	\qquad\qquad 
	V_{\ell}(1,0) = \frac{4^{\lfloor \ell/2\rfloor}-1}{3 \cdot 4^{\lfloor \ell/2\rfloor}}.
	\]
	For even $\ell$ we have
	\[
	V_\ell(1,1) = V_\ell(1,0) = \frac{2^\ell - 1}{3 \cdot 2^\ell}
	\qquad\qquad
	V_\ell(0,1) = \frac{2 \cdot 4^{\ell/2 - 1} + 1}{6 \cdot 4^{\ell/2 - 1}}
	\]
\end{lemma}
\begin{proof}
	See \Cref{lem:IQP-recurrence-closed} and \Cref{corr:lem:IQP-recurrence-closed'} in \Cref{chap:appendix-contract}.
\end{proof}

As a result, we get the following formula for the variance:

\bigskip
\begin{theorem}
	\[
	\text{Var}\left(\frac{\partial \langle H\rangle}{\partial \theta_1}\right)
	= \begin{cases}
	V_{\ell}(\sum a_j,\sum b_j) & \text{if $\ell$ is even and } a_1=...=a_n \\
	V_{\ell}(\sum b_j,\sum a_j) & \text{if $\ell$ is odd and } b_1=...=b_n \\
	0 & \text{otherwise.}
	\end{cases}
	\]
\end{theorem}
\begin{proof}
	See \Cref{thm:IQP-var-formula} in \Cref{chap:appendix-contract}.
\end{proof}

\begin{corollary} \label{corr:iqp1-barren}
	Either $\text{Var}\left(\frac{\partial \langle H\rangle}{\partial \theta_1}\right) = 0$ or $\text{Var}\left(\frac{\partial \langle H\rangle}{\partial \theta_1}\right) \to \frac{1}{3}$ for $\ell \to \infty$.
	In particular, this proves \Cref{hyp:iqb1-barren}.
\end{corollary}
\begin{proof}
	This follows from the fact that all terms in \Cref{lem:IQP-closed} converge to $\frac{1}{3}$ for $\ell \to \infty$.
	See \Cref{corr:IQP-var-converge} in \Cref{chap:appendix-contract} for the full details.
\end{proof}

\begin{corollary}
	For example, in the case $H = Z^{\otimes n}$ we have $a_j = 1$ and $b_j = 0$ for all $j$ such that
	\[  
	\text{Var}\left(\frac{\partial \langle H\rangle}{\partial \theta_1}\right)
	= \begin{cases}
		\frac{2^\ell - 1}{3 \cdot 2^\ell} & \text{if $\ell$ is even.} \\[10pt]
		\frac{2 \cdot 4^{\lfloor \ell/2\rfloor} + 1}{6 \cdot 4^{\lfloor \ell/2 \rfloor}} & \text{if $\ell$ is odd}
	\end{cases}
	 \]
\end{corollary}
Note that this exactly matches the numerical values from \Cref{fig:iqp1-multi-layer-converge}.

\section{Barren Plateau Mitigation Techniques} \label{sec:barren-mitigation}

After identifying barren plateaus in a variety of ansätze, we want to close this chapter with a brief discussion of how to avoid them.
Crucially, the barren plateau analysis in this chapter relied on the assumption made in \Cref{sec:barren-backgound} that the parameters $\theta_i$ are uniformly and independently sampled from $[-\pi,\pi]$.
This means that our results, as well the ones in the literature like McLean et al.~\cite{mcclean2018barren}, no longer apply if one chooses a different parameter initialisation.

To this end, different initialisation strategies have been proposed with the goal of avoiding barren plateaus:
For example, Grant et al.~\cite{grant2019initialization} choose parameters such that the circuit turns into a sequence of shallow blocks that each evaluate to the identity, thus reducing the effective circuit depth.
Kulshrestha and Safro~\cite{kulshrestha2022beinit} experimentally show that initialisation with the Beta distribution reduces the prevalence of barren plateaus.

Apart from initialisation strategies, various other techniques have been proposed:
Rad et al.~\cite{rad2022surviving} use Bayesian learning to find a promising regions in the parameter space which are then explored using local optimisers.
Sack et al.~\cite{sack2022avoiding} introduce a new learning scheme that adapts the learning rate when a barren plateau is detected.
Liu et al.~\cite{liu2022mitigating} propose a novel ansatz family where barren plateaus can be mitigated and Skolik et al.~\cite{skolik2021layerwise} use quantum circuit learning to find ansätze that avoid barren plateaus.
Patti et al.~\cite{patti2021entanglement} discuss a variety of techniques including the addition of noise, reducing entanglement and partitioning the qubit registers depending on the cost function.

\chapter{Discussion} \label{chap:discussion}

\section{Summary of Results}

\subsubsection{Gradient Recipes}

We have refined the diagrammatic differentiation technique by Wang and Yeung \cite{wang2022differentiating} for the special case of parametrised quantum circuits and used it to give diagrammatic proofs of parameter shift rules given by Schuld et al. \cite{schuld2019evaluating} and Anselmetti et al. \cite{anselmetti2021local}.
Furthermore, we derived a novel $2n$-term shift rule for gates that can be represented with $n$ parametrised spiders.
We also discussed the optimality of shift rules, proving an open conjecture by Anselmetti et al.~\cite{anselmetti2021local} by deriving a no-go theorem ruling out shift rules with less than four terms for all gates whose Hermitian generators have eigenvalues of shape $-\lambda,0,\lambda$.

\subsubsection{Barren Plateaus}

We investigated both empirical and formal methods to detect barren plateaus in ansätze using the variance computation framework laid out by Wang and Yeung~\cite{wang2022differentiating}.
For the empirical analysis, we developed a tool that automatically computes $\text{Var}\left({\frac{\partial \langle H\rangle}{\partial \theta_i}}\right)$ which can be used to diagnose barren plateaus without the user having to perform any calculations or mathematical reasoning.
Using this tool, we investigate several ansätze studied by Sim et al. \cite{sim2019expressibility} and empirically concluded that even at a single layer they likely all have barren plateaus.

To showcase the analytical barren plateau analysis powered by ZX, we formally proved this claim for three of the Sim ansätze.
Furthermore, we analysed a range of IQP ansätze, in particular showing that a single layer of the  ansatz used by the quantum natural language processing library \texttt{lambeq}~\cite{kartsaklis2021lambeq} has barren plateaus when measuring in the computational basis.
Additionally, we proved that one of the IQP ansätze does not have barren plateaus, with $\text{Var}\left({\frac{\partial \langle H\rangle}{\partial \theta_i}}\right)$ converging to a constant independent of the number of qubits $n$ when the number of layers $\ell$ goes to infinity.

\section{Discussion and Future Work}

\subsubsection{Gradient Recipes}

One of the initial motivations for using the ZX calculus to study gradient recipes was the hope that the graphical representation of derivatives might make it easier to discover new recipes that go beyond parameter shift rules.
However, as we have mentioned in \Cref{sec:ancilla-recipe}, it proved to be harder than expected to find decompositions of the differentiation gadget that actually yield unitaries when applied to gates.

However, the diagrammatic approach proved to be very fruitful for the analysis of parameter shift rules.
Originally, Schuld et al. \cite{schuld2019evaluating} and Anselmetti et al. \cite{anselmetti2021local} arrived at their shift rules by observing that the Hermitian generators for the gates they consider satisfy $H^2 = I$ and $H^3 = H$, respectively.
From this, they derived systems of equations that yielded the shift rules.
While our approach also involved systems of equations, we arrived at and solved them in a completely different way.
Concretely, we used a diagrammatic approach to find systems of equations that characterise valid shift rules which turned out to be easily solvable using a discrete sine transform.
The benefit of this hybrid approach involving both graphical and algebraic techniques is that it applies to a wider range of gates.
This allowed us to generalise to the $2n$-term shift rule, whereas Schuld et al.'s and Anselmetti et al.'s approach only works for gates that satisfy specific eigenvalue constraints.
Wierichs et al.~\cite{wierichs2022general} obtained their generalised shift rule by expressing the expectation value $\langle H\rangle$ in terms of a discrete Fourier transform (DFT) which is closely related to the discrete sine transform.
This might suggest a possible connection between our diagrammatically obtained system of equations and Wierichs et al.'s DFT reconstruction of the expectation value which would be interesting to investigate in the future.

Another interesting question that showed up at multiples points in our work is the relationship between eigenvalues of a parametrised unitary and the minimum number of parametrised spiders required to implement the unitary in ZX.
To the best of our knowledge, this question has not been investigated before.
We have given a general upper bound, and a lower bound for the special case of eigenvalues $-\lambda$, 0, $\lambda$.
Those bounds were close enough to derive existing parameter shift rules and prove Anselmetti et al.'s conjecture \cite{anselmetti2021local}.
However, it would be interesting to investigate whether there are tighter bounds and if there is a deeper relationship between parametrised spiders and eigenvalues.
Besides being interesting in its own right, this could potentially unify our generalised shift rule with the one given Wierichs et al.~\cite{wierichs2022general}.
Orthogonally, proving lower bounds could lead to general optimality results for shift rules, generalising our no-go theorem to arbitrary parametrised unitaries.

\subsubsection{Barren Plateaus}

We have presented both empirical and analytical methods to detect barren plateaus in ansätze.
Our numerical tool can be used to quickly check a specific combination of ansatz and Hamiltonian for barren plateaus.
While the tool cannot formally \textit{prove} or \textit{disprove} the existence of barren plateaus, it gives a good indication of the behaviour of an ansatz for practically used circuit dimensions.
The potential future use case we envision for software like this is as a part of the QML practitioners' toolbox for evaluating the suitability of ansätze for QML tasks.
For example, exponentially vanishing variance curves like in \Cref{fig:sim-single-layer} could indicate that experimentation with different initialisation strategies might be warranted.
Additionally, the numerical data is very useful for gaining confidence in theoretical results.
In particular, it gave us confidence that the intricate formula we derived for the variance of the multi-layer $\IQP{1}$ ansatz is correct.

While all the experiments in this thesis run relatively quickly, we have some ideas how to further improve the performance of our tool.
For example, we could experiment with decomposing the triangle directly instead of using the representation with four $\frac{\pi}{4}$ spiders.
In that case, we would no longer benefit from the efficient magic state and cat decompositions used in \cite{kissinger2022classical}.
However, we could search for alternative, more efficient decompositions of groups of triangles.
Here, it might also help that the triangles in \Cref{thm:variance} are embedded in a regular, known structure that is possibly easier to decompose.

Moving beyond the methods and looking at the concrete data we obtained, it is surprising that all Sim ansätze we considered seem to already have barren plateaus for a single layer.
The results by McClean et al. \cite{mcclean2018barren} for example only apply if the ansatz has enough layers to approximate a 2-design.
Sim et al. showed that most of their ansätze only gain their expressive power when adding more layers.
But our results indicate that the low expressivity of a single layer already suffices to produce barren plateaus.
The same seems to be true for the IQP circuit used in \texttt{lambeq}~\cite{kartsaklis2021lambeq}.
In fact, the only ansatz we considered that does not have barren plateaus is the very simple $\IQP{1}$.
This might suggest that the expressiveness vs. trainability trade-off described by Holmes et al. \cite{holmes2022connecting} is already significant at fairly low expressive powers with mildly expressible circuits already having poor trainability.
Analysing more circuits in this way is needed to gain a better understanding of this relationship in the future.

The fact that barren plateaus already appear in a single layer actually simplified our analysis.
First, running the numerical experiments for multi-layer circuits is more expensive.\footnote{Note that for the ansätze we considered, adding a qubit adds $O(1)$ parametrised spiders while a new layers adds $O(n)$ new parametrised spiders.}
Secondly, we have already seen for the example of $\IQP{1}$ that the formal analysis of multi-layer ansätze requires significantly more work.
Nonetheless, it would be very interesting to expand on this work and analyse ansätze where the barren plateau phenomenon only appears after adding enough layers.
One example of this would be \textit{non-local} cost functions as discussed in \cite{cerezo2021cost}.

Finally, the biggest limitation of the current ZXW-based barren plateau analysis is the fact that \Cref{thm:variance} only applies if each parameter occurs exactly once in the ZX representation of the ansatz.
In particular, this excludes all anätze that use controlled rotations.
We have discussed a small caveat to this in \Cref{sec:barren-multiple-parameters} where the analysis is possible in certain special cases where the Hamiltonian cancels out additional parameter occurrences.
However, in the general case, such ansätze cannot be handled by our method.
The difficulty of adding support for this differs between our numerical and our analytical approach.
Both would require extending or generalising the integration results by Wang and Yeung~\cite{wang2022differentiating}.
But while the numerical tool could, in principle (barring performance concerns), work with any diagram that represents the variance, in order to prove results by hand we need a diagram that is amenable to manual rewriting and reasoning.
Thus, finding such representations for the gradient variance of circuits with multiple parameter occurrences would serve to significantly generalise the results presented in this work.

\appendix
\chapter{Constructing the Ancilla State} \label{chap:appendix-W}

\def\tikzitpath{chapter4/figs/}

In \Cref{sec:ancilla-recipe}, we discussed a gradient recipe that prepares ancillae in the state $\input{\tikzitpathancilla/state.tikz}$.
In this appendix, we explain how to prepare this state on a quantum device.
First, not that because of the $(\pcy)$ rule, it actually suffices to prepare the state $\input{\tikzitpathancilla/state-2.tikz}$.
As it turns out, this state is an equal superposition of basis states:
\[
\input{\tikzitpathancilla/state-2.tikz}
\eqq{\ref{eqn:green-spider-def}} \input{\tikzitpathancilla/state-3.tikz} ~+~ \input{\tikzitpathancilla/state-4.tikz}
\eqq{\ref{eqn:W2-act}} \input{\tikzitpathancilla/state-5.tikz} ~+~ \input{\tikzitpathancilla/state-6.tikz} ~+~ \input{\tikzitpathancilla/state-7.tikz}
\]
To construct this, we define the following gate:
\begin{definition}
	Let $D(p) := R_Y(2\arccos(\sqrt p))$ for $0 \leq p \leq 1$.
\end{definition}

\begin{lemma} \label{lem:D-split}
	This gate satisfies $D(p) \ket{0} = \sqrt{p}\ket{0} + \sqrt{1-p}\ket{1}$.
\end{lemma}
\begin{proof}
	In general, we have
	\[
	R_Y(\alpha)\ket{0} 
	= \begin{pmatrix}
		\cos(\frac{\alpha}{2}) & -\sin(\frac{\alpha}{2}) \\
		\sin(\frac{\alpha}{2}) & \cos(\frac{\alpha}{2})
	\end{pmatrix} 
	\begin{pmatrix}
		1 \\ 0
	\end{pmatrix}
	=
	\begin{pmatrix}
		\cos(\frac{\alpha}{2}) \\
		\sin(\frac{\alpha}{2})
	\end{pmatrix}
	\]
	Setting $\alpha = 2\arccos(\sqrt p)$, we get
	\[
	D(p) \ket{0}
	= \begin{pmatrix}
		\sqrt p \\
		\sqrt{1-p}
	\end{pmatrix}
	= \sqrt p \ket{0} + \sqrt{1-p}\ket{1}. \qedhere
	\]
\end{proof}

Using $D(p)$, we can construct the state as follows:
\begin{gather*}
	\input{\tikzitpathancilla/calc/1.tikz}
	\eqq{\text{Lem. }\ref{lem:D-split}} \sqrt{\frac{2}{3}}~ \input{\tikzitpathancilla/calc/2-1.tikz} ~+~ \sqrt{\frac{1}{3}}~ \input{\tikzitpathancilla/calc/2-2.tikz} \\[\linesep]
	~=~ \sqrt{\frac{2}{3}}~ \input{\tikzitpathancilla/calc/3-1.tikz} ~+~ \sqrt{\frac{1}{3}}~ \input{\tikzitpathancilla/calc/3-2.tikz} 
	\eqq{\text{Lem. }\ref{lem:D-split}}  \sqrt{\frac{2}{3}}\sqrt{\frac{1}{2}}~ \input{\tikzitpathancilla/calc/4-1.tikz} ~+~ \sqrt{\frac{2}{3}}\sqrt{\frac{1}{2}}~ \input{\tikzitpathancilla/calc/4-2.tikz} ~+~ \sqrt{\frac{1}{3}}~ \input{\tikzitpathancilla/calc/3-2.tikz} \\[\linesep]
	~=~ \sqrt{\frac{1}{3}}\ket{00} + \sqrt{\frac{1}{3}}\ket{01} + \sqrt{\frac{1}{3}}\ket{10}
	~=~ \sqrt{\frac{1}{3}}~ \input{\tikzitpathancilla/state-2.tikz}
\end{gather*}

\chapter{Details on Recursive Contraction}
\label{chap:appendix-contract}

\def\tikzitpath{chapter5/figs/}

Here, we give the full details for the recursive contraction of \Cref{fig:IQP-var-multi}.
For the reader's convenience, we restate the diagram:

\bigskip
\begin{adjustwidth}{-2cm}{-2cm}
	\begin{center}
		$\frac{(-1)^{\sum a_jb_j}}{\sqrt 2^h}~ \input{\tikzitpathIQP/IQP5/var-7.tikz}$
	\end{center}
\end{adjustwidth}

\section{Deriving the Recurrence Relation}

\newcommand{\VE}{\mathit{VE}}
\newcommand{\VO}{\mathit{VO}}

To contract \Cref{fig:IQP-var-multi}, we introduce the notation
\begin{align*}
	k_e &:= k_2 = k_4 = ... = k_\ell = b_1+...+b_n \\
	k_o &:= k_1 = k_3 = ... = k_{\ell-1} = a_1+...+a_n 
\end{align*}
and write $\VE_\ell(k_e,k_o,c_1,...,c_n)$ and $\VO_\ell(k_e,k_o,c_1,...,c_n)$ for the diagram for $\text{Var}\left(\frac{\partial \langle H\rangle}{\partial \theta_i}\right)$ for even and odd $\ell$ respectively, excluding the factor $(-1)^{\sum a_jb_j}$.
Furthermore, we write $\overline x$ for the negation of a Boolean variable, i.e. $\overline 0 = 1$ and $\overline 1 = 0$.

Out goal is to find recursive formulas to compute $\VE_\ell$ and $\VO_\ell$.
For the base case, consider $\VE_0$:
\begin{equation} \label{eqn:VE-base}
	\VE_0(k_e,k_o,c_1,...,c_n)
	~=~ \input{\tikzitpathIQP/IQP5/contract/base.tikz}
	~=~ \begin{cases}
		1 & \text{if } c_1=...=c_n=0 \\
		0 & \text{otherwise.}
	\end{cases} 
\end{equation}
Note that the term becomes zero if the $c_j$ are not all the same.
Hence, from now on we can ignore all terms where this is the case and simplify the notation to $\VO_\ell(k_e,k_o,c)$ and $\VE_\ell(k_e,k_o,c)$.
Now, we proceed recursively:
\begin{itemize}
	\item
	If $i = \ell$ and $\ell$ is odd then
	\begin{align*}
		\VO_\ell(k_e,k_o,c)
		~&=~ \underbrace{\frac{1}{\sqrt 2^h}}_{\substack{\text{Scalar for}\\\text{part up to $\ell$}}} \cdot \underbrace{\frac{1}{\sqrt 2^{\lfloor\ell/2\rfloor}}}_{\substack{\text{Scalar for Had.}\\\text{wires to $L_2,L_4,...$}}} \input{\tikzitpathIQP/IQP5/contract/rec-odd-i-1.tikz} \\[\linesep]
		\eqqa{\ref{eqn:pink-decompose}} \sum_{x\in\{0,1\}} \frac{1}{\sqrt 2^h} \cdot \frac{1}{\sqrt 2^{\lfloor\ell/2\rfloor}} \cdot \frac{1}{2} \cdot (-1)^x ~ \input{\tikzitpathIQP/IQP5/contract/rec-odd-i-2.tikz} \\[\linesep]
		\eqqa{\cc} \sum_{x\in\{0,1\}} \frac{1}{\sqrt 2^h} \cdot \frac{1}{2} \cdot (-1)^x ~ \input{\tikzitpathIQP/IQP5/contract/rec-odd-i-3.tikz} \\[\linesep]
		\eqqa{\sf} \sum_{x\in\{0,1\}} \frac{1}{\sqrt 2^h} \cdot \frac{1}{2} \cdot (-1)^x ~ \input{\tikzitpathIQP/IQP5/contract/rec-odd-i-4.tikz} \\[\linesep]
		~&=~ \begin{cases}
			0 & \text{if } k_o = 0 \\
			\frac{1}{2} (\VE_{\ell-1}(k_e,1,c) - \VE_{\ell-1}(\overline k_e,1,\overline c)) & \text{if } k_o = 1
		\end{cases} \numberthis\label{eqn:VO-i}
	\end{align*}
	
	\item
	If $i = \ell$ and $\ell$ is even then
	\begin{align*}
		\VE_\ell(k_e,k_o,c)
		~&=~ \frac{1}{\sqrt 2^h} \cdot \frac{1}{\sqrt 2^{\ell/2}} ~ \input{\tikzitpathIQP/IQP5/contract/rec-even-i-1.tikz} \\[\linesep]
		\eqqa{\ref{eqn:pink-decompose}} \sum_{x\in\{0,1\}} \frac{1}{\sqrt 2^h} \cdot \frac{1}{\sqrt 2^{\ell/2}} \cdot \frac{1}{2} \cdot (-1)^x ~ \input{\tikzitpathIQP/IQP5/contract/rec-even-i-2.tikz} \\[\linesep]
		\eqqa{\cc,\cp} \sum_{x\in\{0,1\}} \frac{1}{\sqrt 2^h} \cdot \frac{1}{2} \cdot (-1)^x ~ \input{\tikzitpathIQP/IQP5/contract/rec-even-i-3.tikz} \\[\linesep]
		~&=~ \begin{cases}
			0 & \text{if } k_e = 0 \\
			\frac{1}{2}(\VO_{\ell-1}(1,k_o,c) - \VO_{\ell-1}(1,\overline k_o,c)) & \text{if } k_e = 1
		\end{cases} \numberthis\label{eqn:VE-i}
	\end{align*}
	
	\item
	If $i \neq \ell$ and $\ell$ is odd then
	\begin{align*}
		\VO_\ell(k_e,k_o,c)
		~&=~ \frac{1}{\sqrt 2^h} \cdot \frac{1}{\sqrt 2^{\lfloor\ell/2\rfloor}} ~ \input{\tikzitpathIQP/IQP5/contract/rec-odd-x-1.tikz} \\[\linesep]
		\eqqa{\ref{eqn:pink-decompose}} \sum_{x\in\{0,1\}} \frac{1}{\sqrt 2^h} \cdot \frac{1}{\sqrt 2^{\lfloor\ell/2\rfloor}} \cdot \frac{1}{2} ~ \input{\tikzitpathIQP/IQP5/contract/rec-odd-x-2.tikz} \\[\linesep]
		\eqqa{\cc,\sf} \sum_{x\in\{0,1\}} \frac{1}{\sqrt 2^h} \cdot \frac{1}{2} ~ \input{\tikzitpathIQP/IQP5/contract/rec-odd-x-3.tikz} \\[\linesep]
		~&=~ \begin{cases}
			\VE_{\ell-1}(k_e,0,c) & \text{if } k_o = 0 \\
			\frac{1}{2} (\VE_{\ell-1}(k_e,1,c) - \VE_{\ell-1}(\overline k_e,1,\overline c)) & \text{if } k_o = 1
		\end{cases} \numberthis\label{eqn:VO-x}
	\end{align*}
	
	\item
	If $i \neq \ell$ and $\ell$ is even then
	\begin{align*}
		\VE_\ell(k_e,k_o,c)
		~&=~ \frac{1}{\sqrt 2^h} \cdot \frac{1}{\sqrt 2^{\ell/2}} ~ \input{\tikzitpathIQP/IQP5/contract/rec-even-x-1.tikz} \\[\linesep]
		\eqqa{\ref{eqn:pink-decompose}} \sum_{x\in\{0,1\}} \frac{1}{\sqrt 2^h} \cdot \frac{1}{\sqrt 2^{\ell/2}} \cdot \frac{1}{2} ~ \input{\tikzitpathIQP/IQP5/contract/rec-even-x-2.tikz} \\[\linesep]
		\eqqa{\cc,\cp} \sum_{x\in\{0,1\}} \frac{1}{\sqrt 2^h} \cdot \frac{1}{2} ~ \input{\tikzitpathIQP/IQP5/contract/rec-even-x-3.tikz} \\[\linesep]
		~&=~ \begin{cases}
			\VO_{\ell-1}(0,k_o,c) & \text{if } k_e = 0 \\
			\frac{1}{2} (\VO_{\ell-1}(1,k_o,c) - \VO_{\ell-1}(1,\overline k_o,c)) & \text{if } k_e = 1
		\end{cases} \numberthis\label{eqn:VE-x}
	\end{align*}
\end{itemize}
This yields an recursive algorithm for computing $\text{Var}\left(\frac{\partial \langle H\rangle}{\partial \theta_i}\right)$.
One interesting thing to note is that
\begin{gather*}
	\VO_{\ell}(0,0,c) 
	\eqq{\ref{eqn:VO-x}} \VE_{\ell-1}(0,0,c) 
	\eqq{\ref{eqn:VE-x}} \VO_{\ell-2}(0,0,c) 
	= ... \\
	= \begin{cases}
		\VO_{i}(0,0,c) & \text{if $i$ is odd} \\
		\VE_{i}(0,0,c) & \text{if $i$ is even} 
	\end{cases}
	\eqq{\ref{eqn:VO-i},\ref{eqn:VE-i}} 0.
	\numberthis\label{eqn:VO-VE-00}
\end{gather*}
Furthermore, for $i \neq \ell,\ell-1$ we can derive
\begin{align}
	\VO_{\ell}(0,0,c) \eqqa{\ref{eqn:VO-VE-00}} \VO_{\ell-2}(0,0,c) \label{eqn:VO-00} \\
	\VO_{\ell}(0,1,c) \eqqa{\ref{eqn:VO-x},\ref{eqn:VE-x}} \frac{1}{4}(2\VO_{\ell-2}(0,1,c) + \VO_{\ell-2}(1,0,\overline c) - \VO_{\ell-2}(1,1,\overline c)) \label{eqn:VO-01} \\
	\VO_{\ell}(1,0,c) \eqqa{\ref{eqn:VO-x},\ref{eqn:VE-x}}\frac{1}{2}(\VO_{\ell-2}(1,0,c) - \VO_{\ell-2}(1,1,c)) \label{eqn:VO-10} \\
	\VO_{\ell}(1,1,c) \eqqa{\ref{eqn:VO-x},\ref{eqn:VE-x}} \frac{1}{4}(-2\VO_{\ell-2}(0,1,\overline c) - \VO_{\ell-2}(1,0,c) + \VO_{\ell-2}(1,1,c)) \label{eqn:VO-11}
\end{align}
%\begin{align*}
%	\VO_{\ell}(k_e,k_o,c)
%	&= \begin{cases}
%		\VO_{\ell-2}(0,0,c) & \text{if } k_e=0 \text{ and } k_o=0 \\
%		\frac{1}{2}(\VO_{\ell-2}(0,1,c) - \frac{1}{2}(\VO_{\ell-2}(1,1,\overline c) - \VO_{\ell-2}(1,0,\overline c))) & \text{if } k_e=0 \text{ and } k_o=1 \\
%		\frac{1}{2}(\VO_{\ell-2}(1,0,c) - \VO_{\ell-1}(1,1,c)) & \text{if } k_e=1 \text{ and } k_o=0 \\
%		\frac{1}{2}(\frac{1}{2}(\VO_{\ell-2}(1,1,c) - \VO_{\ell-2}(1,0,c)) - \VO_{\ell-2}(0,1,\overline c)) & \text{if } k_e=1 \text{ and } k_o=1
%	\end{cases} \\
%	&= \begin{cases}
%		\VO_{\ell-2}(0,0,c) & \text{if } k_e=0 \text{ and } k_o=0 \\
%		\frac{1}{4}(2\VO_{\ell-2}(0,1,c) + \VO_{\ell-2}(1,0,\overline c) - \VO_{\ell-2}(1,1,\overline c)) & \text{if } k_e=0 \text{ and } k_o=1 \\
%		\frac{1}{2}(\VO_{\ell-2}(1,0,c) - \VO_{\ell-2}(1,1,c)) & \text{if } k_e=1 \text{ and } k_o=0 \\
%		\frac{1}{4}(-2\VO_{\ell-2}(0,1,\overline c) - \VO_{\ell-2}(1,0,c) + \VO_{\ell-2}(1,1,c)) & \text{if } k_e=1 \text{ and } k_o=1
%	\end{cases} \\
%\end{align*}

\section{Solving the Recurrence Relation}

When considering concrete values of $i$, we can derive closed-form solutions for $\text{Var}\left(\frac{\partial \langle H\rangle}{\partial \theta_i}\right)$.
We discuss the case $i = 1$.
First, note that
\begin{align*}
	\VO_1(k_e,k_o,c)
	\eqqa{\ref{eqn:VE-i}} \begin{cases}
		0 & \text{if } k_o = 0 \\
		\frac{1}{2} (\VE_0(k_e,0,c) - \VE_0(\overline k_e,0,\overline c)) & \text{if } k_o = 1
	\end{cases} \\
	\eqqa{\ref{eqn:VE-base}} \begin{cases}
		0 & \text{if } k_o = 0 \\
		\frac{1}{2} \cdot (-1)^c & \text{if } k_o = 1
	\end{cases} \numberthis\label{eqn:VO-base}
\end{align*}
As a consequence, we have
\begin{align}
	\VE_\ell(k_e,k_o,\overline c) &= -\VE_\ell(k_e,k_o,c) \label{eqn:VE-c-neg} \\ 
	\VO_\ell(k_e,k_o,\overline c) &= -\VO_\ell(k_e,k_o,c). \label{eqn:VO-c-neg}
\end{align}
Therefore,
\begin{align*}
	\VO_{\ell}(1,1,c)
	\eqqa{\ref{eqn:VO-11}} \frac{1}{4}(-2\VO_{\ell-2}(0,1,\overline c) - \VO_{\ell-2}(1,0,c) + \VO_{\ell-2}(1,1,c)) \\
	\eqqa{\ref{eqn:VO-c-neg}} \frac{1}{4}(2\VO_{\ell-2}(0,1,c) + \VO_{\ell-2}(1,0,\overline c) - \VO_{\ell-2}(1,1,\overline c)) \\
	\eqqa{\ref{eqn:VO-01}} \VO_{\ell}(0,1) \numberthis\label{eqn:VO-11=VO-01}
	\\[10pt]
	\VO_{\ell}(0,1,c) \eqqa{\ref{eqn:VO-01},\ref{eqn:VO-11=VO-01}} \frac{1}{4}(3\VO_{\ell-2}(0,1,c) - \VO_{\ell-2}(1,0,c)) \numberthis\label{eqn:VO-01'} \\
	\VO_{\ell}(1,0,c) \eqqa{\ref{eqn:VO-01},\ref{eqn:VO-11=VO-01}} \frac{1}{2}(\VO_{\ell-2}(1,0,c) - \VO_{\ell-2}(0,1,c)). \numberthis\label{eqn:VO-10'}
\end{align*}
Now, we just need to derive a closed form for this recurrence relation:
\begin{lemma} \label{lem:IQP-recurrence-closed}
	$\VO$ has the following closed-form representation:
	\[ \VO_{2l+1}(0,1,c) = (-1)^c \cdot \frac{2\cdot 4^l + 1}{6\cdot 4^l} \qquad\qquad \VO_{2l+1}(1,0,c) = (-1)^{1-c} \cdot \frac{4^l-1}{3\cdot 4^l} \]
\end{lemma}
\begin{proof}
	By induction on $l$:
	\begin{itemize}
		\item We have $\VO_1(0,1,c) \eqq{\ref{eqn:VO-base}} (-1)^c \cdot \frac{1}{2} = (-1)^c \cdot \frac{2\cdot 4^0 + 1}{6\cdot 4^0}$ and $\VO_1(1,0,c) \eqq{\ref{eqn:VO-base}} 0 = (-1)^{1-c} \cdot \frac{4^0-1}{3\cdot 4^0}$.
		
		\item
		We have
		\begin{align*}
			\VO_{2l+3}(0,1,c) 
			\eqqa{\ref{eqn:VO-01'}} \frac{1}{4}\left( 3\VO_{2l+1}(0,1,c) - \VO_{2l+1}(1,0,c) \right) \\
			\eqqa{\text{IH}} \frac{1}{4}\left( 3\cdot(-1)^c\cdot\frac{2\cdot 4^l + 1}{6\cdot 4^l} - (-1)^{1-c} \cdot \frac{4^l-1}{3\cdot 4^l} \right) \\
			&= (-1)^c \cdot \frac{1}{4} \left( \frac{2\cdot 4^l + 1}{6\cdot 4^l} + \frac{4^l-1}{3\cdot 4^l} \right) \\
			&= (-1)^c \cdot \frac{1}{4} \cdot \frac{8\cdot 4^l-1}{6\dot 4^l} \\
			&= (-1)^c \cdot \frac{2\cdot 4^{l+1} + 1}{6\cdot 4^{l+1}} \\
			\\
			\VO_{2l+3}(1,0,c)
			\eqqa{\ref{eqn:VO-10}} \frac{1}{2} (\VO_{2l+1}(1,0,c) - \VO_{2l+1}(0,1,c)) \\
			\eqqa{\text{IH}} \frac{1}{2} \left( (-1)^{1-c} \cdot \frac{4^l-1}{3\cdot 4^l} - (-1)^c \cdot \frac{2\cdot 4^l + 1}{6\cdot 4^l} \right) \\
			&= (-1)^{1-c} \cdot \frac{1}{2} \left( \frac{4^l-1}{3\cdot 4^l} + \frac{2\cdot 4^l + 1}{6\cdot 4^l} \right) \\
			&= (-1)^{1-c} \cdot \frac{1}{2} \cdot \frac{4 \cdot 4^l -1}{6 \cdot 4^l} \\
			&= (-1)^{1-c} \cdot \frac{4^{l+1}-1}{3\cdot 4^{l+1}} \qedhere
		\end{align*}
	\end{itemize}
\end{proof}

\begin{corollary} \label{corr:lem:IQP-recurrence-closed'}
	$\VE$ has the following closed-form representation:
	\begin{gather*}
		\VE_{2l}(0,1,c) \eqq{\ref{eqn:VE-x}} \VO_{2l-1}(0,1,c) \eqq{\text{Lem. }\ref{lem:IQP-recurrence-closed}} (-1)^c \cdot \frac{2\cdot 4^{l-1} + 1}{6\cdot 4^{l-1}} 
		\\
		\VE_{2l}(1,0,c) \eqq{\ref{eqn:VO-x}} \VO_{2l+1}(1,0,c) \eqq{\text{Lem. }\ref{lem:IQP-recurrence-closed}} (-1)^{1-c} \cdot \frac{4^l-1}{3\cdot 4^l}
	\end{gather*}
	\begin{align*}
		\VE_{2l}(1,1,c) 
		\eqqa{\ref{eqn:VE-x}} \frac{1}{2}(\VO_{2l-1}(0,1,c) - \VO_{2l-1}(1,0,c)) \\
		\eqqa{\text{Lem.} \ref{lem:IQP-recurrence-closed}} \frac{1}{2}(-1)^c \left(\frac{2\cdot 4^{l-1}+1}{6\cdot 4^{l-1}} + \frac{4^{l-1}-1}{3 \cdot 4^{l-1}}  \right) \\
		&= (-1)^c \frac{4^l - 1}{3 \cdot 4^l}
	\end{align*}
\end{corollary}

Now, recalling the definition of $k_e$, $k_o$, and $c$, we finally have
\begin{theorem} \label{thm:IQP-var-formula}
	We have
	\[
		\text{Var}\left(\frac{\partial \langle H\rangle}{\partial \theta_1}\right)
		= \begin{cases}
			(-1)^{\sum a_jb_j} \cdot \VE_{\ell}(\sum a_j,\sum b_j, a_1) & \text{if $\ell$ is even and } a_1=...=a_n \\
			(-1)^{\sum a_jb_j} \cdot \VO_{\ell}(\sum b_j,\sum a_j, b_1) & \text{if $\ell$ is odd and } b_1=...=b_n \\
			0 & \text{otherwise.}
		\end{cases}
	\]
\end{theorem}

\begin{corollary} \label{corr:IQP-var-converge}
	Either $\text{Var}\left(\frac{\partial \langle H\rangle}{\partial \theta_1}\right) = 0$ or $\text{Var}\left(\frac{\partial \langle H\rangle}{\partial \theta_1}\right) \to \frac{1}{3}$ for $\ell \to \infty$.
\end{corollary}
\begin{proof}
	This essentially follows from the fact that all terms in \Cref{lem:IQP-recurrence-closed} and \Cref{corr:lem:IQP-recurrence-closed'} converge to $\pm\frac{1}{3}$.
	To be precise, we can show that the negation always cancels out by considering the different cases:
	Suppose $\ell$ is odd and $b_1 = ... = b_n = 0$.
	Then 
	\[ 
	\text{Var}\left(\frac{\partial \langle H\rangle}{\partial \theta_1}\right) 
	= \VO_\ell(0,\sum a_j,0) 
	\eqq{\ref{eqn:VO-00},\text{Lem. }\ref{lem:IQP-recurrence-closed}} \begin{cases}
		0 & \text{if } \sum a_j = 0 \\
		 \frac{2\cdot 4^{\lfloor \ell/2\rfloor} + 1}{6\cdot 4^{\lfloor \ell/2\rfloor}} \to \frac{1}{3} & \text{if } \sum a_j = 1
	\end{cases} 
	\]
	If $b_1=...=b_n=1$, then we have $\sum b_j = 1$ since we assume that $n$ is odd.
	Thus, 
	\[
	\text{Var}\left(\frac{\partial \langle H\rangle}{\partial \theta_1}\right) 
	= \VO_\ell(1,\sum a_j,1)
	\eqq{\ref{eqn:VO-c-neg},\text{Lem. }\ref{lem:IQP-recurrence-closed}} \begin{cases}
		\frac{4^{\lfloor \ell/2\rfloor}-1}{3\cdot 4^{\lfloor \ell/2\rfloor}} \to \frac{1}{3} & \text{if } \sum a_j = 0 \\
		\frac{2\cdot 4^{\lfloor \ell/2\rfloor} + 1}{6\cdot 4^{\lfloor \ell/2\rfloor}} \to \frac{1}{3} & \text{if } \sum a_j = 1
	\end{cases}
	\]
	We do not get a negation in the second case since $\sum a_j = 1$ implies that $(-1)^{\sum a_jb_j} = 1$.
	The case for even $\ell$ and $a_1 = ... = a_n$ is symmetric.
	Otherwise, $\text{Var}\left(\frac{\partial \langle H\rangle}{\partial \theta_1}\right) = 0$.
\end{proof}

\chapter{Additional Lemmas and Proofs} \label{chap:appendix-proofs}

\def\tikzitpath{chapter-appendix/figs/}

\begin{lemma}
	For all $x \in \{0, 1\}$, we have
	\begin{equation} \label{eq:pi-copy}
		\input{\tikzitpathlem:pi-copy/statement-1.tikz} ~=~ \frac{1}{\sqrt 2}~ \input{\tikzitpathlem:pi-copy/statement-2.tikz}
	\end{equation}
\end{lemma}
\begin{proof}
	\[
	\input{\tikzitpathlem:pi-copy/statement-1.tikz}
	\eqq{\sf,\pi} \input{\tikzitpathlem:pi-copy/proof-1.tikz}
	\eqq{\sc} \input{\tikzitpathlem:pi-copy/proof-2.tikz}
	\eqq{\cc} \frac{1}{\sqrt 2}~ \input{\tikzitpathlem:pi-copy/statement-2.tikz} \qedhere
	\]
\end{proof}

\begin{lemma}
	\begin{equation} \label{eqn:hopf-had}
		\input{\tikzitpathlem:hopf-had/statement-1.tikz} ~=~ 2~ \input{\tikzitpathlem:hopf-had/statement-2.tikz}
	\end{equation}
\end{lemma}
\begin{proof}
	\begin{gather*}
	\input{\tikzitpathlem:hopf-had/statement-1.tikz}
	\eqq{\sf} \input{\tikzitpathlem:hopf-had/proof-1.tikz}
	\eqq{\cc} \sqrt{2}~ \input{\tikzitpathlem:hopf-had/proof-2.tikz} \\[\linesep]
	\eqq{\ho} \sqrt{2}~ \input{\tikzitpathlem:hopf-had/proof-3.tikz}
	\eqq{\cc,\sf} 2~ \input{\tikzitpathlem:hopf-had/statement-2.tikz} \qedhere
	\end{gather*}
\end{proof}

\printProofs

%\addtocontents{toc}{\vspace{5mm}}
%
%\cleardoublepage\include{chapterAppendix}
%

\renewcommand{\emph}{\textit}
\bibliographystyle{unsrt}
\bibliography{bibliography}

\end{document}